\documentclass[aps,prb,amsmath,twocolumn,amssymb,titlepage,superscriptaddress,showpacs,floatfix]{revtex4-1}
\usepackage[english]{babel}
\usepackage{graphicx}
\usepackage{subcaption}
\usepackage{cleveref}
\usepackage{transparent}
\usepackage{amsmath}
\usepackage{amssymb}
\usepackage{epstopdf}
\usepackage{amsfonts}
\usepackage{bm}
\usepackage{mathdots}
\usepackage{times}
\usepackage{pdfpages}
\usepackage{mathtools}
\usepackage{stmaryrd}
\usepackage{tabularx}
\usepackage{hhline}
\usepackage{lipsum}
\usepackage{empheq}

\allowdisplaybreaks

\makeatletter
\AtBeginDocument{\let\LS@rot\@undefined}
\makeatother

\newcommand*{\citen}[1]{%
  \begingroup
    \romannumeral-`\x % remove space at the beginning of \setcitestyle
    \setcitestyle{numbers}%
    \cite{#1}%
  \endgroup   
}

\usepackage[
   justification=raggedright,
   format=plain]{caption}
\newcommand{\ave}[1]{\langle #1 \rangle}
\newcommand{\bolds}[1]{\boldsymbol #1}

\begin{document}
\title{SO(4) FLEX+DMFT formalism with SU(2)$\otimes$SU(2)-symmetric impurity solver for superconductivity in the repulsive Hubbard model}
\author{Sharareh Sayyad}
\affiliation{Institute for Solid State Physics, University of Tokyo, Kashiwanoha,
Kashiwa, 277-8581 Chiba, Japan}
\author{Naoto Tsuji}
\affiliation{RIKEN Center for Emergent Matter Science (CEMS), Wako 351-0198, Japan}
\author{Massimo Capone}
\affiliation{International School for Advanced Studies (SISSA), Via Bonomea 265, I-34136 Trieste, Italy}
\affiliation{CNR-IOM Democritos, Via Bonomea 265, I-34136 Trieste, Italy}
\author{Hideo Aoki}
\affiliation{National Institute of Advanced Industrial Science and Technology (AIST), Tsukuba 305-8568, Japan}
\affiliation{Department of Physics, University of Tokyo, Tokyo 113-0033, Japan}

\pacs{71.27.+a,05.10.-a,02.70.-c}

\begin{abstract}
Here we have developed a FLEX+DMFT formalism, where the symmetry properties of the system are incorporated by 
constructing a SO(4) 
generalization of the conventional fluctuation-exchange 
approximation (FLEX) coupled self-consistently to the 
dynamical mean-field theory (DMFT). Along with this line, we emphasize that the SO(4) symmetry 
is the lowest group-symmetry that enables
us to investigate superconductivity and antiferromagnetism on an equal footing. 
We have imposed this by 
decomposing the electron operator into 
auxiliary fermionic and slave-boson constituents 
that respect SU(2)$_{\rm spin}\otimes$SU(2)$_{\eta{\rm spin}}$.  This is used not in a mean-field 
treatment as in the usual slave-boson formalisms, but instead 
in the DMFT impurity solver 
with an SU(2)$_{\rm spin}\otimes$SU(2)$_{\eta{\rm spin}}$ hybridization function to
incorporate the FLEX-generated bath information into DMFT iterations.  
While there have been attempts such as the doublon-less SU(2) slave-boson formalism, 
the present ``full-SU(2)" slave-boson 
formalism is expected to provide a new platform 
for addressing the underlying
physics for various quantum orders, which compete with each other 
and can coexist.

\end{abstract}
\maketitle

\section{Introduction}\label{sec:whySB}

It has been recognized that several high-temperature superconductor families 
share a universal property of the phase diagram with superconductivity 
in proximity to magnetic 
phases~\cite{Dai2015,Keimer2015,Stewart2017,Capone2009}. 
Hence a theoretical formalism that treats the spin-fluctuation-mediated 
pairing and 
magnetic or other phases  arising from strong repulsive interactions 
on an equal footing is highly desired. 
In the hole-doped cuprates, the phase diagram features spin- and charge-density waves, 
the pseudogap (PG) region, and the strange metal 
on top of the $d$-wave superconductivity.  
Although there exist various materials with different crystal 
structures within the cuprate family, the crucial ingredient 
in the whole family is the two-dimensional $\rm CuO_{2}$ plane~\cite{Tsukada2005,Wesche2015}.  
The three-band model for the copper oxide is usually 
simplified into a two dimensional square lattice, which is widely believed to be the building block from which high-temperature superconductivity originates~\cite{Schrieffer2007}. 
Countless theoretical investigations have addressed various experimental observations, 
specifically the structure of gap-functions in both superconducting phases~\cite{Mesot1999,Fettes1999,Kitatani2015}, pseudogap regions~\cite{Ping2002,Matsuura2017,Gull2013,Wu2018}, and the antiferromagnetic phase~\cite{Schrieffer2007,Krahl2009,Kuroki1999}. Many-body numerical algorithms including, among others, (i) extensions of the mean-field approximation~\cite{Krull2016,Sentef2017,Potthoff2003}, (ii) various generalizations of the dynamical mean-field theory~(DMFT)~\cite{Gull2013,Kitatani2015,Jarrell2001,Maier2002,Staar2013,Ayral2013,Capone2006,Kotliar2001}, 
(iii) diagrammatic extensions~\cite{Rubtsov2008,Vucicevic2017,Kitatani2018}, 
and (iv) quantum Monte Carlo (QMC) methods~\cite{Tahara2008,Kuroki1996,Ying2014}. These 
approaches mostly involve various approximations, but there is no general consensus about the capability of the present approaches to investigate the multiple phases on an equal footing. 
This is imperative, 
since the glue for the pair formation as well as the origin of pseudogap phase should emerge by treating superconducting (SC) and antiferromagnetic (AF) phases in a unified framework.    

Along with the advances of the above-mentioned numerical toolboxes, 
the phenomenologies approaches in terms of competing and/or synergistic order parameters
based on generalizations of the Ginzburg-Landau~(GL) theory have been 
developed for providing insights into various phases. 
Such approaches introduce the 
order parameters of the system based on symmetry considerations~\cite{Zacher2000,Sun2005,Sun2008,Demler2004,Zhang1997,Podolsky2004}. 
In the GL theory, the respected symmetries of the relevant 
order parameters indeed govern the diversity of the associated phase diagram.
As a result, for systems whose phase diagram is largely known from experiments,
a reverse strategy is to find a large symmetry group which will be reduced to one of its subgroups upon the emergence of distinct phases. It is also crucial to terminate these subgroups such that all the smallest subgroups obey the conservation laws, namely conservation of charge~($\rm U(1)$) and spin~(SU(2)$_{\rm s}$). 
It has been discussed that for families of high-temperature superconductors, the generators of the supergroup should be determined such that p- and d-wave superconductivity, staggered magnetization, charge density waves, spin, charge, and number operators can be expressed in their terms~\cite{Wu2001}.  
For this reason, it has been argued that evoked symmetries can be classified~\cite{Wu2003,Wu2001} 
in terms of subgroups of SU(4) as
\begin{empheq}[left={{\rm SU(4)} \supset \empheqlbrace}]{alignat*=2}
    &{\rm SO(4)}\otimes {\rm U(1)} &\quad  \supset  {\rm SU(2)}_{\rm s} \otimes U(1) ,\\
    &{\rm SO(5)}  &\quad \supset  {\rm SU(2)}_{\rm s} \otimes U(1)  ,\\
    &{\rm SU(2)}_{\rm s} \otimes {\rm SU(2)}_{\eta} &\quad \supset  {\rm SU(2)}_{\rm s} \otimes U(1) .
\end{empheq}
Here, ${\rm SO(4)}\otimes {\rm U(1)}$ represents the AF order~($\rm SO(4)$) generated by the total and relative spin operators on even and odd sites, along with ${\rm U(1)}$ the charge group. SU(2)$_{\eta}$ is the second SU(2) symmetry inherent 
in the Hubbard model involving what is called $\eta$ spin~\cite{Yang1989}.  Indeed, 
${\rm SU(2)}_{\rm s}  \otimes  {\rm SU(2)}_{\eta}$ introduces an algebra 
for describing the d-wave SC order as well as the AF order~\cite{Sachdev2009, Chatterjee2017, Sachdev2019, Wu2018}. (We briefly recall the symmetries associate with the different subgroups. More details will be given in the following.)
The well-known SO(5) theory~\cite{Zhang1997,Demler2004} 
is one of the prominent 
examples, where the two-dimensional superconducting gap function and the three-component antiferromagnetic order constitute the generators. Introducing the primary 
orders in any subgroups of the SU(4) theory, one can interpret and predict the emergence of different orders. Although this approach can elegantly explain various aspects of the phase diagram as far as the Ginzburg-Landau~(GL) type phenomenologies are concerned, a microscopic theory is necessary to give a solid basis to the GL approach and to provide information about the origin of the order parameters. 

Thus it is crucial to combine a microscopic approach with 
a symmetry-group theoretic approach.  
One class of methods which is suitable to achieve this goal is formed by the slave-particle 
methods~\cite{Kim2007,Hermele2007,Lechermann2007,Lee2006,Lee2001,Wen1996,Medici2005,Ruegg2010,Yu2012,Medici2017}. 
In this framework, an electron operator is regarded to be comprising 
various auxiliary particles, where each component conveys a 
specific symmetry property of the original electron. 
As this procedure enlarges the Hilbert space, the factorized particles 
are subject to a constraint for suppressing unphysical states. 
Under such constraints, the method is dubbed the 
{\it slave-particle decomposition}. 

A caution in advance: usual slave-particle works 
primarily use mean-field approaches. In the present work, 
by contrast, 
we use the slave-particles in the impurity solver for 
the FLEX-DMFT framework as we shall elaborate.  
Having said that, 
the slave building blocks so far considered are typically SU(2) slave-rotor~\cite{Kim2007,Hermele2007}, or SU(2) doublon-less slave-boson~\cite{Lee2006,Lee2001,Wen1996}, can then shed light on the origin of different order parameters, where each of the order parameters involves 
one of the introduced auxiliary particles.
One should note that in these approaches
it is a challenge to characterize the nature of order parameters
which are not necessarily associated with one particular auxiliary particle. 
While the computational cost will obviously blow up as one 
increases the number of involved slave-particles, 
an appropriate choice may control the added cost.
Of particular interest is the doublon-less SU(2) slave-boson decomposition, which was initially introduced to treat the $t$-$J$ model under the Gutzwiller projection that eliminates doubly-occupied states. 
This decomposition respects the rotational symmetry 
of real spins (which we denote SU(2)$_{\rm s}$). 
When one performs the mean-field treatment for this model, 
in the doublon-less SU(2) slave-boson picture, the  
solution shows that the pseudogap and superconducting transition temperatures can be attributed to the auxiliary spin-less and charge-less particles, respectively~\cite{Lee2001,Wen1996}. Align with the slave-particle approaches, SU(2) gauge theory of fluatuating antiferromagnetism~\cite{Sachdev2009, Chatterjee2017, Sachdev2019, Wu2018} attempts to address the underlying physics of by fractionalizing the physical operators into spin-less and charge-less particles with SU(2) symmetry which can be spontanously broken by the condensation of the Higgs field.  

However, while the results, obtained using the SU(2) gauge theory or the slave-particle methods, give a satisfactory picture of the various competing phases, 
these have been obtained within the mean-field treatment, which has to 
be examined if we are seriously interested in the correlation 
physics.  For instance, one conspicuous feature in the phase diagram 
for the SU(2) doublon-less slave-boson is the superconducting Tc dome 
that is 
entirely covered by the pseudogap regime~\cite{Lee2001}, which 
agrees with some experimental pieces of evidence~\cite{Taillefer2010,Kordyuk2015}, 
but it contrasts with other experiments~\cite{Ramshaw2015}. 
Thus this makes us to question if the result would be an artifact of 
mean-field approximations, and whether a more appropriate treatment of 
the electron correlation should be required for the d-wave SC 
and other orders. 
This has precisely motivated us to propose in the present paper 
a new approach, in which we introduce 
two concepts: (i) We first improve the slave-particle decomposition itself to allow double occupancies to study moderate repulsive interactions 
away from the strong-coupling limit, 
and (ii) we then apply the formalism not to  a mean-field treatment, 
but to the FLEX+DMFT algorithm, 
which combines the fluctuation-exchange approximation~(FLEX) and the dynamical mean-field theory~(DMFT). We opt here for the FLEX+DMFT formalism~\cite{Gukelberger2015, Kitatani2015},
where 
FLEX can treat the 
momentum-dependent pairing interaction for d-wave SC, while DMFT can treat the Mott transition, thereby allowing 
us to incorporate both spatial (FLEX) and dynamical (DMFT) fluctuations.

Specifically, inspired by the phenomenological SU(4) theory of superconductivity~\cite{Sun2005,Sun2008,Wu2001}, here we extend the slave-boson decomposition to capture 
\begin{itemize}
\item[(i)] the bipartite nature of the AF and d-wave SC phase, along with 
\item[(ii)] SU(2)$_{\rm s} \bigotimes$ SU(2)$_{\eta}$ symmetry [section(\ref{sec:Ham}) below].
\end{itemize}
We call the formalism a ``bipartite full-SU(2) slave-boson", 
which enables us to investigate AF and d-wave SC in the Hubbard model on an equal footing.  
Our approach substantially improves a previous mean-field work~\cite{Sentef2017} which indeed invoked the SO(4) symmetry. 
While the present bipartite treatment maintains the SU(2)$_{\rm s}$ and SU(2)$_{\eta}$ on a bipartite lattice as in Ref.~\citen{Hermele2007}, the latter is a mean-field treatment.  
Another difference is, instead of the O(4) rotor as the slave-particle 
employed in Ref.~\citen{Hermele2007}, we introduce here two species of bosons to convey charge-related information. 

The organization of this paper is as follows. In Sec.~\ref{sec:Ham} we introduce the model Hamiltonian and relevant symmetry properties. Section~\ref{sec:numerics} starts with constructing a new SO(4) slave-boson formalism 
for Green's function and other quantities 
that respects full-SU(2) symmetry and the 
bipartite structure. We then present the SO(4) DMFT+FLEX formalism by explaining the SO(4) DMFT method, followed by introducing the SO(4) FLEX. We shall then explain how the DMFT self-consistency loop is performed in the slave-particle space for the DMFT 
impurity solver. 
We conclude the paper in Sec.~\ref{sec:conclusion}.

\section{Model Hamiltonian}\label{sec:Ham}
The one-band Hubbard Hamiltonian on the square lattice is given by
\begin{align}
 H =
-
&\sum\limits_{i,j,\sigma}
\Big(
v_{|i-j|}
c^{\dagger}_{i\sigma}
c_{j\sigma}
+ {\rm h.c.}
\Big)
-\mu \sum\limits_{i,\sigma} n_{i\sigma} \nonumber \\
+
&\sum\limits_{i} U
\Big( n_{i\uparrow}-\frac{1}{2} \Big)
\Big( n_{i\downarrow}-\frac{1}{2} \Big)
,\label{eq:Hubbard}
\end{align}
where $c^{\dagger}_{i \sigma}$ creates an electron with spin $\sigma$ at site $i$, and $n_{i\sigma}= c^{\dagger}_{i\sigma} c_{i \sigma}$. 
The interaction~($U$) describes the on-site Coulomb repulsion between 
electrons with opposite spins. The density of electrons controlled by 
the doping is specified by the chemical potential~($\mu$). 
Electrons hop from site $i$ to $j$ on the lattice with the amplitude $v_{|i-j|}$, here taken into account up to the third-neighbors. 
We take the nearest-neighbor $v_{1}$ as the unit of energy hereafter. 
 
This model has not only the usual SU(2) spin-rotational 
symmetry, but another SU(2) symmetry.  
For the latter, we can define a pseudospin (often called 
$\eta$-spin), 
\begin{align}
    &{\boldsymbol \eta}_{i} = (\eta^x_{i}, \eta^y_{i}, \eta^z_{i}) \nonumber \\
&=\frac{1}{2} \left( 
\epsilon_{i} c^{\dagger}_{i \downarrow } c^{\dagger}_{i \uparrow }
   +\epsilon_{i} c_{ i \uparrow } c_{i \downarrow }, 
i \epsilon_{i} c^{\dagger}_{i \downarrow }c^{\dagger}_{i \uparrow }
   -i \epsilon_{i} c_{ i \uparrow } c_{i \downarrow }, 
1- n_{i \uparrow} - n_{i \downarrow}
   \right), \nonumber \\
%  & {\boldsymbol T}^{2}_{i} = \big( T^{z} \big)^{2} + \frac{1}{2} \big( T^{+}T^{-}+ T^{-}T^{+} \big), \\
  & \eta^{+}_{i}= \eta^{x}_{i} + i \eta^{y}_{i}= \epsilon_{i} c_{ i \uparrow }c_{i \downarrow },
   \nonumber \\
  & \eta^{-}_{i}= \eta^{x}_{i} - i \eta^{y}_{i}= \epsilon_{i} c^{\dagger}_{i \downarrow }c^{\dagger}_{i \uparrow },
 \end{align}
where $\epsilon_{i}$, the bipartite factor, is 
defined on two (A,B) sublattices as
\begin{equation}
 \epsilon_{i}= 
 \begin{cases}
  1 \quad i\in {\rm A},\\
  -1 \quad i\in {\rm B}.
 \end{cases}
\end{equation}
If we rewrite the Hubbard interaction as
 \begin{align}
H_{\rm U} = &U \sum\limits_{i}  \left( n_{i \downarrow}
 -\frac{1}{2} \right)\left( n_{i \uparrow} -\frac{1}{2} \right)= \nonumber \\
 = & 
\sum\limits_{i}  
\frac{2U}{3} \Big[ \frac{1}{4}
 \big( 1- n_{i \uparrow} -n_{i \downarrow} \big)
 \big( 1-n_{i \uparrow} -n_{i \downarrow} \big) \nonumber \\
 &+\frac{U}{2}
 \big( c^{\dagger}_{i \uparrow} c^{\dagger}_{i \downarrow} c_{i \downarrow} c_{i \uparrow}
 +c_{i \downarrow} c_{i \uparrow} c^{\dagger}_{i \uparrow} c^{\dagger}_{i \downarrow} \big)
 \Big] \nonumber \\
&= \sum\limits_{i} \frac{2U}{3} {\boldsymbol {\eta}}_{i}^{2} + \sum\limits_{i} 2 \mu {\eta}_{i}^{z},
\label{eq:elel}
\end{align}
we can immediately see that, at half-filling~($\mu=0$), $\sum_i{\boldsymbol \eta}_{i}$ 
commutes with the Hamiltonian. 

In order to treat both of antiferromagnetism and d-wave superconductivity, 
namely, to incorporate the two SU(2) symmetries simultaneously, we then 
introduce, following Ref.~\cite{Hermele2007}, a representation of an electron operator at site $i$,
\begin{equation}
 \widetilde{C}_{i}=\begin{pmatrix}
  c_{ i \uparrow } & \epsilon_{i} c^{\dagger}_{i \downarrow } \\
  c_{i \downarrow } & - \epsilon_{i}c^{\dagger}_{i \uparrow }
  \end{pmatrix},\label{eq:C2x2}
\end{equation}
where we express $2\times 2$ matrices with tilde hereafter.  
Then the Hubbard Hamiltonian can be cast into a manifestly SU(2)$\otimes$SU(2) 
representation as
\begin{align}
 H_{\rm SU2}=&- \sum\limits_{i}\sum\limits_{j (\neq i)} v_{|i-j|}
 {\rm Tr}\big( \tau^{z} \widetilde{C}^{\dagger}_{i}\widetilde{C}_{j} \big)
\nonumber \\
 &- \left(\sum\limits_{i }^{\rm A} \sum\limits_{j (\neq i)}^{\rm A}+\sum\limits_{i }^{\rm B} \sum\limits_{j( \neq i)}^{\rm B}\right) v_{|i-j|}
 {\rm Tr}\big( \widetilde{C}^{\dagger}_{i}\widetilde{C}_{j} \big)
 + H_{\rm U},\label{eq:HamHubSU2}
\end{align}
where we have decomposed the hopping terms into those 
across different sublattices~(the nearest-neighbor hopping) 
as represented by the first term, and those within the same 
sublattice (second- and third-neighbor hoppings) 
represented by the second term.  
The factor $\epsilon_{i}\epsilon_{j}$, arising from $\widetilde{C}^{\dagger}_{i}\widetilde{C}_{j}$ in these terms, depends on whether the hopping is 
inter- or intra-sublattice, so that for the former we have 
inserted $\tau^{z}$ to compensate $\epsilon_{i}\epsilon_{j}=-1$ in terms of 
Pauli matrices 
$\bolds{\tau}=(\tau^{x}, \tau^{y},\tau^{z})$ on the $2\times 2$ space in Eq.\ref{eq:C2x2}. 
In this representation, ${\boldsymbol \eta}_{i}$ is the right-generator of the  SU(2)$_{\eta}$\cite{Yang1989,Kitamura2016} as
 \begin{align}
    &{\boldsymbol \eta}_{i} = \frac{1}{4} {\rm Tr} \big( \widetilde{C}_{i} 
{\boldsymbol \tau} \widetilde{C}^{\dagger}_{i} \big).
 \end{align}
 Similar to the $SU(2)_{\rm s}$ symmetry which allows transforming the spin-up electrons into spin-down electrons, SU(2)$_{\eta}$ symmetry enables doubly occupied states to be converted into empty states. 
  Thus, at finite doping, with more electrons than number of sites, the SU(2)$_{\eta}$ symmetry will be lowered to the $U(1)$ charge symmetry. 
However, to enable our formalism to treat all doping regimes, including the half-filling, 
we incorporate the SU(2)$_{\eta}$ symmetry 
in our formalism.  Even when we 
express the Hamiltonian in terms of $\widetilde{C}$ operators, 
the formalism does not enforce the symmetry and can describe 
the case of broken SU(2)$_\eta$ symmetry away from half-filling.

The usual spin-rotational SU(2)$_{s}$, on the other hand, is the 
left-generator as
 \begin{align}
&{\boldsymbol S}_{i} = 
\frac{1}{4} {\rm Tr} \big( \widetilde{C}^{\dagger}_{i} {\boldsymbol \tau} 
\widetilde{C}_{i} \big) \nonumber \\
& = \frac{1}{2} \left( c^{\dagger}_{i \downarrow } c_{ i \uparrow }
   + c^{\dagger}_{i \uparrow } c_{i \downarrow }, 
i c^{\dagger}_{i \downarrow }c_{ i \uparrow }
   -i \epsilon_{i} c^{\dagger}_{i \uparrow } c_{i \downarrow }, 
n_{i \uparrow } - n_{i \downarrow }
   \right), \nonumber \\
 &  S^{+}_{i}= S^{x}_{i} + i S^{y}_{i}=  c^{\dagger}_{i \uparrow }c_{i \downarrow },
  \nonumber \\
  & S^{-}_{i}= S^{x}_{i} - i S^{y}_{i}=  c^{\dagger}_{i \downarrow }c_{ i \uparrow }.
 \end{align}
Thus Eq.~\ref{eq:HamHubSU2}, at half-filling, enjoys a symmetry~\cite{Feldbacher2003},
\begin{equation}
{\rm SU(2)_s} \otimes {\rm SU(2)_\eta} / \mathbb{Z}_{2} \approx {\rm SO(4)},
\end{equation}
where $\mathbb{Z}_{2}$ is the sublattice symmetry.  

\begin{figure}
 \includegraphics[width=0.4\textwidth]{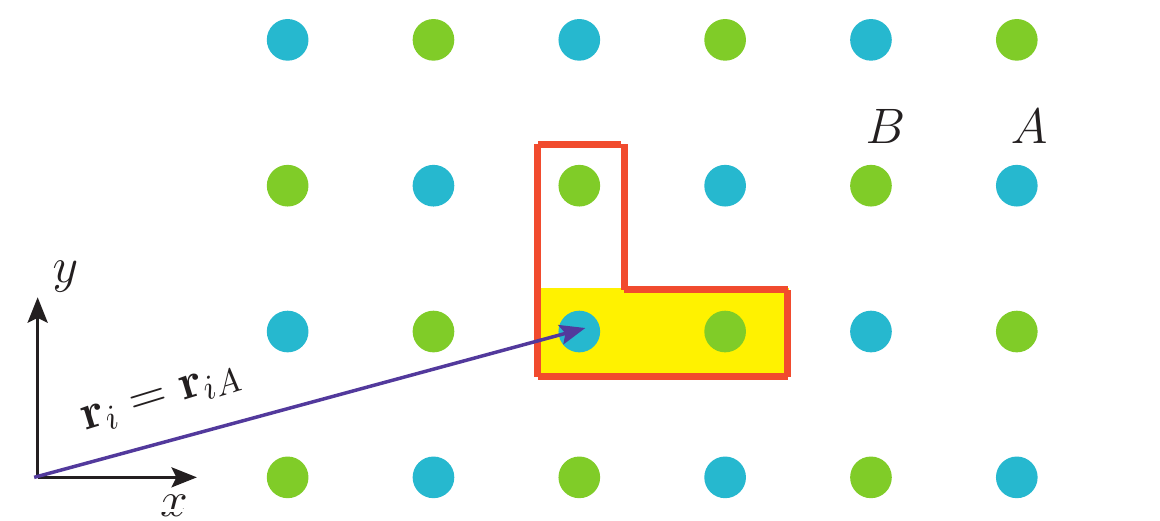}
 \caption{\label{Fig:cluster}
Square lattice decomposed into sublattices A~(blue circles) and B~(green). 
Red line delineates the cluster which we employ in Eq.~\ref{eq:CNb}. We characterize the coordinate ($\bolds{r}_{i}$) of the 
$i$th cluster by $\bolds{r}_{iA}$ of the 
$A$ site in the cluster. The lattice has the four-fold rotational symmetry, so that the cluster 
is equivalent to a bipartite cluster shown in yellow.
 }
\end{figure}

\section{Numerical method}\label{sec:numerics}

In order to introduce a numerical formalism that treats AF and SC phases on an equal footing, we now formulate 
the FLEX+DMFT method~\cite{Gukelberger2015} with our novel bipartite full-SU(2) slave-boson impurity solver in the 
${\rm SO(4)} (\approx {\rm SU(2)_s} \otimes {\rm SU(2)_\eta} /
\mathbb{Z}_{2})$ representation. 
To preserve the SO(4) symmetry in the whole formalism, we first 
introduce a spinor operator in the four-component Nambu (N) representation as
\begin{align}
 C^{\dagger}_{{\rm N}i}=
 \begin{pmatrix}
c^{\dagger}_{i \uparrow}, &
-\epsilon_{i} c^{\dagger}_{i \uparrow}, &
 c_{i \downarrow}, &
\epsilon_{i}^{*} c_{i \downarrow}
\end{pmatrix},\label{eq:CNambu}
\end{align}
where we again incorporate the bipartite factor $\epsilon_{i}$.  
This representation puts the two Nambu doublets together, namely 
\begin{eqnarray}
\Psi_{1i}= (0, \,  -\epsilon_{i}c^{\dagger}_{i \uparrow }, \, c_{i \downarrow }, \, 0),\nonumber \\
\Psi_{2i}= ( c_{i \uparrow },  \, 0, \, 0, \, \epsilon_{i}c^{\dagger}_{i \downarrow } ),
\end{eqnarray}
with which Eq.~\ref{eq:CNambu} 
is expressed as 
\[
{C}_{{\rm N}i}^{\dagger} =                                             
 {\Psi_{2i}^{\dagger} } +
 {\Psi_{1 i} }
.
\]
As the superconducting gap function in the high-Tc cuprates is known to be singlet~\cite{Keimer2015}, we represent Eq.~\ref{eq:CNambu} in such a way that the singlet pairing correlation is included in the two-point propagator~(see Eq.~\ref{eq:genGk} below). 

Since antiferromagnetism and d-wave pairing both 
involve neighboring sites, which we call $i$ and $i+\hat{x}$ (and $i+\hat{y}$) 
where $\hat{x} (\hat{y})$ is the nearest-neighbor 
vector along $x (y)$ axis, we can cast, for treating them on an equal footing, the ${C}_{{\rm N}i}$ defined in Eq.(\ref{eq:CNambu}) 
into a bipartite representation where, at cluster
 $i$, ${C}_{{\rm N}i}$ involves 
both of $(A,B)$ sublattice sites. We thus redefine Eq.~\ref{eq:CNambu} as
 \begin{align}
 C^{\dagger}_{{\rm N}i}=
 \begin{pmatrix}
c^{\dagger}_{iA \uparrow}, &
-\widetilde{\epsilon}_{B} c^{\dagger}_{iB \uparrow}, &
 c_{iA \downarrow}, &
\widetilde{\epsilon}_{B}^{*} c_{iB \downarrow} 
\end{pmatrix}.\label{eq:CNb}
\end{align}
where $c_{i A \sigma}$ denotes electron at sublattice $A$ on cluster site $i$ with spin $\sigma$.
To incorporate the rotational symmetry of our square lattice, we take  $c_{iB \sigma} \equiv ( c_{i+\hat{x} A \sigma} + c_{i+\hat{y} A \sigma} ) /2$, where the factor $2$ imposes having only two sites per unit cell, and
$\widetilde{\epsilon}_{ B} \equiv ( \epsilon_{i+\hat{x}} + \epsilon_{i+\hat{y}} )/2 $. 
Note that the coordinate of cluster $i$~($\bolds{r}_{\rm i}$) is the same as its $A$ sublattice~($\bolds{r}_{\rm i A}$), see also Fig.~\ref{Fig:cluster}. 
Applying the Fourier transform to the electron operators on each sublattice 
as 
$c^{\dagger}_{\bolds{k}A(B)}= (1/N)\sum\limits_{l}^{\rm A(B)} e^{i\bolds{k}\cdot\bolds{r}_{l}}
c^{\dagger}_{l A(B) }$ with momentum $\bolds{k}\equiv (k_{x},k_{y})$ and 
$N$ being the total number of lattice sites, we obtain
\begin{align}
 &C^{\dagger}_{{\rm N}\bolds{k}}(\tau)= \label{eq:CNBk} \\
 &\begin{pmatrix}
  c^{\dagger}_{\bolds{k} A \uparrow}(\tau),
  &
  -c^{\dagger}_{\bolds{k}+\bolds{Q} B \uparrow}(\tau),
  &
  c_{-\bolds{k} A \downarrow}(\tau),
  &
  c_{- (\bolds{k}+\bolds{Q}) B \downarrow}(\tau)
 \end{pmatrix}.\nonumber
\end{align}
Here we have used $\widetilde{\epsilon}_{ B}=e^{ i \bolds{Q}\cdot\bolds{r}_{\rm iB}}$ with $\bolds{Q}=(\pi, \pi )$ being the AF (Brillouin-zone corner) wave vector, and $\bolds{r}_{\rm iB}$ being the coordinate of the sublattice B on bipartite site $i$.  

To calculate Green's function, we use the the Heisenberg representation of electrons, $c(\tau)= e^{\tau H} c e^{-\tau H}$, where $H$ is given by Eq.~\ref{eq:Hubbard}, and $\tau$ is the Matsubara time. 
In the present four-component Nambu representation, the 
Green's function is a $4\times 4$ matrix, ${\cal G}_{4 \times 4 }$, 
which is given in terms of the normal component defined as
\begin{align}
G_{ \alpha \beta \sigma \sigma }(\bolds{k},\bolds{k}'; \overline{\tau}) =
- \ave{{\cal T}_{\tau} c_{ \bolds{k} \alpha \sigma}(\tau) c^{\dagger}_{ \bolds{k}' \beta \sigma}(\tau')},
 \end{align}
along with the anomalous component for treating SC phases, 
\begin{align}
F_{ \alpha \beta \sigma \overline{\sigma} }(\bolds{k},\bolds{k}';\overline{\tau})=
 - \ave{{\cal T}_{\tau}  c_{ \bolds{k} \alpha \sigma}(\tau) c_{ \bolds{k}' \beta \overline{\sigma}}(\tau')}. 
 \end{align}
Here $\overline{\tau}=\tau-\tau'$, $\alpha$ and $\beta$ denote sublattice indices, ${\cal T}_{\tau}$ is the time-ordering operator, 
and $\overline{\sigma}=-\sigma$. 
In the momentum space the $4\times 4$ Green's function is concisely expressed as
\begin{align}
{\cal G}_{4 \times 4 }(\bolds{k}, \overline{\tau}) &= - \ave{ {\cal T}_{\tau} [\widetilde{C}_{{\rm N}\bolds{k}}]^{\intercal}(\tau) \widetilde{C}^{\dagger}_{{\rm N}\bolds{k}}(\tau') },
\end{align}
with $\overline{\tau} \equiv \tau-\tau'$, 
for which 
the matrix elements are explicitly given as
  \onecolumngrid
 \begin{align}
 &{\cal G}_{4 \times 4}(\bolds{k}; \overline{\tau}) = \label{eq:genGk} \\
 &
\begin{pmatrix}
    {G}_{AA\uparrow \uparrow }(\bolds{k},\bolds{k}; \overline{\tau})  
  & -{G}_{AB\uparrow \uparrow }(\bolds{k},\bolds{k}+\bolds{Q}; \overline{\tau}) 
  & {F}_{AA\uparrow \downarrow }(\bolds{k},-\bolds{k}; \overline{\tau})  
  & {F}_{AB\uparrow \downarrow }(\bolds{k},-(\bolds{k}+\bolds{Q}); \overline{\tau})  
  \\
    -{G}_{BA \uparrow \uparrow } (\bolds{k}+\bolds{Q}, \bolds{k}; \overline{\tau})  
  &  {G}_{BB \uparrow \uparrow } (\bolds{k}+\bolds{Q}, \bolds{k}+\bolds{Q}; \overline{\tau})
  & -{F}_{BA \uparrow \downarrow} (\bolds{k}+\bolds{Q}, -\bolds{k}; \overline{\tau})
  & -{F}_{BB \uparrow \downarrow } (\bolds{k}+\bolds{Q}, -( \bolds{k}+\bolds{Q}); \overline{\tau})
  \\
    {F}^{*}_{AA \downarrow \uparrow }(-\bolds{k},\bolds{k}; \overline{\tau})  
  & -{F}^{*}_{AB \downarrow \uparrow}(-\bolds{k},\bolds{k}+\bolds{Q}; \overline{\tau})  
  & -{G}_{AA \downarrow \downarrow }(-\bolds{k},-\bolds{k}; \overline{\tau})  
  & -{G}_{AB \downarrow \downarrow }(-\bolds{k},-(\bolds{k}+\bolds{Q}); \overline{\tau}) 
 \\
    {F}^{*}_{BA \downarrow \uparrow  } (-(\bolds{k}+\bolds{Q}), \bolds{k}; \overline{\tau})
  & -{F}^{*}_{BB \downarrow \uparrow } (-(\bolds{k}+\bolds{Q}),  \bolds{k}+\bolds{Q}; \overline{\tau})
  & -{G}_{BA \downarrow \downarrow } (-(\bolds{k}+\bolds{Q}), -\bolds{k}; \overline{\tau}) 
  & -{G}_{BB \downarrow \downarrow } (-(\bolds{k}+\bolds{Q}), -( \bolds{k}+\bolds{Q}); \overline{\tau})
\end{pmatrix}.\nonumber
\end{align}
\twocolumngrid

 Equivalently one can transform the imaginary-time Green's function into Matsubara-frequency space as
 \begin{equation}
  {\cal G}_{4 \times 4}(\bolds{k}; i\omega_{n})= \int_{0}^{\beta} {\rm d}{\tau} e^{i \omega_{n}\tau} {\cal G}_{4 \times 4}(\bolds{k}; \tau),\label{eq:Gkomega}
 \end{equation}
with $\beta=1/T$, where the fermionic Matsubara 
frequencies are given by $\omega_{n}=\beta(2n-1)/\pi$.

The lattice Green's function in Eq.~\ref{eq:Gkomega} satisfies the Nambu-Dyson equation,
\begin{align}
 {\cal G}_{4 \times 4}(k)^{-1}= i \omega_{n}\mathbb{I}_{4 \times 4} -\varepsilon_{4 \times 4}(\bolds{k}) -{ \Sigma}^{\rm eff}_{4\times 4 }(k),\label{eq:Dyson44}
\end{align}
where $k\equiv (\bolds{k}, i\omega_{n})$, and $\mathbb{I}_{4 \times 4}$ denotes the identity matrix.
In the above, $\varepsilon_{4 \times 4}(\bolds{k})$ 
is the dispersion due to the hopping terms in Eq.~\ref{eq:Hubbard},
\begin{align}
 &\varepsilon_{4 \times 4}(\bolds{k})=\nonumber \\
 &\quad \begin{pmatrix}
  \varepsilon'(\bolds{k})& \varepsilon(\bolds{k}) & 0 & 0 \\
  \varepsilon(\bolds{k}) &\varepsilon'(\bolds{k}+\bolds{Q}) & 0 &  0 \\
  0 & 0 &  -\varepsilon'(-\bolds{k}) & -\varepsilon(-\bolds{k}) \\
  0 & 0 & -\varepsilon(-\bolds{k}) & -\varepsilon'(-\bolds{k}-\bolds{Q})
 \end{pmatrix},
\end{align}
with
\begin{align}
 \varepsilon(\bolds{k}) &=-2 v_{1} \big[\cos(k_{x}) + \cos(k_{y})\big] ,
\nonumber \\
  \varepsilon'(\bolds{k}) &= \mu -4 v_{2} \cos(k_{x})  \cos(k_{y}) \nonumber \\
  &\qquad  -2 v_{3} \big[\cos(2k_{x}) + \cos(2k_{y})\big],
 \end{align}
where $v_1, v_2, v_3$ are first, second, and third neighbor hopping amplitudes, respectively.  

Another term, ${ \Sigma}^{\rm eff}_{4\times 4 }(k)$,  in Eq.~\ref{eq:Dyson44} is 
the self-energy, and in FLEX+DMFT formalism which we adopt here, 
this comprises a combination of the self-energies in DMFT and FLEX, 
 \begin{align}\label{eq:sigmaeff}
  \Sigma_{4 \times 4}^{\rm eff}(k) = \Sigma_{4 \times 4}^{\rm DMFT}( i\omega_{n}) + \Sigma_{4 \times 4}^{\rm nl}(k),
 \end{align}
where we have defined the non-local part of the FLEX self-energy as
 \begin{equation}
  \Sigma_{4 \times 4}^{\rm nl}(k) = \Sigma_{4 \times 4}^{\rm FLEX}(k) - \Sigma_{4 \times 4}^{\rm FLEX, local}(i\omega_{n}), 
 \end{equation}
where we subtract the local part of the FLEX contribution, 
$\Sigma_{4 \times 4}^{\rm FLEX, local}$, to avoid double counting of local Feynman diagrams.\cite{Kitatani2015} 
This local contribution is obtained by following the 
FLEX self-energy prescription as introduced below
[Eq.(\ref{eq:flexselfmatrix})] after
substituting the lattice Green's functions with the local Green's function,
 \begin{equation}
   {\cal G}^{\rm loc}_{4\times 4}(i \omega_{n})= \frac{1}{N}\sum\limits_{\bolds{k}}{\cal G}_{4 \times 4}(k). 
   \label{eq:Gloc0}
 \end{equation}
The local Green's function has a matrix representation,
 \begin{equation}
 {\cal G}^{\rm loc}_{4 \times 4}=
\begin{pmatrix}
    { G}_{AA\uparrow \uparrow }  
  & { G}_{AB\uparrow \uparrow } 
  & { F}_{AA\uparrow \downarrow }  
  & -{ F}_{AB\uparrow \downarrow }  
  \\
    { G}_{BA\uparrow \uparrow }  
  & { G}_{BB\uparrow \uparrow }  
  & { F}_{BA\uparrow \downarrow }  
  & -{ F}_{BB\uparrow \downarrow } 
 \\
    { F}^{*}_{AA\downarrow \uparrow }   
  & { F}^{*}_{AB\downarrow \uparrow } 
  & -{ G}_{AA\downarrow \downarrow } 
  & { G}_{AB\downarrow \downarrow } 
  \\
    -{ F}^{*}_{BA\downarrow \uparrow } 
  & -{ F}^{*}_{BB\downarrow \uparrow }  
  & +{ G}_{BA\downarrow \downarrow } 
  & -{ G}_{BB\downarrow \downarrow } 
\end{pmatrix},\label{eq:Gloc}
 \end{equation}
where $G_{AB\sigma \sigma '}$, and $F_{AB\sigma \sigma '}$  are defined as  
\begin{align}
 G_{AB\sigma \sigma '}(\tau-\tau') &=- \langle c_{A \sigma}(\tau) f_{B \sigma'}^{\dagger}(\tau') \rangle, \\
 F_{AB\sigma \sigma '}(\tau-\tau') &=- \langle c_{A \sigma}(\tau) f_{B \sigma'}(\tau') \rangle.
\end{align}
Here $c_{A(B) \sigma})$ denotes the annihilation operator of an electron with spin $\sigma$ residing on sublattice $A~(B)$ of the impurity site.

\begin{figure*}[ht] 

\includegraphics[width=0.9\textwidth]{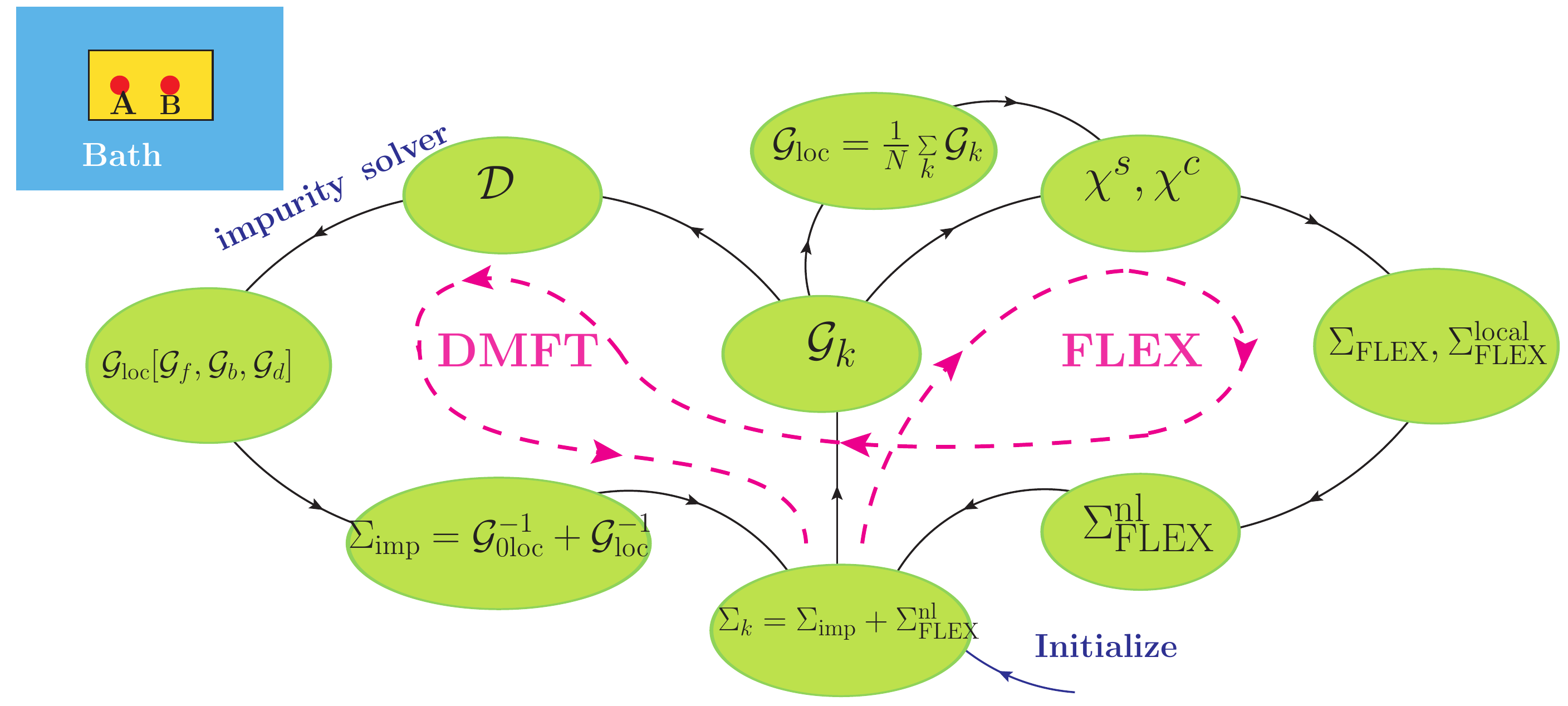}
\caption{
The FLEX+DMFT double self-consistency loops. The iteration starts with initializing the nonlocal FLEX self-energy with the DMFT self-energy set to zero. The obtained effective self-energy 
is then employed in computing the lattice Green's function 
${\cal G}_{4 \times 4}(k)$ and its local counterpart ${\cal G}_{4 \times 4}(i \omega_{n})$. 
With these Green's functions, local and momentum-dependent spin ($\chi^{s}$) and charge  ($\chi^{c}$) susceptibilities are computed, which are fed in the evaluation of the new FLEX self-energy. The nonlocal part of 
the self-energy is then employed to calculate the hybridization function, 
$\cal{D}$, which is in turn inserted into the DMFT impurity problem to obtain the local Green's function. Next, we determine the impurity self-energy using the interacting local Green's functions. Combine the DMFT self-energy with the nonlocal FLEX self-energy to start the next $\rm DFMT+FLEX$ iteration. The double loops are continued until the convergence of the lattice Green's functions is achieved. 
The inset shows a schematic a bipartite impurity~(yellow rectangle) which is in contact with the bath~(blue).  
}
\label{Fig:DMFTFLEX}
\end{figure*}

The FLEX+DMFT self-consistency iteration is described in detail in the 
following sections, where the outline, see Fig.~\ref{Fig:DMFTFLEX}, is 
as follows: 
\begin{enumerate}
\item Initialize $\Sigma^{\rm nl}_{4 \times 4}$, and obtain the lattice Green's function~(${\cal G}_{4 \times 4}$) without the impurity self-energy.
\item Insert the Green's function into the FLEX loop to compute a new $\Sigma^{\rm FLEX}$, which is then plugged in the ${\rm DMFT}$ iteration 
for computing the hybridization function. 
\item With the hybridization function, solve the impurity problem to evaluate the impurity self-energy. 
\item The effective lattice self-energy is then determined by summing the impurity and nonlocal FLEX self-energies. 
\item With $\Sigma^{\rm eff}$, Eq.~\ref{eq:sigmaeff},  
we solve the Dyson equation, Eq.~\ref{eq:Dyson44}, to update the lattice Green's function.  
\item Repeat the above double (FLEX+DMFT) self-consistency 
loops until the convergence is attained.
\end{enumerate}

A difference from the FLEX+DMFT in Ref.\cite{Kitatani2015} is that 
here we explicitly treat the hybridization function ($\cal{D}$ in 
Fig.~\ref{Fig:DMFTFLEX}).

\subsection{DMFT with the bipartite full-SU(2) slave-boson solver}

In the usual DMFT, the many-body problem is mapped onto an impurity problem which is embedded in a self-consistent medium (bath). In this procedure, we have thus a mean-field 
treatment in real space, while 
we do retain temporal (dynamical) quantum fluctuations, which enables the method to treat Mott transitions.  
Thus the approach is a computationally 
less demanding algorithm.  
We shall later improve the method by combining 
with the FLEX framework to incorporate spatial fluctuations.  
An implementation of DMFT requires to solve the impurity model and compute the Green's function.
Several impurity solvers have been implemented, each with its own advantages and disadvantages. 
Numerically exact approaches including continuous-time Quantum Monte Carlo, exact diagonalization and Numerical Renormalization Group provide numerically accurate results at least for single-site DMFT~\cite{Georges1996}, but the numerical cost increases rapidly with the number of orbitals and sites in the unit cell and/or in approaches where DMFT is combined with non-local effects. Approximate analytical methods can provide the desired information at a smaller cost.

To allow our DMFT framework to treat AF and SC on an 
equal footing, we should consider a two-site impurity cluster (inset 
of Fig.1).
Thus we expound here the SO(4) DMFT steps to solve the 
two-site impurity problem using the bipartite full-SU(2) slave-boson solver. 

The impurity action for the Hamiltonian (\ref{eq:Hubbard}) is given by
\begin{align}
  {\cal S} &= \int_{0}^{\beta}  {\rm d}\tau 
 \Big(
  \sum\limits_{ {\mathfrak i}}^{A,B} \sum\limits_{\sigma} c^{\dagger}_{{\mathfrak i} \sigma}  \partial_{\tau}  c_{{\mathfrak i} \sigma}
 +\mu \sum\limits_{ {\mathfrak i} \in \{A,B\},\sigma}  n_{{\mathfrak i} \sigma} +H_{\rm imp}
 \Big)
 \nonumber \\
 &
 +\iint_{0}^{\beta} {\rm d}\tau {\rm d}\tau'
\sum\limits_{{\mathfrak i}{\mathfrak j} }^{A,B} \sum\limits_{\sigma}
c^{\dagger}_{{\mathfrak i} \sigma} (\tau) 
{\mathfrak{G}}_{\sigma \sigma {\mathfrak  i}{\mathfrak j }}(\tau-\tau')
c_{{\mathfrak j} \sigma} (\tau') 
\nonumber \\
 &
 +\iint_{0}^{\beta} {\rm d}\tau {\rm d}\tau'
\sum\limits_{{\mathfrak i}{\mathfrak j}}^{A,B} \sum\limits_{ \sigma}
c^{\dagger}_{{\mathfrak i} \sigma} (\tau) 
{\mathfrak{F}}_{\sigma \overline{\sigma} {\mathfrak  i}{\mathfrak j }}(\tau-\tau')
c^{\dagger}_{{\mathfrak j} \overline{\sigma} } (\tau') 
 ,\label{eq:Sel}
\end{align}
where, in this section, the electron annihilation ($c$) and number ($n$) 
operators refer to the impurity cluster, and 
the mathfrac site index ${\mathfrak  i}$ refers to A 
and B sublattice sites in the impurity cluster.  
The impurity Hamiltonian reads
\begin{align}
 H_{\rm imp} =& - \sum\limits_{\sigma}
 v_{1}  \Big( c^{\dagger}_{A \sigma}c_{B \sigma} +  c^{\dagger}_{B \sigma} c_{A \sigma} \Big)
 +\mu \sum\limits_{ {\mathfrak i} }^{A,B} \sum\limits_{ \sigma} n_{{\mathfrak i} \sigma} 
 \nonumber \\
 &
 + U \sum\limits_{ {\mathfrak i}}^{A,B} 
 \left( n_{{\mathfrak i} \uparrow} -\frac{1}{2} \right)
 \left( n_{{\mathfrak i} \downarrow} -\frac{1}{2} \right).
\label{eq:Hamimp}
\end{align}
$\mathfrak G$ and $\mathfrak F$ appearing in Eq.(\ref{eq:Sel}) 
are normal and anomalous hybridization functions, respectively, which convey the bath information, and are expressed, in a $4\times4$ matrix form 
in the present formalism, as
\begin{align}
{\cal D}_{4 \times 4} \equiv
\begin{pmatrix}
    {\mathfrak G}_{AA\uparrow \uparrow }  
  & {\mathfrak G}_{AB\uparrow \uparrow } 
  & {\mathfrak F}_{AA\uparrow \downarrow }  
  & -{\mathfrak F}_{AB\uparrow \downarrow }  
  \\
    {\mathfrak G}_{BA\uparrow \uparrow }  
  & {\mathfrak G}_{BB\uparrow \uparrow }  
  & {\mathfrak F}_{BA\uparrow \downarrow }  
  & -{\mathfrak F}_{BB\uparrow \downarrow } 
 \\
    {\mathfrak F}^{*}_{AA\downarrow \uparrow }   
  & {\mathfrak F}^{*}_{AB\downarrow \uparrow } 
  & -{\mathfrak G}_{AA\downarrow \downarrow } 
  & {\mathfrak G}_{AB\downarrow \downarrow } 
  \\
    -{\mathfrak F}^{*}_{BA\downarrow \uparrow } 
  & -{\mathfrak F}^{*}_{BB\downarrow \uparrow }  
  & {\mathfrak G}_{BA\downarrow \downarrow } 
  & -{\mathfrak G}_{BB\downarrow \downarrow } 
\end{pmatrix},\label{eq:Hyb}
\end{align}
where $\mathfrak G (\mathfrak F)$ appear in the 
diagonal (off-diagonal) blocks.  
In the FLEX+DMFT iteration we determine this matrix through 
\begin{align}
  &\Big( 1+  {\cal A}_{4 \times 4} \Big)
  * {\cal D}_{4 \times 4}(i\omega_{n}) 
  = {\cal B}_{4 \times 4}(i\omega_{n}), \label{eq:Hyb0} \\
  &{\cal A}_{4 \times 4}(i\omega_{n})=  \sum\limits_{\bolds{k}} \Big[\varepsilon_{4 \times 4}(\bolds{k})+ \Sigma^{\rm nl}_{4 \times 4}(k) \Big], 
  \\
  &{\cal B}_{4 \times 4}(i\omega_{n}) \nonumber\\
&\quad=\sum\limits_{\bolds{k}} \Big[
 \big( \varepsilon_{4 \times 4}+ \Sigma^{\rm nl}_{4 \times 4} \big)
 *
 {\cal G}_{4 \times 4} 
 *
 \big( \varepsilon_{4 \times 4}+ \Sigma^{\rm nl}_{4 \times 4} \big) \Big](k),
\end{align}
where ``$*$" denotes the convolution integral on the Matsubara frequency, and we have employed Eq.(\ref{eq:Gloc}) to obtain Eq.(\ref{eq:Hyb}) (see Appendix 
for 
detailed derivation).  
If we define, as in ${\cal G}^{\rm loc}_{4 \times 4}$ in Eq.(\ref{eq:Gloc}), the impurity Green's function in a  $4 \times 4$ matrix form with 
\begin{equation}
{\cal G}_{\rm imp} = -\ave{C_{{\rm N} {\mathfrak i}} C^{\dagger}_{{\rm N} {\mathfrak i}} },
\end{equation}
we can then impose that 
\begin{equation}
{\cal G}_{\rm imp}(i\omega_{n}) = {\cal G}_{4 \times 4}^{\rm loc}(i \omega_{n}) 
\end{equation}
in the DMFT scheme. To solve the impurity problem, we shall introduce  in the next section the {\it bipartite full-SU(2) slave-boson} impurity solver.

\subsubsection{Bipartite full-SU(2) slave-boson impurity solver}

Within the slave-particle formalism, we decompose the electron operator into 
fermionic and bosonic particles such that the associated matrix elements of physical states be equal to their counterparts in the auxiliary slave-particle Hilbert space. Within this prescription, the fermion statistics for the physical particle may or may not be satisfied depending on what kind of slave-particles we adopt. The doublon-less SU(2) slave-boson, in particular, represents the electron operator as a composite of charge-less fermions~(spinons) with up and down spins, and 
two-flavored spin-less bosons~(holons) with flavor indices~($1,2$). The two species of holons are required to maintain the SU(2)$_{\rm s}$ symmetry of the system. The doublon-less SU(2) slave-boson representation is actually introduced to examine the limit of the large repulsive interaction, where we eliminate double occupancies 
by applying the Gutzwiller projection~\cite{Lee2001,Lee2006,Wen1996}. One consequence of this is that the fermion statistics is violated. To remedy this, here we introduce another 
set of spin-less bosons~(doublons) that take care of double 
occupancies. This procedure will also incorporate the SU(2)$_{\eta}$ symmetry into the formalism. 
Then the original electron creation and annihilation operators are expressed as
\begin{align}
 & c_{ \mathfrak{i} \sigma}^{\dagger} = \frac{1}{\sqrt{2}} (f^{\dagger}_{ \mathfrak{i} \sigma} b_{\mathfrak{i} 1}+
 \sigma \epsilon_{\mathfrak{i}} f_{\mathfrak{i} \overline{\sigma} } b_{\mathfrak{i} 2} ) 
 + \frac{1}{\sqrt{2}} (\overline{\sigma} \epsilon_{\mathfrak{i}} f_{\mathfrak{i} \overline{\sigma}} d^{\dagger}_{\mathfrak{i} 1}+
 f^{\dagger}_{ \mathfrak{i} \sigma} d^{\dagger}_{\mathfrak{i} 2} ),\nonumber \\
 &
 c_{ \mathfrak{i} \sigma} = \frac{1}{\sqrt{2}} (b^{\dagger}_{\mathfrak{i} 1}f_{ \mathfrak{i} \sigma}+ 
 \sigma \epsilon_{\mathfrak{i}} b_{ \mathfrak{i} 2}^{\dagger} f^{\dagger}_{\mathfrak{i} \overline{\sigma} }) 
 + \frac{1}{\sqrt{2}} (\overline{\sigma} \epsilon_{\mathfrak{i}} d_{1i}f^{\dagger}_{\mathfrak{i} \overline{\sigma}}+d_{\mathfrak{i} 2} 
 f_{\mathfrak{i} \sigma}),
 \label{eq:c_SU2}
\end{align}
where we have introduced the spinon operator~($f$) that 
satisfies the fermionic commutation relation, 
along with the holon operators $(b_1, b_2)$ and doublon operators ($d_1, d_2$) that respectively 
satisfy the bosonic commutation relation. 
In the above we have maintained the bipartite formalism, where 
the index in mathfrac (${\mathfrak{i}} \in \{A,B\}$) 
refers to impurity sites, and 
$\overline{\sigma} \equiv -\sigma$.   
Then the 
fermionic commutation relation for the original electron operator is preserved.
This decomposition thus realizes our perception of the real-spin 
SU(2) symmetry described by spinons 
along with the charge ($\eta$) SU(2) symmetry 
taken care of by creation and annihilation of holon 
and doublon operators.  On top of this, 
the $\mathbb{Z}_{2}$ symmetry is imposed by exploiting the bipartite 
factor ($\epsilon_{\mathfrak{i}}$) in the above representation. 
We can make its structure more transparent 
by adopting the $2 \times 2$ matrix forms as in Eq.\ref{eq:C2x2} 
to have
 \begin{align}
 \widetilde{C}_{\mathfrak{i}} &= \frac{1}{\sqrt{2}} \widetilde{F}_{\mathfrak{i}} (\widetilde{B}_{\mathfrak{i}}+\widetilde{D}_{\mathfrak{i}}),\\
 \widetilde{F}_{\mathfrak{i}} &=
   \begin{pmatrix}
 f_{  \mathfrak{i} \uparrow } &  \epsilon_{\mathfrak{i}} f^{\dagger}_{ \mathfrak{i} \downarrow} \\
 f_{ \mathfrak{i} \downarrow} & -\epsilon_{\mathfrak{i}} f^{\dagger}_{ \mathfrak{i} \uparrow} 
 \end{pmatrix},\\
 \widetilde{B}_{\mathfrak{i}} &=
 \begin{pmatrix}
  b^{\dagger}_{\mathfrak{i} 1} & -b_{\mathfrak{i} 2} \\
  b_{ \mathfrak{i} 2}^{\dagger} & b_{\mathfrak{i} 1}
 \end{pmatrix} , \quad
 \widetilde{D}_{\mathfrak{i}} =
 \begin{pmatrix}
  d_{ \mathfrak{i} 2} & d^{\dagger}_{\mathfrak{i} 1} \\
  -d_{\mathfrak{i} 1} & d_{\mathfrak{i} 2}^{\dagger}
 \end{pmatrix}.
 \label{eq:C22}
 \end{align}

The bipartite full-SU(2) slave-boson representation has a Hilbert space that is larger than the physical one,
so we need to eliminate the unphysical states by a constraint, 
which can be obtained as follows~\cite{You2012}.
For the SU(2) gauge-invariance of the electron doublet~($\widetilde{C}$ above) 
we can construct the SU(2) generators, ${\widetilde{K}}_{x/y/z}$, 
in such a way that we have commutation relations, 
\[
[{\widetilde{K}}_{x/y/z}, \widetilde{C}]=0.
\]
We can show that this equation has a solution 
in a vector form,
  \begin{align}
  {\widetilde{\boldsymbol K}}_{\mathfrak{i}}::\quad \widetilde{B}_{\mathfrak{i}}{\boldsymbol \tau}\widetilde{B}_{\mathfrak{i}}^{\dagger}
  - \frac{1}{2} {\rm Tr}\left(\widetilde{F}_{\mathfrak{i}} {\boldsymbol \tau} \widetilde{F}^{\dagger}_{\mathfrak{i}}\right)- \widetilde{D}_{\mathfrak{i}}^{\dagger} {\boldsymbol \tau}\widetilde{D}_{\mathfrak{i}} = {\bf 0},
  \label{eq:Kvec}
 \end{align}
 where the three components have explicit forms of
 \begin{align}
   &{\widetilde{K}}^{z}_{\mathfrak{i}}::f^{\dagger}_{ \mathfrak{i} \uparrow}f_{ \mathfrak{i} \uparrow}+f^{\dagger}_{ \mathfrak{i} \downarrow}f_{ \mathfrak{i} \downarrow}
   +d^{\dagger}_{\mathfrak{i} 1}d_{\mathfrak{i} 1}-d^{\dagger}_{\mathfrak{i} 2}d_{\mathfrak{i} 2}
   +b^{\dagger}_{\mathfrak{i} 1}b_{\mathfrak{i} 1}-b^{\dagger}_{\mathfrak{i} 2}b_{\mathfrak{i} 2}=1,
%    \label{eq:Kz}
\nonumber\\
   &{\widetilde{K}}^{+}_{\mathfrak{i}}=K^{x}_{\mathfrak{i}}+iK^{y}_{\mathfrak{i}}:: -2\epsilon_{\mathfrak{i}} f_{ \mathfrak{i} \uparrow}f_{ \mathfrak{i} \downarrow}
   + 2 b_{\mathfrak{i} 2}^{\dagger}b_{\mathfrak{i} 1} +2 d^{\dagger}_{\mathfrak{i} 2}d_{\mathfrak{i} 1}=0,
%    \label{eq:K+}
\nonumber
   \\
   &{\widetilde{K}}^{-}_{\mathfrak{i}}=K^{x}_{\mathfrak{i}}-\mathfrak{i}K^{y}_{\mathfrak{i}}:: -2\epsilon_{\mathfrak{i}} f^{\dagger}_{ \mathfrak{i} \downarrow}f^{\dagger}_{ \mathfrak{i} \uparrow}
   +2 b^{\dagger}_{\mathfrak{i} 1}b_{\mathfrak{i} 2}+2d^{\dagger}_{\mathfrak{i} 1}d_{\mathfrak{i} 2}=0.
   \label{eq:K}
 \end{align}

The noninteracting 
states ($|\rangle$'s) of the system on a bipartite unit cell can also be translated in the SU(2)
slave-boson decomposition as
 \begin{align}
 |\rm vac_{A}, vac_{B} \rangle & = \frac{1}{\sqrt{2}} \big( b^{\dagger}_{\mathfrak{i}1}
 + \epsilon_{\mathfrak{i}} b^{\dagger}_{\mathfrak{i} 2}f^{\dagger}_{ \mathfrak{i} \downarrow} f^{\dagger}_{ \mathfrak{i} \uparrow}\big)
 |\rm vac_{A}, vac_{B} \rangle_{\rm SB}, 
%\label{eq:emptystate} 
\nonumber \\
 c^{\dagger}_{ \mathfrak{i} \uparrow}|\rm vac_{A}, vac_{B} \rangle &=
 f^{\dagger}_{ \mathfrak{i} \uparrow}  |\rm vac_{A}, vac_{B} \rangle_{\rm SB},
%\label{eq:fupstate} 
\nonumber \\
 c^{\dagger}_{ \mathfrak{i} \downarrow}|\rm vac_{A}, vac_{B} \rangle &=
 f^{\dagger}_{ \mathfrak{i} \downarrow}
 |\rm vac_{A}, vac_{B} \rangle_{\rm SB},
%  \label{eq:fdownstate}
 \nonumber \\
 c^{\dagger}_{ \mathfrak{i} \downarrow} c^{\dagger}_{ \mathfrak{i} \uparrow} |\rm vac_{A}, vac_{B} \rangle & 
 = \frac{1}{\sqrt{2}} \big(\epsilon_{\mathfrak{i}} d_{i1}^{\dagger} 
 + d^{\dagger}_{\mathfrak{i} 2} f^{\dagger}_{ \mathfrak{i} \downarrow}
 f^{\dagger}_{ \mathfrak{i} \uparrow} 
  \big)|\rm vac_{A}, vac_{B} \rangle_{\rm SB}\label{eq:fupfdownstate},
 \end{align}
 where $\mathfrak{i} \in \{A,B\} $, and the vacuum of the bipartite impurity site is defined as 
 $|{\rm vac}_{A}, {\rm vac}_{B}\rangle \equiv | {\rm vac} \rangle_{A} \otimes | {\rm vac} \rangle_{B} $. 
Here the slave-boson vacuum is decomposed as 
\[
| {\rm vac} \rangle_{\mathfrak{i}}= | {\rm vac} \rangle_{f\mathfrak{i}} \otimes | {\rm vac} \rangle_{b\mathfrak{i}} +
  | {\rm vac} \rangle_{f\mathfrak{i}} \otimes | {\rm vac} \rangle_{d\mathfrak{i}},
\]
where $| {\rm vac} \rangle_{f\mathfrak{i}(b\mathfrak{i},d\mathfrak{i})}$ is the associated vacuum of the spinons~(bosons) at impurity site $\mathfrak{i}$. 
 One should note that the bipartite SU(2) impurity solver, with the bipartite 
factor $\epsilon_{\mathfrak{i}}$ inserted, results
 in a sublattice dependence of the projected states 
 in Eq.(\ref{eq:fupfdownstate}). 
 In addition, one should note that the transformed states of the
 SU(2) slave-boson can be occupied by more than one slave-particles.
 Obviously, removing $d$-bosons from our slave-boson picture would exclude the doubly-occupied sites,
 which is precisely the distinction between
 the present formalism and the conventional 
doublon-less SU(2) slave-boson~\cite{Wen1996,Lee2001}.

The total density of holes at $\mathfrak{i}$ is equivalent to the total number of holons, 
\begin{equation}
1- n_{\mathfrak{i}} =  b^{\dagger}_{\mathfrak{i} 1} b_{\mathfrak{i} 1} 
+b^{\dagger}_{\mathfrak{i} 2} b_{\mathfrak{i} 2}.
\end{equation}
Similarly, the double occupancy is expressed as
\begin{align}
 n_{\mathfrak{i} \uparrow} n_{\mathfrak{i} \downarrow} = &\frac{1}{2} \left(
 d^{\dagger}_{\mathfrak{i} 1} d_{\mathfrak{i} 1} + \epsilon_{\mathfrak{i}} d^{\dagger}_{\mathfrak{i} 1} d_{\mathfrak{i}2} f_{\mathfrak{i}\uparrow} f_{ \mathfrak{i}\downarrow } \right.
 \nonumber \\
   & \left. + \epsilon_{\mathfrak{i}} d^{\dagger}_{\mathfrak{i}2} d_{\mathfrak{i}1} f^{\dagger}_{\mathfrak{i}\uparrow} f^{\dagger}_{ \mathfrak{i}\downarrow }
  + d^{\dagger}_{ \mathfrak{i} 2} d_{\mathfrak{i} 2} f^{\dagger}_{ \mathfrak{i} \downarrow} f^{\dagger}_{ \mathfrak{i} \uparrow} f_{ \mathfrak{i} \uparrow} f_{ \mathfrak{i} \downarrow}
  \right) .
\end{align}
The impurity Hamiltonian (\ref{eq:Hamimp}) can be 
expressed up to a constant as
\begin{align}
 H_{\rm imp} = &
H_{\rm hop}+ 
\left( \mu-\frac{U}{2} \right) \sum\limits_{\mathfrak{i}}^{A,B}(
b^{\dagger}_{\mathfrak{i} 1} b_{\mathfrak{i} 1}
+
b^{\dagger}_{\mathfrak{i} 2} b_{\mathfrak{i} 2}
)
 \nonumber \\ &
+ \frac{U}{2} \sum\limits_{\mathfrak{i}}^{A,B} \left(
 d^{\dagger}_{\mathfrak{i} 1} d_{\mathfrak{i} 1} + \epsilon_{\mathfrak{i}} d^{\dagger}_{\mathfrak{i} 1} d_{\mathfrak{i} 2} f_{\mathfrak{i}\uparrow} f_{ \mathfrak{i}\downarrow } \right.
  \nonumber \\& \qquad
\left. 
+ \epsilon_{\mathfrak{i}} d^{\dagger}_{\mathfrak{i} 2} d_{\mathfrak{i} 1} f^{\dagger}_{\mathfrak{i}\downarrow} f^{\dagger}_{ \mathfrak{i}\uparrow }
  + d^{\dagger}_{ \mathfrak{i} 2} d_{\mathfrak{i} 2 } 
  f^{\dagger}_{ \mathfrak{i} \downarrow} f^{\dagger}_{ \mathfrak{i} \uparrow} f_{ \mathfrak{i} \uparrow} f_{ \mathfrak{i} \downarrow}
  \right) 
  \nonumber \\ &
   -\sum\limits_{\mathfrak{i}}^{A,B}
   \lambda^{+}_{\mathfrak{i}}
\Big(-2\epsilon_{\mathfrak{i}} f_{ \mathfrak{i} \uparrow}f_{ \mathfrak{i} \downarrow}
   + 2 b_{\mathfrak{i} 2}^{\dagger}b_{\mathfrak{i} 1} +2 d^{\dagger}_{\mathfrak{i} 2}d_{\mathfrak{i} 1}
  \Big)
  \nonumber \\ &
  -\sum\limits_{\mathfrak{i}}^{A,B}
  \lambda^{-}_{\mathfrak{i}} 
\Big(-2\epsilon_{\mathfrak{i}} f^{\dagger}_{ \mathfrak{i} \downarrow}f^{\dagger}_{ \mathfrak{i} \uparrow}
   +2 b^{\dagger}_{\mathfrak{i} 1}b_{\mathfrak{i} 2}+2d^{\dagger}_{\mathfrak{i} 1}d_{\mathfrak{i} 2}
   \Big)
   \nonumber \\ &
  -\sum\limits_{\mathfrak{i}}^{A,B}
  \lambda^{z}_{\mathfrak{i}} \Big(
  f^{\dagger}_{ \mathfrak{i} \uparrow}f_{ \mathfrak{i} \uparrow}+f^{\dagger}_{ \mathfrak{i} \downarrow}f_{ \mathfrak{i} \downarrow}
  \nonumber \\ & \qquad 
   +d^{\dagger}_{\mathfrak{i} 1}d_{\mathfrak{i} 1}-d^{\dagger}_{\mathfrak{i} 2}d_{\mathfrak{i} 2}
   +b^{\dagger}_{\mathfrak{i} 1}b_{\mathfrak{i} 1}-b^{\dagger}_{\mathfrak{i} 2}b_{\mathfrak{i} 2} 
   \Big)
  ,\label{eq:Himp}
\end{align}
where $H_{\rm hop}$ is the hopping term in Eq.~\ref{eq:Hamimp} translated into the slave-boson language, see Supplementary material, while 
($\lambda^{z}_{A/B}, \lambda^{+}_{A/B}, \lambda^{-}_{A/B}$) are 
Lagrange multipliers introduced to impose the SU(2) constraints in Eqs.~\ref{eq:K} on each sublattice. 
Now we can express the action in Eq.~\ref{eq:Sel} in the slave-boson language as 
\begin{align}
 {\cal S}_{\rm imp} &= 
 \int_{0}^{\beta} {\rm d}{\tau} \Big[ 
 \sum\limits_{\mathfrak{i}}^{A,B}\sum\limits_{\sigma} f^{\dagger}_{\mathfrak{i} \sigma} \partial_{\tau} f_{i \sigma} +H_{\rm imp} \nonumber \\
 &+ \sum\limits_{\mathfrak{i}}^{A,B}
 \big(
 b^{\dagger}_{\mathfrak{i} 1} \partial_{\tau} b_{\mathfrak{i} 1} 
 +d^{\dagger}_{\mathfrak{i} 1} \partial_{\tau} d_{\mathfrak{i} 1} 
 \big)\nonumber \\
  &+ \sum\limits_{\mathfrak{i}}^{A,B}
 \big(
 b^{\dagger}_{\mathfrak{i} 2} \partial_{\tau} b_{\mathfrak{i} 2} 
 +d^{\dagger}_{\mathfrak{i} 2} \partial_{\tau} d_{\mathfrak{i} 2} 
 \big)
  \Big]
  \nonumber \\
 &
 +\iint_{0}^{\beta} {\rm d}\tau {\rm d}\tau'
\sum\limits_{\mathfrak{i}}^{A,B}\sum\limits_{\sigma}
c^{\dagger}_{\mathfrak{i} \sigma} (\tau) 
{\mathfrak{G}}_{\sigma  \mathfrak{i} \sigma\mathfrak{j}}(\tau-\tau')
c_{\mathfrak{j} \sigma} (\tau') 
\nonumber \\
&
 +\iint_{0}^{\beta} {\rm d}\tau {\rm d}\tau'
\sum\limits_{\mathfrak{i}}^{A,B}\sum\limits_{\sigma}
c^{\dagger}_{\mathfrak{i} \sigma} (\tau) 
{\mathfrak{F}}_{\sigma \mathfrak{i} \overline{\sigma} \mathfrak{j}}(\tau-\tau')
c^{\dagger}_{\mathfrak{j} \overline{\sigma} } (\tau').
\end{align}
Employing this action we can derive the equations of motion for the spinon, holon, and doublon operators, respectively, for details see Supplementary material.  
If Lagrange multipliers are correctly determined, the obtained results should satisfy $\ave{\bolds K}_{i}={\bf 0}$ in terms of expectation values, which approximately satisfy Eq.~\ref{eq:Kvec}.
Alternatively, 
the auxiliary equations of motion can be iteratively solved by a new set~($\lambda_{A/B}^{+/-/z}$) until constraints are satisfied.
With the final slave-particle Green's functions, we can update the electron's Green's functions~(${\cal G}_{4 \times 4}^{\rm loc}$), see again Supplementary material.  Finally, the DMFT self-energy is obtained as
\begin{align}
 \Sigma^{\rm DMFT}_{4 \times 4}(i \omega_{n}) &= [{\cal G}_{4 \times 4}^{0 {\rm loc}}]^{-1}(i \omega_{n}) - [{\cal G}_{4 \times 4}^{\rm loc}]^{-1}(i \omega_{n}), \nonumber \\
  [{\cal G}_{4 \times 4}^{0 {\rm loc}}]^{-1}(i \omega_{n}) &= i \omega_{n} \mathbb{I}_{4 \times 4}- {\cal D}_{4 \times 4}.
\end{align}

\subsection{Fluctuation-exchange approximation (FLEX)}

The fluctuation-exchange approximation provides a self-energy of an interacting system by summing over bubble and ladder diagrams. The self-energy in this formalism can be obtained from close-linked diagrams known as the Luttinger-Ward functional $\Phi$, so that the scheme is a conserving approximation~\cite{Wang2015}.  
Studies of the normal states of the Hubbard model reveal that including only the 
particle-hole 
channels as dominating contributors enables us to study the properties of the incommensurate antiferromagnetic spin structure as well as the superconducting instabilities in the overdoped regime~\cite{Kuroki1999,Yanase2001,Oka2010,Kitatani2015}. It has been discussed that, due to insufficiently 
treated dynamical and pairing fluctuations in this formalism, the superconducting dome in the phase diagram can only be partially captured with the AF region overestimated~\cite{Yanase2001}.

In the following, we shall propose an extended SO(4) FLEX self-energy, which treats the superconducting pairing and spin fluctuations on an equal footing. 
Hence not only the fluctuations in the electron pairs having 
momenta $\bolds{k}$ and $-\bolds{k}$, as treated in Ref.~\citen{Yan2005}, but also the pairing between $\bolds{k}$ and $-(\bolds{k}+\bolds{Q})$, known as $\eta$-pairing~\cite{Sentef2017}, are incorporated into the present formalism.

Now we delve into the present FLEX formalism 
by introducing the FLEX self-energy in a $4 \times 4$ matrix form as
\onecolumngrid
\begin{align}
 \Sigma&^{\rm FLEX}_{4 \times 4}(k) = \Sigma^{\rm H}_{4 \times 4} + \nonumber \\
 &\begin{pmatrix}
  \Sigma^{G}_{\uparrow \uparrow} (kk) 
  & \Sigma^{G}_{\uparrow \uparrow}( k,k+\bolds{Q}) 
  & \Sigma^{ F}_{\uparrow \downarrow}( k ,-k) 
  & \Sigma^{ F}_{\uparrow \downarrow}( k ,-(k+\bolds{Q}))  
  \\
  \Sigma^{G}_{ \uparrow \uparrow}( k +\bolds{Q},k) 
  & \Sigma^{G}_{\uparrow \uparrow}( k+\bolds{Q},k+\bolds{Q})
  & \Sigma^{F}_{\uparrow \downarrow}( k+\bolds{Q},-k) 
  & \Sigma^{F}_{\uparrow \downarrow}( k+\bolds{Q},-(k+\bolds{Q}))
  \\
  \Sigma^{*F}_{\downarrow \uparrow}( -k,k)  &
  \Sigma^{*F}_{\downarrow \uparrow}( -k,k+\bolds{Q})  &
  \Sigma^{*G}_{\downarrow \downarrow }(-k,-k) &
  \Sigma^{*G}_{\downarrow \downarrow}( -k,-(k+\bolds{Q})
  \\
  \Sigma^{*F}_{\downarrow \uparrow}( -(k+\bolds{Q}),k)  &
  \Sigma^{*F}_{\downarrow \downarrow}( -(k+\bolds{Q}), (k+\bolds{Q})) &
  \Sigma^{*G}_{\downarrow \downarrow}(-(k+\bolds{Q}),-k) &
  \Sigma^{*G}_{\downarrow \downarrow}(-(k+\bolds{Q}), -(k+\bolds{Q}))
 \end{pmatrix}. \label{eq:flexselfmatrix}
\end{align}
Here $k+\bolds{Q}\equiv (\bolds{k}+\bolds{Q},i\omega_{n})$,
%\twocolumngrid
the Hartree self-energy~($\Sigma^{\rm H}$) is a diagonal matrix with momentum-independent elements as
\onecolumngrid
\begin{align}
 \Sigma^{\rm H}_{4 \times 4 } 
&= U \sum\limits_{\bolds{k}}  \begin{pmatrix}
 G_{\uparrow \uparrow }(\bolds{k}, \bolds{k}; \beta) & 0 & 0 & 0 \\
 0 & G_{\uparrow \uparrow }(\bolds{k} + \bolds{Q}, \bolds{k}+ \bolds{Q}; \beta) & 0 & 0  \\
  0 & 0 & -G_{\downarrow \downarrow }(-\bolds{k}, -\bolds{k}; \beta) & 0  \\
 0& 0 & 0 & -G_{\downarrow \downarrow }(-(\bolds{k} + \bolds{Q}), -(\bolds{k}+ \bolds{Q}); \beta)   \\
\end{pmatrix},
\end{align}
\twocolumngrid
which is just equal to $U n_{4 \times 4}$ 
with $n_{4 \times 4}$ being the diagonal elements of 
${\cal G}_{4 \times 4}$ in Eq.~\ref{eq:genGk} at $\tau=\beta$.  

\begin{figure}[htp]

\begin{subfigure}[t]{.2\textwidth}
\includegraphics[width=\linewidth]{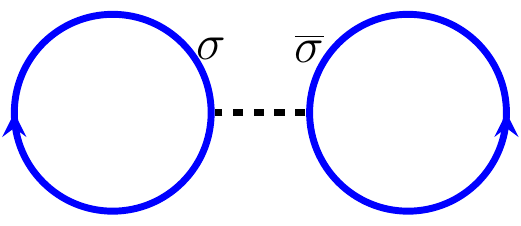}
        \caption{}\label{fig:fig_a}
\end{subfigure}
\begin{subfigure}[t]{.4\textwidth}
\includegraphics[width=\linewidth]{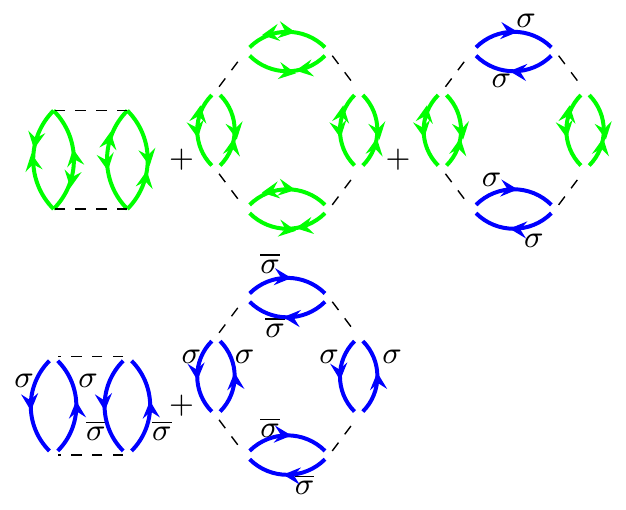}
\caption{}\label{fig:fig_b}
\end{subfigure}

\begin{subfigure}[t]{.4\textwidth}
\includegraphics[width=\linewidth]{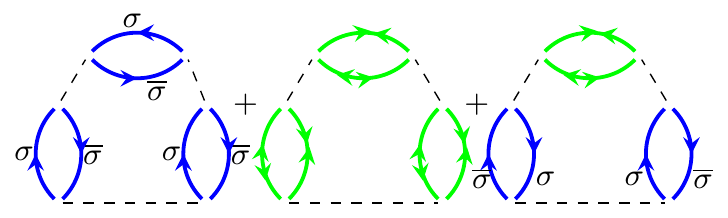}
\caption{}\label{fig:fig_c}
\end{subfigure}

\medskip

\caption{\label{Fig:LWFLEX} Diagrammatic expansion of the Luttinger-Ward functional~($\Phi$) in the SO(4) FLEX. 
(a) The Hartree self-energy. 
(b,c) are the bubble diagrams for (b) the longitudinal spin, charge, and pairing fluctuations, with anomalous 
Green's functions included, and 
(c) the third-order terms in the Luttinger-Ward functional 
from the transverse-spin contributions. 
Here, solid single-arrowed (blue) lines stand for normal Green's functions $G$, solid inward double-arrowed (green) lines the anomalous Green's functions $F$, 
outward double arrows $F^{*}$, and dashed lines the Hubbard interaction~$U$.}
\end{figure}

\begin{figure}[htp]

\begin{subfigure}[t]{.4\textwidth}
\vspace{0pt}
\includegraphics[width=\linewidth]{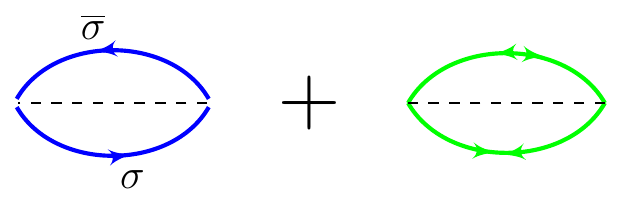}
 \caption{}\label{fig:fig_a1}
\end{subfigure}

\medskip

\begin{subfigure}[t]{.4\textwidth}
\vspace{0pt}
\includegraphics[width=\linewidth]{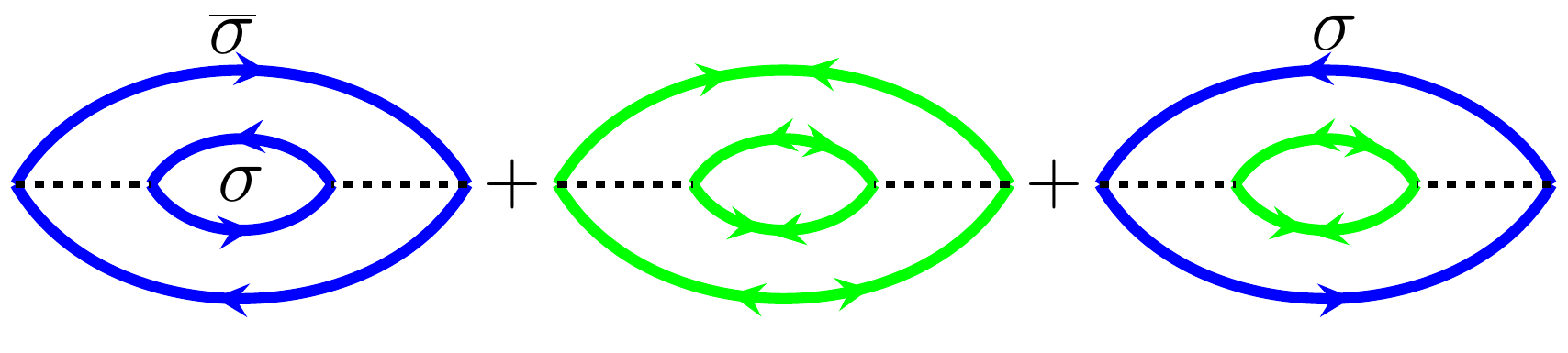}
 \caption{}\label{fig:fig_b1}
\end{subfigure}

\medskip

\begin{subfigure}[t]{.4\textwidth}
\vspace{0pt}
\includegraphics[width=0.8\linewidth]{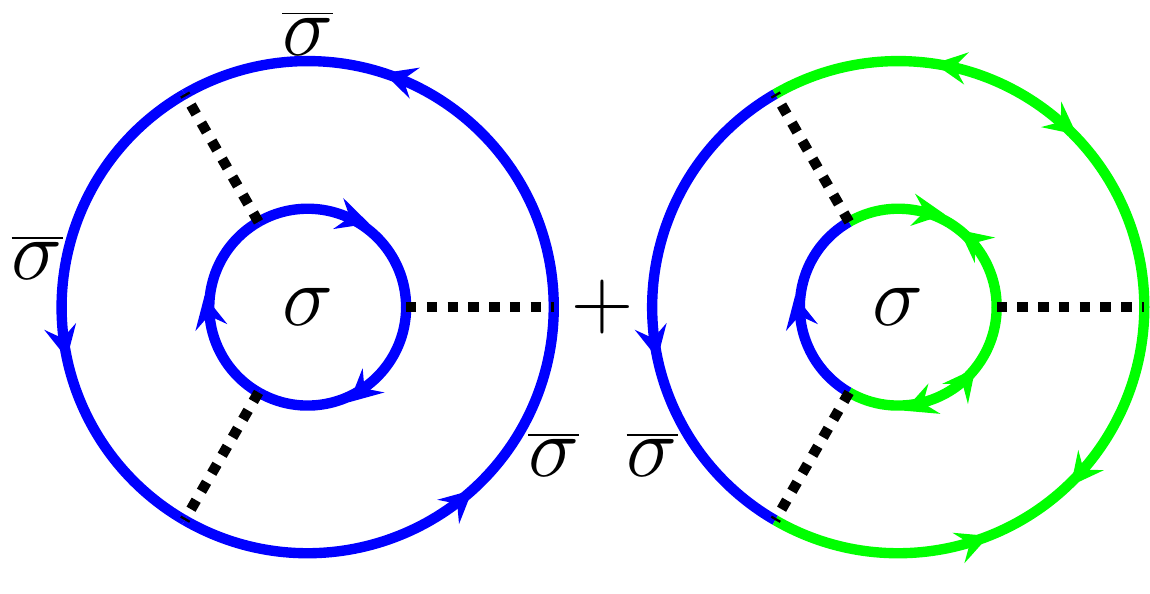}
 \caption{}\label{fig:fig_c1}
\end{subfigure}

\medskip

\begin{subfigure}[t]{.4\textwidth}
\vspace{0pt}
\includegraphics[width=\linewidth]{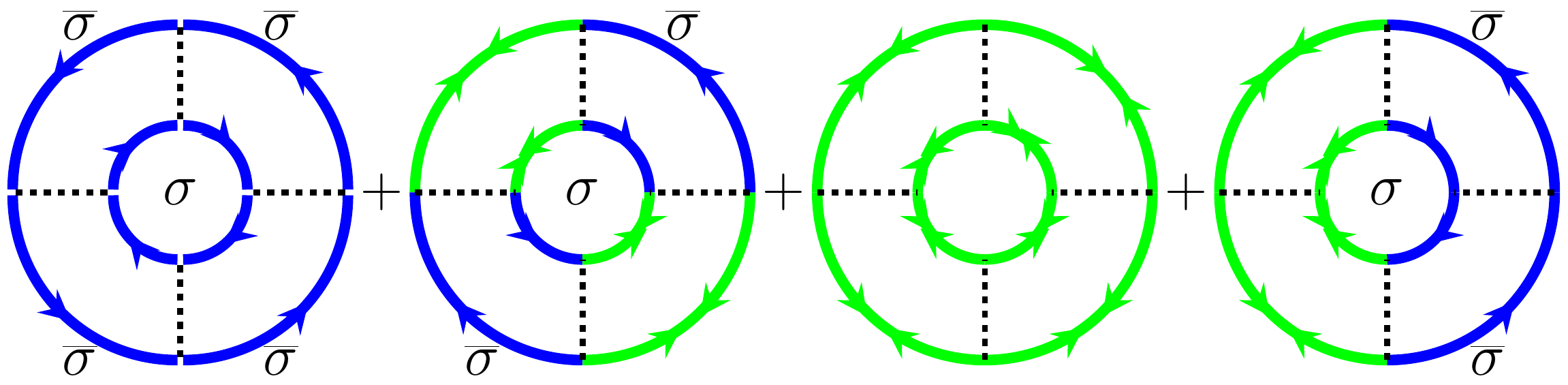}
 \caption{}\label{fig:fig_d1}
\end{subfigure}

\caption{\label{Fig:LWFLEX_phpp} The ladder-diagram contributions 
to the Luttinger-Ward functional~($\Phi$) in the SO(4) FLEX in the particle-hole channel in the first-order (a), second (b), 3rd (c) 
and 4th-order (d) in $U$. 
As in the previous figure, solid single-arrowed (blue) lines stand for normal Green's functions $G$, solid inward double-arrowed (green) lines the anomalous Green's functions $F$, 
outward double arrows $F^{*}$, and dashed lines the Hubbard interaction~$U$. 
The full SO(4) FLEX self-energy is constructed from the functional derivative of the Luttinger-Ward functionals which exemplified in Fig.~\ref{Fig:LWFLEX} and here.
}
\end{figure}

The normal self-energy, 
\begin{align}
\Sigma^{G}_{\bolds{q}} &= \delta \Phi/\delta G_{-\bolds{q}},  
\end{align}
and the anomalous self-energy, 
\begin{align}
\Sigma^{F}_{\bolds{q}} &= \delta \Phi/\delta F^{\dagger}_{-\bolds{q}},   \label{eq:anself}
\end{align}
are obtained from the functional derivative of the Luttinger-Ward functional 
$\Phi$, for which 
the expansion up to the forth-order in $U$ is displayed in Fig.~\ref{Fig:LWFLEX}.
Here, the normal component of the self-energy consisting of the 
particle-hole and transverse-spin contributions is given by
\begin{align}
 \Sigma^{G}_{ \sigma \sigma }(\overline{k} \overline{k}') =& 
 -\frac{1}{\beta N} \sum\limits_{\bolds{q}} {G}_{\sigma \sigma }(\overline{k}-q,\overline{k}'-q) \Gamma^{\rm ph}_{\sigma \sigma, \overline{k}\overline{k}'}(q)
%  \nonumber  \\
%   &-\frac{1}{\beta N} \sum\limits_{\bolds{q}} {G}_{\sigma \sigma }(q-\overline{k},q-\overline{k}') \Gamma_{\sigma \sigma, \overline{k}\overline{k}'}^{\rm pp}(q)
  ,
 \end{align}
while the anomalous component has
 \begin{align}
 \Sigma^{F}_{ \sigma \overline{\sigma} }(\overline{k} \overline{k}')= 
 -\frac{1}{\beta N} \sum\limits_{\bolds{q}} {F}_{\sigma \overline{\sigma} }(\overline{k}-q,\overline{k}'-q) \Gamma^{\rm scp}_{\sigma \overline{\sigma}, \overline{k}\overline{k}'}(q).
\end{align}
Here, $N$ is the total number of lattice sites, and 
\[
\overline{k}, \overline{k}' \equiv (\bolds{k},i\omega_{n})\; {\rm or}\; (\bolds{k}+\bolds{Q},i\omega_{n}),
\]
which comes from the structure of Eqs(\ref{eq:genGk},\ref{eq:flexselfmatrix}) 
and we have to make an appropriate choice depending on which 
matrix element in Eq(\ref{eq:flexselfmatrix}) is considered, 
and $q \equiv (\bolds{q},i\nu_{n}) $ with $\nu_{n}=2\pi n/ \beta$ being the Matsubara frequency for bosons.

Since the Luttinger-Ward functional here incorporates the anomalous part with the anomalous self-energy given in Eq.~\ref{eq:anself}, 
we should consider the local correction 
to the anomalous self-energy $\Sigma^{F}$ 
as $\Sigma^{F}_{\rm FLEX+DMFT} = \Sigma^{F}_{\rm FLEX} + \Sigma^{F}_{\rm loc}$.  
Now, our interest here is the anisotropic, 
$d$-wave pairing instability in the repulsive model, 
for which we can ignore the local correction to the anomalous self-energy $\Sigma^{F}_{\rm loc}$ which does not depend on momentum.
The remaining term, $\Sigma^{F}_{\rm FLEX}= \delta \Phi_{\rm FLEX}[G,F^{\dagger},F]/\delta F^{\dagger}$, is the
same as the right-hand side of the linearized Eliashberg equation 
(\ref{eq:Eliashberg} below) 
if we linearize the anomalous part.
Then our formalism treats the normal and anomalous self-energies consistently, 
as functional derivatives of the same Luttinger-Ward functional $\Phi_{\rm FLEX+DMFT}$~\cite{Kamenev2011, Kitatani2015}.

The particle-hole~(ph),
and superconducting pairing~(scp)
vertex functions are computed as~\cite{Yan2005,Yan2006}
\begin{align}
 \Gamma_{\overline{k} \overline{k}' , \sigma \sigma}^{\rm ph}(q) = & 
 \frac{3}{2} \frac{U^{2} \chi^{s0}_{\overline{k} \overline{k}' , \sigma \sigma}(q)}{1+U\chi^{s0}_{\overline{k} \overline{k}' , \sigma \sigma}(q)} \nonumber \\ &
 + \frac{1}{2} \frac{U^{2} \chi^{c0}_{\overline{k} \overline{k}' , \sigma \sigma}(q)}{1-U\chi^{c0}_{\sigma \sigma}(q)}
  -U^{2} \chi^{G}_{\overline{k} \overline{k}' , \sigma \sigma}(q),
  \nonumber \\
  = & \frac{3}{2} U^{2} \chi^{s}_{\sigma \sigma, \overline{k} \overline{k}'}(q) 
  +\frac{1}{2} U^{2} \chi^{c}_{\sigma \sigma, \overline{k} \overline{k}'}(q) 
  \nonumber \\ & \qquad \qquad
  - U^{2} \chi^{G}_{\sigma \sigma, \overline{k} \overline{k}'}(q).
  \end{align}
  \begin{align}
 \Gamma_{\overline{k} \overline{k}' ,\sigma \overline{\sigma}}^{\rm scp}(q) = & 
 \frac{3}{2} \frac{U^{2} \chi^{s0}_{\overline{k} \overline{k}' , \sigma \sigma}(q)}{1+U\chi^{s0}_{\overline{k} \overline{k}' , \sigma \sigma}(q)} \nonumber \\ &
 - \frac{1}{2} \frac{U^{2} \chi^{c0}_{\overline{k} \overline{k}' , \sigma \sigma}(q)}{1-U\chi^{c0}_{\overline{k} \overline{k}' , \sigma \sigma}(q)}
 -U^{2} \chi^{F}_{\overline{k} \overline{k}' ,\sigma \overline{\sigma}}(q),
 \nonumber \\
  = & \frac{3}{2} U^{2} \chi^{s}_{\sigma \sigma, \overline{k} \overline{k}'}(q) 
  -\frac{1}{2} U^{2} \chi^{c}_{\sigma \sigma, \overline{k} \overline{k}'}(q) 
  \nonumber \\& \qquad \qquad
  - U^{2} \chi^{F}_{\sigma \sigma, \overline{k} \overline{k}'}(q).
 \end{align}
There, the spin ($s$) and charge ($c$) susceptibilities are given by
\begin{align}
\chi^{s}_{\overline{k} \overline{k}' , \sigma \sigma}(q)= &  \frac{ \chi^{s0}_{\overline{k} \overline{k}' , \sigma \sigma}(q)}{1+U\chi^{s0}_{\overline{k} \overline{k}' , \sigma \sigma}(q)} ,\\
\chi^{c}_{\overline{k} \overline{k}' , \sigma \sigma}(q)= &   \frac{  \chi^{c0}_{\overline{k} \overline{k}' , \sigma \sigma}(q)}{1-U\chi^{c0}_{\overline{k} \overline{k}' , \sigma \sigma}(q)}, \\
\end{align}
with
\begin{align}
 \chi^{s0}_{\overline{k} \overline{k}' , \sigma \sigma}(q)= & \chi^{G}_{\overline{k} \overline{k}' , \sigma \sigma}(q) -\chi^{F}_{\overline{k} \overline{k}' ,\sigma \overline{\sigma}}(q) ,\\
 \chi^{c0}_{\overline{k} \overline{k}' , \sigma \sigma}(q)= & \chi^{G}_{\overline{k} \overline{k}' , \sigma \sigma}(q) +\chi^{F}_{\overline{k} \overline{k}' ,\sigma \overline{\sigma}}(q), \\
 \chi^{G}_{\overline{k} \overline{k}' , \sigma \sigma}(q)= &-\frac{1}{\beta N} \sum\limits_{k}  
 G_{\sigma \sigma} (\overline{k}+q,\overline{k}'+q) G_{\sigma \sigma}( \overline{k} \overline{k}'),\\
 \chi^{F}_{\overline{k} \overline{k}' ,\sigma \overline{\sigma}}(q) =& -\frac{1}{\beta N} \sum\limits_{k}  F^{*}_{\sigma \overline{\sigma} }(\overline{k}+q,\overline{k}'+q) F_{\sigma \overline{\sigma} }(\overline{k} \overline{k}') .
 \end{align}

In order to obtain the transition temperature for the converged Green's functions, we solve the linearized Eliashberg equation,
 \begin{align}
  \lambda_{\rm EL} & \Delta_{\sigma \overline{\sigma}}(\overline{k})=\nonumber \\
   & -\frac{1}{N \beta} \sum\limits_{\bolds{q}} V^{\rm eff}_{\sigma,\overline{k} \overline{k}'}(q)
  G_{\sigma \sigma}(\overline{k}', \overline{k}')G_{\overline{\sigma} \overline{\sigma}}(-\overline{k}', -\overline{k}')
  \Delta_{\sigma \overline{\sigma}}(\overline{k}'),
 \end{align} \label{eq:Eliashberg}
where 
$\lambda_{\rm EL}$ denotes the largest eigenvalue of the linearized Eliashberg equation, $\Delta_{\sigma \overline{\sigma}}$ is the the gap function, and 
$q=\overline{k}-\overline{k}'$ with choices of $\overline{k}$ and $\overline{k}'$ depending on the involved normal~($G$) matrix elements of the Green's function ${\cal G}_{4 \times 4}$ in Eq.~\ref{eq:Eliashberg}. The effective singlet-pairing interaction $V^{\rm eff}$ is
 \begin{equation}
  V^{\rm eff}_{\sigma, \overline{k} \overline{k}'}(q) = U + \frac{3}{2} U^{2} \chi^{s}_{\sigma \sigma, \overline{k} \overline{k}'}(q)
  - \frac{1}{2} U^{2} \chi^{c}_{\sigma \sigma, \overline{k} \overline{k}'}(q).
 \end{equation}
 The superconducting transition occurs when the maximum eigenvalue of the diagonalized Eliashberg equation reaches $\lambda_{\rm EL}=1$.

\section{Conclusion}\label{sec:conclusion}

In conclusion, we have proposed a novel formalism to explore 
correlated systems as exemplified by the one-band repulsive Hubbard model.
We have presented the SO(4) generalization of the FLEX+DMFT method so that both antiferromagnetism and superconductivity can be treated on an equal footing.
Namely, in the FLEX+DMFT formalism, we solve the bipartite impurity problem in the SO(4) DMFT by introducing a novel ``full-SU(2)" slave-boson impurity solver. This impurity solver respects the group-symmetry properties of the Hubbard model, namely spin SU(2) and pseudospin SU(2) symmetries. We have introduced bosonic and fermionic auxiliary particles to convey all the charge- and spin-related information in our impurity solver. 
This approach is particularly suitable to study the interplay between AFM and d-wave SC treating
the two phases on equal footing, while previous calculations based on cluster extension of DMFT~\cite{Capone2006}
can suffer of some bias due to the choice of the cluster.
Going over to FLEX+DMFT with the slave-boson framework with 
double self-consistent loops then incorporates 
improved k-dependent fluctuations. 
As our slave-particle decomposition is treated within the FLEX+DMFT approach, addressing correlated physics of antiferromagnetism and d-wave superconductivity is feasible. 
Furthermore, the decomposed nature of our impurity solver may shed light on the origin of, still puzzling, pseudogap physics. Numerical study will be desirable as a future work.

Extending our formalism to address multi-band physics~\cite{Bunemann2011} is another exciting 
direction to pursue. This will enable us not only explore the three-band model, more suitable for cuprates, but also enable us to examine multi-band superconductors, e.g., iron-based superconductors~\cite{Dai2015}, where the superconducting phase also sits adjacent to the magnetic phase.

\acknowledgments
We thank Martin Eckstein and Abolhassan Vaezi for a collaboration in early stages of this work.  
We also appreciate fruitful discussions with Motoharu Kitatani. 
H.A. wishes to thank hospitality of 
Department of Physics, ETH Z\"{u}rich, where the present manuscript 
was started.  
Sh.S. would like to thank Patrick A. Lee for pointing out the possibility of 
adding one more boson to the formalism. 
She also appreciates Makiko Nio and Tatsumi Aoyama for helpful instructions on performing diagrammatic
calculations with FORM~\cite{Vermaseren2000,Aoyama2012}.
Sh.S. and H.A. acknowledge a support from the ImPACT Program of
the Council for Science, Technology and Innovation, Cabinet
Office, Government of Japan (Grant No. 2015-PM12-05-01)
from JST. H.A. is also supported
by JSPS KAKENHI Grant No. JP26247057. M.C.  acknowledges support from H2020 Frame-work Programme, under ERC Advanced Grant No.692670~“FIRSTORM” and MIUR PRIN 2015~(Prot.2015C5SEJJ001) and SISSA/CNR project ”Superconductivity, Ferroelectricity and Magnetism in bad metals”~(Prot.232/2015).

\bibliography{Man_SB}

\pagebreak
\widetext
\begin{center}
\textbf{\large Supplemental Materials:\\
SO(4) FLEX+DMFT formalism with SU(2)$\otimes$SU(2)-symmetric impurity solver for superconductivity in the repulsive Hubbard model}
\end{center}

\setcounter{equation}{0}
\setcounter{section}{0}
\setcounter{figure}{0}
\setcounter{table}{0}
\setcounter{page}{1}
\makeatletter
\renewcommand{\theequation}{S\arabic{equation}}
\renewcommand{\thefigure}{S\arabic{figure}}
\renewcommand{\bibnumfmt}[1]{[S#1]}
\renewcommand{\citenumfont}[1]{S#1}

\onecolumngrid

\section{Bipartite full-SU(2) slave-boson impurity solver for the DMFT}

Employing the notation of the bipartite full-SU(2) slave-boson, we can rewrite the impurity Hamiltonian in Eq.~\eqref{eq:Hamimp} as
\begin{align}
 H_{\rm imp}& =
 \Big( \mu-\frac{U}{2} \Big) \big(  b^{\dagger}_{i 1} b_{i 1} + b^{\dagger}_{i 2} b_{i 2} \big)
+ \frac{U}{2} \sum\limits_{i }^{A,B}  \left(
 d^{\dagger}_{i 1} d_{i 1} + \epsilon_{i} d^{\dagger}_{i 1} d_{i 2} f_{i\uparrow} f_{ i\downarrow } 
 + \epsilon_{i} d^{\dagger}_{i 2} d_{i 1} f^{\dagger}_{i\downarrow} f^{\dagger}_{ i\uparrow }
  + d^{\dagger}_{2 i} d_{i 2 } f^{\dagger}_{i \downarrow} f^{\dagger}_{i \uparrow } f_{i \uparrow } f_{i \downarrow}
  \right) 
  \nonumber \\ &
   -\sum\limits_{i}^{A,B} \lambda^{+}_{i} \big(
  -2\epsilon_{i} f_{i \uparrow }f_{i \downarrow}
   + 2 b_{i 2}^{\dagger}b_{i 1} +2 d^{\dagger}_{i 2}d_{i 1}
  \big)
  -\sum\limits_{i }^{A,B} \lambda^{-}_{i} \big(
   -2\epsilon_{i} f^{\dagger}_{i \downarrow}f^{\dagger}_{i \uparrow }
   +2 b^{\dagger}_{i 1}b_{i 2}+2d^{\dagger}_{i 1}d_{i 2}\big)
   \nonumber \\ &
  -\sum\limits_{i }^{A,B} \lambda^{z}_{i}
  \big(f^{\dagger}_{i \uparrow }f_{i \uparrow }+f^{\dagger}_{i \downarrow}f_{i \downarrow}
   +d^{\dagger}_{i 1}d_{i 1}-d^{\dagger}_{i 2}d_{i 2}
   +b^{\dagger}_{i 1}b_{i 1}-b^{\dagger}_{i 2}b_{i 2} \big)
   \nonumber \\ &
 +\frac{ v_{1}}{2} \Big[
  \big( f_{A \uparrow }f^{\dagger}_{B \uparrow }  
 + f_{A \downarrow }f^{\dagger}_{B \downarrow }   \big)
 \big( b_{A 2}b^{\dagger}_{B 2} - b_{A 2}d_{B 1} 
 - d^{\dagger}_{A 1}b^{\dagger}_{B 2} + d^{\dagger}_{A 1}d_{B 1} \big)
 \nonumber \\& \qquad
+ \big( f^{\dagger}_{B \uparrow }f_{A \uparrow }  + f^{\dagger}_{B \downarrow }f_{A \downarrow }  \big)
\big( - b^{\dagger}_{A 1}b_{B 1} - b^{\dagger}_{A 1}d^{\dagger}_{B 2} - d_{A 2}b_{B 1} - d_{A 2}d^{\dagger}_{B 2} \big)
\nonumber \\& \qquad
+ \big( f^{\dagger}_{A \uparrow }f_{B \uparrow } + f^{\dagger}_{A \downarrow }f_{B \downarrow }  \big) 
\big(  - b_{A 1}b^{\dagger}_{B 1} - b_{A 1}d_{B 2} - d^{\dagger}_{A 2}b^{\dagger}_{B 1} - d^{\dagger}_{A 2}d_{B 2} \big) 
 \nonumber \\& \qquad
+ \big( f_{B \uparrow }f^{\dagger}_{A \uparrow }  + f_{B \downarrow }f^{\dagger}_{A \downarrow } \big)
\big( b^{\dagger}_{A 2}b_{B 2} - b^{\dagger}_{A 2}d^{\dagger}_{B 1} - d_{A 1}b_{B 2} + d_{A 1}d^{\dagger}_{B 1}\big)
\nonumber \\& \qquad
+ \big( f^{\dagger}_{A \uparrow }f^{\dagger}_{B \downarrow } -f^{\dagger}_{A \downarrow }f^{\dagger}_{B \uparrow } \big)
( b_{A 1}b^{\dagger}_{B 2} - b_{A 1}d_{B 1} + d^{\dagger}_{A 2}b^{\dagger}_{B 2} - d^{\dagger}_{A 2}d_{B 1} )
 \nonumber \\& \qquad
+ \big( f^{\dagger}_{B \downarrow }f^{\dagger}_{A \uparrow } - f^{\dagger}_{B \uparrow }f^{\dagger}_{A \downarrow } \big) 
\big( b^{\dagger}_{A 2}b_{B 1} + b^{\dagger}_{A 2}d^{\dagger}_{B 2} - d_{A 1}b_{B 1} - d_{A 1}d^{\dagger}_{B 2} \big)
\nonumber \\& \qquad
+ \big( f_{B \downarrow }f_{A \uparrow }  - f_{B \uparrow }f_{A \downarrow } \big)
\big( b^{\dagger}_{A 1}b_{B 2} - b^{\dagger}_{A 1}d^{\dagger}_{B 1} + d_{A 2}b_{B 2} - d_{A 2}d^{\dagger}_{B 1} \big)
 \nonumber \\ & \qquad
+ \big( f_{A \uparrow }f_{B \downarrow } - f_{A \downarrow }f_{B \uparrow } \big) 
\big( b_{A 2}b^{\dagger}_{B 1} + b_{A 2}d_{B 2} - d^{\dagger}_{A 1}b^{\dagger}_{B 1} - d^{\dagger}_{A 1}d_{B 2} \big)
\Big]
  ,\label{eq:Himp2}
\end{align}
where ($\lambda^{z}, \lambda^{+}, \lambda^{-}$) are Lagrange multipliers for enforcing the SU(2) vector constraints in Eq.~\eqref{eq:Kvec}. The SU(2) nature of the doubly-occupied state imposes the sextic terms (the last term in the first line) in the impurity Hamiltonian.
As this term is local in time for simplicity  we would apply a mean-field contraction in the spinon sector such that
\begin{align}
f^{\dagger}_{i \downarrow } f^{\dagger}_{i \uparrow } f_{i \uparrow } f_{i \downarrow} =
n^{*F}_{ ii \downarrow \uparrow } f_{i \uparrow } f_{i \downarrow}
+n^{F}_{ ii \uparrow  \downarrow } f^{\dagger}_{i \uparrow } f^{\dagger}_{i \downarrow}
+n_{ii \downarrow  \downarrow } f^{\dagger}_{i \uparrow } f_{i \uparrow }
+ n_{ii \uparrow  \uparrow } f^{\dagger}_{i \downarrow}f_{i \downarrow},
\end{align}
where density operators read
\begin{align}
 &n^{*F}_{ ii \downarrow \uparrow } = \ave{f^{\dagger}_{i \downarrow} f^{\dagger}_{i \uparrow}},\quad
 n^{F}_{ii \uparrow  \downarrow } = \ave{f_{i \uparrow} f_{i \downarrow}},\nonumber \\
 &n_{ii \downarrow  \downarrow } = \ave{f^{\dagger}_{i \downarrow} f_{i \downarrow}},\quad
 n_{ ii \downarrow  \uparrow } = \ave{f^{\dagger}_{i \downarrow} f_{i \uparrow}}.
\end{align}
This impurity Hamiltonian is associated with the action in the slave-boson language, 
\begin{align}
 {\cal S}_{\rm imp}=& 
 \int_{0}^{\beta} \Big[ 
 \sum\limits_{ij }^{A,B} \sum\limits_{\sigma} f^{\dagger}_{i \sigma} \partial_{\tau} f_{i \sigma} +H_{\rm imp} \nonumber \\
 &+ \sum\limits_{ij }^{A,B} 
 \big(
 b^{\dagger}_{i 1} \partial_{\tau} b_{i 1} 
 +b^{\dagger}_{i 2} \partial_{\tau} b_{i 2} 
 +d^{\dagger}_{i 1} \partial_{\tau} d_{i 1} 
 +d^{\dagger}_{i 2} \partial_{\tau} d_{i 2} 
 \big)
  \Big]
  \nonumber \\
 &
 +\iint_{0}^{\beta} {\rm d}\tau {\rm d}\tau'
\sum\limits_{ij }^{A,B} \sum\limits_{\sigma}
c^{\dagger}_{i \sigma} (\tau) 
{\mathfrak{G}}_{ij \sigma \sigma }(\tau-\tau')
c_{j \sigma} (\tau') 
\nonumber \\
 &
 +\iint_{0}^{\beta} {\rm d}\tau {\rm d}\tau'
\sum\limits_{ij }^{A,B} \sum\limits_{\sigma}
c^{\dagger}_{i \sigma} (\tau) 
{\mathfrak{F}}_{ij \sigma \overline{\sigma} }(\tau-\tau')
c^{\dagger}_{j \overline{\sigma} } (\tau').\label{eq:S2}
\end{align}
Here electron operators are defined in Eq.(\ref{eq:c_SU2}) 
in the main text, and the $4\times 4$ matrix representation of the hybridization function has the form of

 \begin{align}
 {\cal D}(\tau-\tau') = 
\begin{pmatrix}
    {\mathfrak G}_{AA \uparrow \uparrow }(\tau-\tau')  
  & {\mathfrak G}_{AB \uparrow \uparrow }(\tau-\tau') 
  & {\mathfrak F}_{AA \uparrow  \downarrow }(\tau-\tau')  
  & -{\mathfrak F}_{AB \uparrow  \downarrow }(\tau-\tau')  
  \\
    {\mathfrak G}_{BA \downarrow \uparrow }(\tau-\tau')  
  & {\mathfrak G}_{BB \downarrow \uparrow }(\tau-\tau')  
  & {\mathfrak F}_{BA \downarrow \downarrow }(\tau-\tau')  
  & -{\mathfrak F}_{BB \downarrow \downarrow }(\tau-\tau') 
 \\
    {\mathfrak F}^{*}_{AA \uparrow  \uparrow } (\tau-\tau')  
  & {\mathfrak F}^{*}_{AB \uparrow  \uparrow } (\tau-\tau')
  & {\mathfrak G}^{*}_{AA \uparrow  \downarrow } (\tau-\tau')
  & -{\mathfrak G}^{*}_{AB \uparrow  \downarrow} (\tau-\tau')
  \\
    -{\mathfrak F}^{*}_{BA \downarrow \uparrow  } (\tau-\tau')
  & -{\mathfrak F}^{*}_{BB \downarrow  \uparrow } (\tau-\tau') 
  & -{\mathfrak G}^{*}_{BA \downarrow  \downarrow } (\tau-\tau')
  & {\mathfrak G}^{*}_{BB \downarrow  \downarrow } (\tau-\tau')
\end{pmatrix},\label{eq:Hybk}
\end{align}

\begin{figure}[htp]

\includegraphics[width=\linewidth]{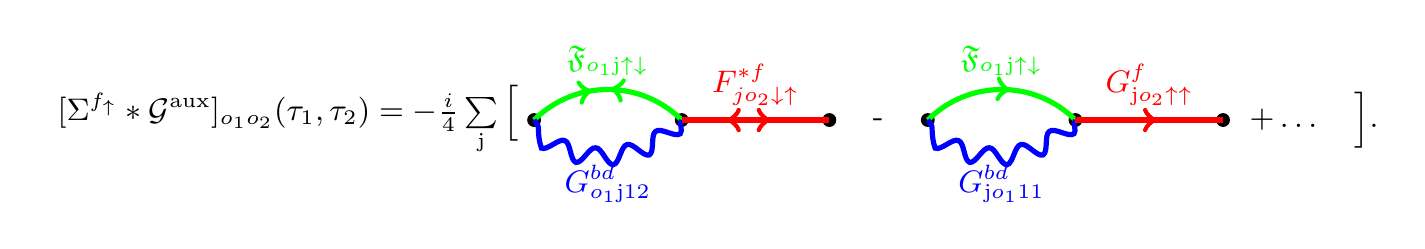}

\caption{\label{Fig:sigma_fup} The self-energy diagrams for the $f_{\uparrow}$ particles. ${\cal G}^{\rm aux}$ stands for $G^{aa'}$ and $F^{aa'}$. Green~(red) arrows denotes hybridization functions~(spinon Green's functions), while blue wiggly lines are the holon/doublon Green's functions.}
\end{figure}

\begin{figure}[htp]

\includegraphics[width=\linewidth]{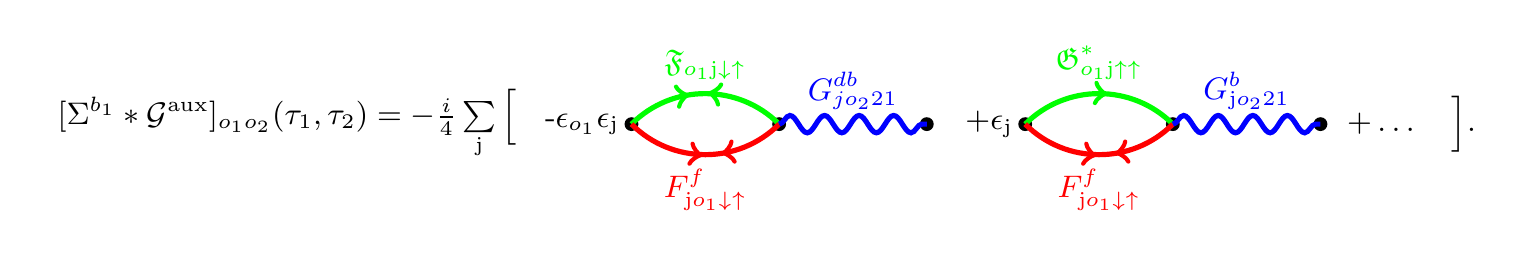}

\caption{\label{Fig:sigma_b1} The self-energy diagrams for the $b_{1}$ particles.  ${\cal G}^{\rm aux}$ stands for $G^{aa'}$ and $F^{aa'}$. Green~(red) arrows denotes hybridization functions~(spinon Green's functions), while blue wiggly lines are the holon/doublon Green's functions.}
\end{figure}

\begin{figure}[htp]

\begin{subfigure}[t]{.9\textwidth}
\vspace{0pt}
\includegraphics[width=\linewidth]{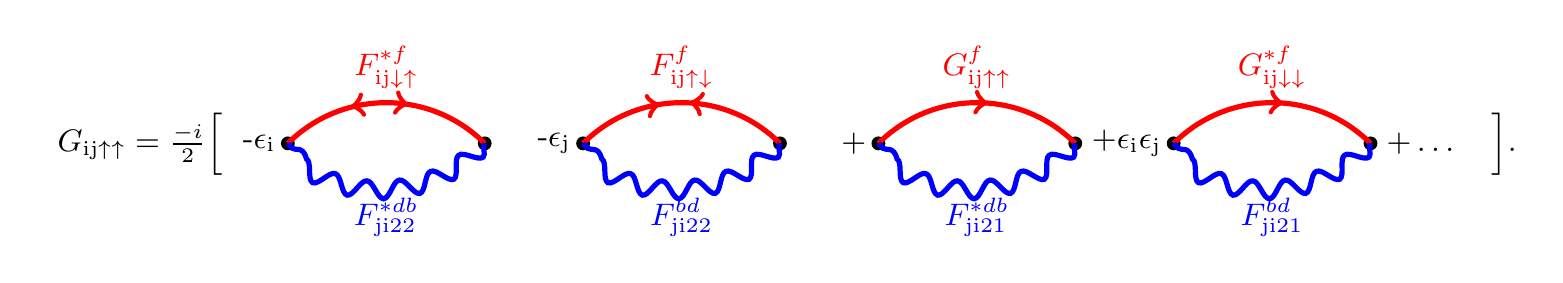}
 \caption{}\label{fig:fig_a1}
\end{subfigure}

\medskip

\begin{subfigure}[t]{.9\textwidth}
\vspace{0pt}
\includegraphics[width=\linewidth]{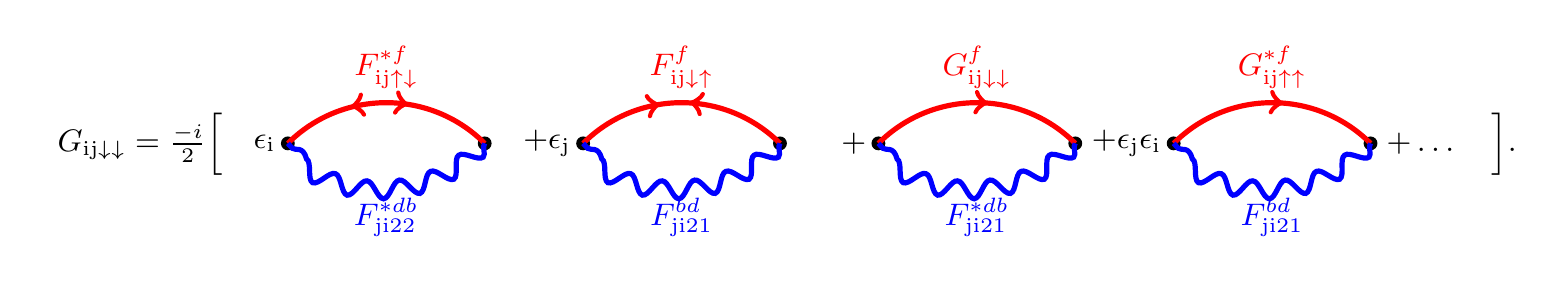}
 \caption{}\label{fig:fig_b1}
\end{subfigure}

\medskip

\begin{subfigure}[t]{.9\textwidth}
\vspace{0pt}
\includegraphics[width=\linewidth]{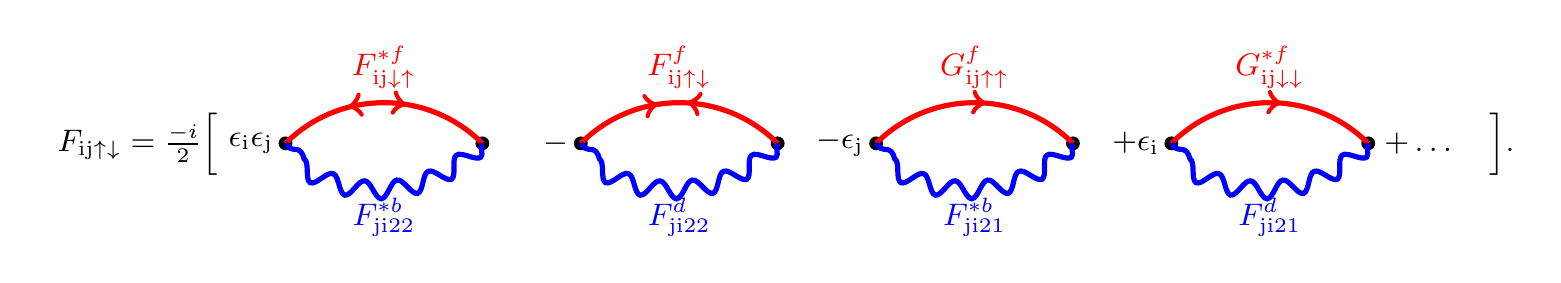}
 \caption{}\label{fig:fig_c1}
\end{subfigure}

 \medskip

\begin{subfigure}[t]{.9\textwidth}
\vspace{0pt}
\includegraphics[width=\linewidth]{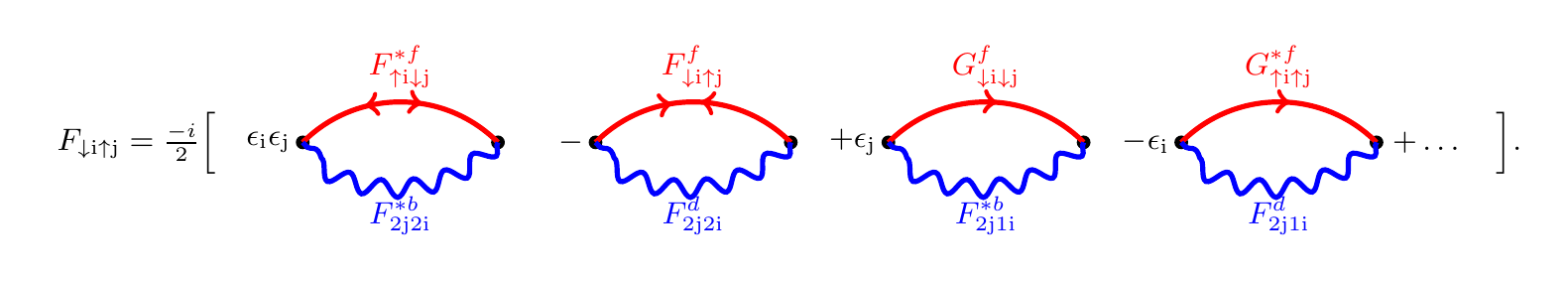}
 \caption{}\label{fig:fig_d1}
\end{subfigure}

\caption{\label{Fig:Gel_SB} Green's functions in terms of slave-particle Green's functions. Red arrows denote spinon Green's functions, while blue wiggly lines are the holon/doublon Green's functions.}
\end{figure}

Using Eq.~\eqref{eq:S2}, we derive the equations of motion for all of spinon, holon, and doublon flavors.
Here we employ the generic notation for the auxiliary Green's functions, 
\begin{align}
 G^{a a'}_{ij \alpha \alpha' }(\tau-\tau') &= - \ave{ a_{i \alpha }(\tau) a'^{\dagger}_{j \alpha' }(\tau') },\nonumber \\
 F^{a a'}_{ij \alpha \alpha' }(\tau-\tau') &= - \ave{ a_{i \alpha }(\tau) a'_{j \alpha' }(\tau') },
\end{align}
where $a\text{ and }a' \in \{f,b,d\}$ label the auxiliary particles, $\alpha$ is the flavor of the particular slave-particle which is $\{\uparrow, \downarrow\}$ for spinons and $\{1,2\}$ for bosonic operators, and $i,j$ denote sublattice indices. When Green's functions consist only one species of the auxiliary particle, namely both $a$ and $a'$ are spinon, holon, or doublon, we refrain from repeating these labels for the Green's functions as $G^{a a'}~(F^{a a'})$, 
with a shorthand $G^{a a}\equiv G^{a}, 
F^{a a}\equiv F^{a}$ when $a=a'$.  
The self-energy diagrams for these auxiliary Green's function, denoted generically as $\cal G^{\rm aux}$, are exemplified in Figs.~\ref{Fig:sigma_fup} and \ref{Fig:sigma_b1}. Using the auxiliary particles, we can construct the Green's functions as presented in Sec.~\ref{Sec:Gel} and plotted in Fig.~\ref{Fig:Gel_SB}.

\subsection{Equation of motion for $f_{A \uparrow}$ particles }

%   \subsection{Equation of motion for $f_{\uparrow A}$ particle}

  \begin{align}
%   derivative
  & \partial_{ \tau}  G^{f}_{AA \uparrow \uparrow }(\tau-\tau_{2})
  -\lambda_{A}^{-} F^{*f}_{AA \downarrow \uparrow }(\tau-\tau_{2}) 
  +\lambda_{A}^{z} G^{f}_{AA \uparrow \uparrow }(\tau-\tau_{2}) 
  -\frac{U}{2}  n^{dd}_{AA 21}(\tau) F^{*f}_{AA \downarrow \uparrow }(\tau-\tau_{2}) 
  \nonumber \\ &
  -\frac{U}{2} \Big[
  + n^{dd}_{AA 22}(\tau)n^{F}_{AA \uparrow \downarrow }(\tau)  F^{*f}_{AA \downarrow \uparrow }(\tau-\tau_{2}) 
  + n^{dd}_{AA 22}(\tau) n_{AA \downarrow \downarrow }(\tau)  G^{f}_{AA \uparrow \uparrow }(\tau-\tau_{2}) 
  \Big]
  \nonumber \\ &
  \frac{v_{1}}{2} \Big[
  + F^{*f}_{BA \downarrow \uparrow }(\tau-\tau_{2}) \Big(  
  - n^{bb}_{BA 21}(\tau) 
  + n^{db}_{AB 11}(\tau)
  - n^{*db}_{AB 22}(\tau) 
  + n^{dd}_{AB 21}(\tau) 
  - n^{bb}_{AB 21}(\tau) 
  + n^{bd}_{BA 11}(\tau) 
  - n^{*db}_{BA 22}(\tau) 
  + n^{d}_{BA 21}(\tau) 
  \Big)
  \nonumber \\ & \qquad
  + G^{f}_{BA \uparrow \uparrow }(\tau-\tau_{2}) \Big( 
  n^{bb}_{BA 11}(\tau)
  + n^{bd}_{AB 12}(\tau) 
  + n^{*db}_{AB 21}(\tau) 
  + n^{dd}_{AB 22}(\tau) 
  - n^{bb}_{AB 22}(\tau) 
  + n^{bd}_{BA 21}(\tau) 
  + n^{*db}_{BA 12}(\tau) 
  - n^{d}_{BA 11}(\tau) 
  \Big)
  \Big]
   \nonumber \\ &
%    derivative with t, integration over tau'
 + \frac{1}{2} \int {\rm d}{\tau'}\Big[
+ F^{*f}_{BA \downarrow \uparrow }(\tau-\tau_{2}) \Big(  
% - {\mathfrak F}^{*}_{AB \downarrow \uparrow }(\tau-\tau') G^{bd}_{AB 12}(\tau-\tau') 
% + {\mathfrak F}^{*}_{AB \downarrow \uparrow }(\tau-\tau') F^{dd}_{AB 11}(\tau-\tau') 
% \nonumber \\ & \qquad \qquad
% + {\mathfrak F}^{*}_{AB \downarrow \uparrow }(\tau-\tau')  F^{*bb}_{AB 22}(\tau-\tau') 
% - {\mathfrak G}^{*}_{AB \downarrow \downarrow }(\tau-\tau') F^{*bd}_{AB 22}(\tau-\tau')
% \nonumber \\ & \qquad \qquad
% + {\mathfrak G}^{*}_{AB \downarrow \downarrow }(\tau-\tau') F^{bd}_{AB 11}(\tau-\tau') 
% + {\mathfrak G}^{*}_{AB \downarrow \downarrow }(\tau-\tau') G^{dd}_{AB 12}(\tau-\tau') 
% \nonumber \\ & \qquad \qquad
+ {\mathfrak F}_{AB \uparrow \downarrow }(\tau-\tau') G^{bd}_{AB 12}(\tau-\tau') 
+ {\mathfrak F}_{AB \uparrow \downarrow }(\tau-\tau') F^{*dd}_{AB 22}(\tau-\tau') 
\nonumber \\ & \qquad \qquad
+ {\mathfrak F}_{AB \uparrow \downarrow }(\tau-\tau') F^{bb}_{AB 11}(\tau-\tau')
- {\mathfrak G}_{AB \uparrow \uparrow }(\tau-\tau') F^{*bd}_{AB 22}(\tau-\tau') 
\nonumber \\ & \qquad \qquad
+ {\mathfrak G}_{AB \uparrow \uparrow }(\tau-\tau') F^{bd}_{AB 11}(\tau-\tau') 
- {\mathfrak G}_{AB \uparrow \uparrow }(\tau-\tau') G^{bb}_{AB 12}(\tau-\tau')
\nonumber \\ & \qquad \qquad
+ {\mathfrak G}_{AB \downarrow \downarrow }(\tau-\tau') G^{*bb}_{AB 21}(\tau-\tau')
+ {\mathfrak F}_{AB \downarrow \uparrow }(\tau-\tau') G^{*bd}_{AB 21}(\tau-\tau')
% \nonumber \\ & \qquad \qquad
% + {\mathfrak F}^{*}_{AB \uparrow \downarrow }(\tau-\tau') G^{*db}_{AB 21}(\tau-\tau')
% + {\mathfrak G}^{*}_{AB \uparrow \uparrow }(\tau-\tau')  G^{*dd}_{AB 21}(\tau-\tau')
  \Big)
 \nonumber \\ & \qquad
+ G^{f}_{BA \uparrow \uparrow }(\tau-\tau_{2}) \Big( 
%   {\mathfrak F}^{*}_{AB \downarrow \uparrow }(\tau-\tau') G^{bd}_{AB 11}(\tau-\tau') 
% + {\mathfrak F}^{*}_{AB \downarrow \uparrow }(\tau-\tau')F^{dd}_{AB 12}(\tau-\tau') 
% \nonumber \\ & \qquad \qquad
% - {\mathfrak F}^{*}_{AB \downarrow \uparrow }(\tau-\tau')F^{*bb}_{AB 21}(\tau-\tau') 
% + {\mathfrak G}^{*}_{AB \downarrow \downarrow }(\tau-\tau')F^{*bd}_{AB 21}(\tau-\tau') 
% \nonumber \\ & \qquad \qquad
% + {\mathfrak G}^{*}_{AB \downarrow \downarrow }(\tau-\tau')F^{bd}_{AB 12}(\tau-\tau')
% - {\mathfrak G}^{*}_{AB \downarrow \downarrow }(\tau-\tau')G^{dd}_{AB 11}(\tau-\tau') 
% \nonumber \\ & \qquad \qquad
- {\mathfrak F}_{AB \uparrow \downarrow }(\tau-\tau') G^{bd}_{AB 11}(\tau-\tau') 
- {\mathfrak F}_{AB \uparrow \downarrow }(\tau-\tau')F^{*dd}_{AB 21}(\tau-\tau') 
\nonumber \\ & \qquad \qquad
+ {\mathfrak F}_{AB \uparrow \downarrow }(\tau-\tau')F^{bb}_{AB 12}(\tau-\tau')
+ {\mathfrak G}_{AB \uparrow \uparrow }(\tau-\tau')F^{*bd}_{AB 21}(\tau-\tau') 
\nonumber \\ & \qquad \qquad
+ {\mathfrak G}_{AB \uparrow \uparrow }(\tau-\tau')F^{bd}_{AB 12}(\tau-\tau')
+ {\mathfrak G}_{AB \uparrow \uparrow }(\tau-\tau') G^{bb}_{AB 11}(\tau-\tau') 
\nonumber \\ & \qquad \qquad
- {\mathfrak G}_{AB \downarrow \downarrow }(\tau-\tau') G^{*bb}_{AB 22}(\tau-\tau')
- {\mathfrak F}_{AB \downarrow \uparrow }(\tau-\tau') G^{bd}_{AB 22}(\tau-\tau')
% \nonumber \\ & \qquad \qquad
% + {\mathfrak F}^{*}_{AB \uparrow \downarrow }(\tau-\tau') G^{db}_{AB 22}(\tau-\tau')
% + {\mathfrak G}^{*}_{AB \uparrow \uparrow }(\tau-\tau') G^{dd}_{AB 22}(\tau-\tau') 
  \Big)
\nonumber \\ & \qquad
+ F^{*f}_{AA \downarrow \uparrow }(\tau'-\tau_{2}) \Big( 
%   {\mathfrak F}^{*}_{AA \downarrow \uparrow }(\tau-\tau') B^{bd}_{AA 12}(\tau-\tau') 
% - {\mathfrak F}^{*}_{AA \downarrow \uparrow }(\tau-\tau') F^{dd}_{AA 11}(\tau-\tau') 
% \nonumber \\ & \qquad \qquad
% - {\mathfrak F}^{*}_{AA \downarrow \uparrow }(\tau-\tau') F^{*bb}_{AA 22}(\tau-\tau') 
% - {\mathfrak G}^{*}_{AA \downarrow \downarrow }(\tau-\tau') F^{*bd}_{AA 22}(\tau-\tau')
% \nonumber \\ & \qquad \qquad
% + {\mathfrak G}^{*}_{AA \downarrow \downarrow }(\tau-\tau') F^{bd}_{AA 11}(\tau-\tau') 
% + {\mathfrak G}^{*}_{AA \downarrow \downarrow }(\tau-\tau') G^{dd}_{AA 12}(\tau-\tau') 
% \nonumber \\ & \qquad \qquad
+ {\mathfrak F}_{AA \uparrow \downarrow }(\tau-\tau') B^{bd}_{AA 12}(\tau-\tau') 
+ {\mathfrak F}_{AA \uparrow \downarrow }(\tau-\tau') F^{*dd}_{AA 22}(\tau-\tau') 
\nonumber \\ & \qquad \qquad
+ {\mathfrak F}_{AA \uparrow \downarrow }(\tau-\tau') F^{bb}_{AA 11}(\tau-\tau')
+ {\mathfrak G}_{AA \uparrow \uparrow }(\tau-\tau') F^{*bd}_{AA 22}(\tau-\tau') 
\nonumber \\ & \qquad \qquad
- {\mathfrak G}_{AA \uparrow \uparrow }(\tau-\tau') F^{bd}_{AA 11}(\tau-\tau') 
+ {\mathfrak G}_{AA \uparrow \uparrow }(\tau-\tau') G^{bb}_{AA 12}(\tau-\tau') 
\nonumber \\ & \qquad \qquad
+ {\mathfrak F}_{AA \downarrow \uparrow }(\tau-\tau') G^{*bd}_{AA 21}(\tau-\tau')
- {\mathfrak G}_{AA \downarrow \downarrow }(\tau-\tau') G^{*bb}_{AA 21}(\tau-\tau')
% \nonumber \\ & \qquad \qquad
% - {\mathfrak G}^{*}_{AA \uparrow \uparrow }(\tau-\tau') G^{*dd}_{AA 21}(\tau-\tau')
% + {\mathfrak F}^{*}_{AA \uparrow \downarrow }(\tau-\tau') G^{*db}_{AA 21}(\tau-\tau')
  \Big)
\nonumber \\ & \qquad
+ G^{f}_{AA \uparrow \uparrow }(\tau'-\tau_{2}) \Big( 
%   {\mathfrak F}^{*}_{AA \downarrow \uparrow }(\tau-\tau') G^{bd}_{AA 11}(\tau-\tau') 
% + {\mathfrak F}^{*}_{AA \downarrow \uparrow }(\tau-\tau') F^{dd}_{AA 12}(\tau-\tau')
% \nonumber \\ & \qquad \qquad
% - {\mathfrak F}^{*}_{AA \downarrow \uparrow }(\tau-\tau') F^{*bb}_{AA 21}(\tau-\tau') 
% - {\mathfrak G}^{*}_{AA \downarrow \downarrow }(\tau-\tau') F^{*bd}_{AA 21}(\tau-\tau') 
% \nonumber \\ & \qquad \qquad
% - {\mathfrak G}^{*}_{AA \downarrow \downarrow }(\tau-\tau') F^{bd}_{AA 12}(\tau-\tau') 
% + {\mathfrak G}^{*}_{AA \downarrow \downarrow }(\tau-\tau') G^{dd}_{AA 11}(\tau-\tau') 
% \nonumber \\ & \qquad \qquad
+ {\mathfrak F}_{AA \uparrow \downarrow }(\tau-\tau') G^{bd}_{AA 11}(\tau-\tau') 
+ {\mathfrak F}_{AA \uparrow \downarrow }(\tau-\tau') F^{*dd}_{AA 21}(\tau-\tau') 
\nonumber \\ & \qquad \qquad
- {\mathfrak F}_{AA \uparrow \downarrow }(\tau-\tau') F^{bb}_{AA 12}(\tau-\tau')
+ {\mathfrak G}_{AA \uparrow \uparrow }(\tau-\tau') F^{*bd}_{AA 21}(\tau-\tau') 
\nonumber \\ & \qquad \qquad
+ {\mathfrak G}_{AA \uparrow \uparrow }(\tau-\tau') F^{bd}_{AA 12}(\tau-\tau') 
+ {\mathfrak G}_{AA \uparrow \uparrow }(\tau-\tau') G^{bb}_{AA 11}(\tau-\tau') 
\nonumber \\ & \qquad \qquad
- {\mathfrak F}_{AA \downarrow \uparrow }(\tau-\tau') G^{*bd}_{AA 22}(\tau-\tau')  
+ {\mathfrak G}_{AA \downarrow \downarrow }(\tau-\tau') G^{*bb}_{AA 22}(\tau-\tau') 
% \nonumber \\ & \qquad \qquad
% + {\mathfrak G}^{*}_{AA \uparrow \uparrow }(\tau-\tau') G^{*dd}_{AA 22}(\tau-\tau')  
% - {\mathfrak F}^{*}_{AA \uparrow \downarrow }(\tau-\tau') G^{*db}_{AA 22}(\tau-\tau')  
  \Big)
  \Big]
   \nonumber \\ &
%    derivative with t', integration over tau
+ \frac{1}{2} \int {\rm d}{\tau}\Big[
+ F^{*f}_{BA \downarrow \uparrow }(\tau-\tau_{2}) \Big(  
% - {\mathfrak F}^{*}_{BA \uparrow \downarrow}(\tau-\tau') G^{bd}_{BA 12}(\tau-\tau') 
% + {\mathfrak F}^{*}_{BA \uparrow \downarrow}(\tau-\tau') F^{dd}_{BA 11}(\tau-\tau')
% \nonumber \\ & \qquad \qquad
% + {\mathfrak F}^{*}_{BA \uparrow \downarrow}(\tau-\tau') F^{*bb}_{BA 22}(\tau-\tau') 
% - {\mathfrak G}^{*}_{BA \uparrow \uparrow }(\tau-\tau') F^{*bd}_{BA 22}(\tau-\tau') 
% \nonumber \\ & \qquad \qquad
% + {\mathfrak G}^{*}_{BA \uparrow \uparrow }(\tau-\tau') F^{bd}_{BA 11}(\tau-\tau') 
% + {\mathfrak G}^{*}_{BA \uparrow \uparrow }(\tau-\tau') G^{dd}_{BA 12}(\tau-\tau') 
% \nonumber \\ & \qquad \qquad
+ {\mathfrak F}_{BA \downarrow \uparrow }(\tau-\tau') G^{bd}_{BA 12}(\tau-\tau') 
+ {\mathfrak F}_{BA \downarrow \uparrow }(\tau-\tau') F^{*dd}_{BA 22}(\tau-\tau') 
\nonumber \\ & \qquad \qquad
+ {\mathfrak F}_{BA \downarrow \uparrow }(\tau-\tau') F^{bb}_{BA 11}(\tau-\tau') 
- {\mathfrak G}_{BA \downarrow \downarrow}(\tau-\tau') F^{*bd}_{BA 22}(\tau-\tau') 
\nonumber \\ & \qquad \qquad
+ {\mathfrak G}_{BA \downarrow \downarrow}(\tau-\tau') F^{bd}_{BA 11}(\tau-\tau') 
- {\mathfrak G}_{BA \downarrow \downarrow}(\tau-\tau') G^{bb}_{BA 12}(\tau-\tau') 
\nonumber \\ & \qquad \qquad
+ {\mathfrak G}_{BA \downarrow \downarrow}(\tau-\tau') G^{*dd}_{BA 21}(\tau-\tau')  
+ {\mathfrak F}_{BA \downarrow \uparrow }(\tau-\tau') G^{*db}_{BA 21}(\tau-\tau') 
% \nonumber \\ & \qquad \qquad
% - {\mathfrak F}^{*}_{BA \uparrow \downarrow}(\tau-\tau') G^{*bd}_{BA 21}(\tau-\tau')  
% - {\mathfrak G}^{*}_{BA \uparrow \uparrow }(\tau-\tau') G^{*bb}_{BA 21}(\tau-\tau')  
\Big)
\nonumber \\ & \qquad
+ G^{f}_{BA \uparrow \uparrow }(\tau-\tau_{2}) \Big(  
% - {\mathfrak F}^{*}_{BA \uparrow \downarrow}(\tau-\tau') G^{bd}_{BA 22}(\tau-\tau') 
% + {\mathfrak F}^{*}_{BA \uparrow \downarrow}(\tau-\tau') F^{dd}_{BA 21}(\tau-\tau') 
% \nonumber \\ & \qquad \qquad
% - {\mathfrak F}^{*}_{BA \uparrow \downarrow}(\tau-\tau') F^{*bb}_{BA 12}(\tau-\tau') 
% + {\mathfrak G}^{*}_{BA \uparrow \uparrow }(\tau-\tau') F^{*bd}_{BA 12}(\tau-\tau') 
% \nonumber \\ & \qquad \qquad
% + {\mathfrak G}^{*}_{BA \uparrow \uparrow }(\tau-\tau') F^{bd}_{BA 21}(\tau-\tau') 
% + {\mathfrak G}^{*}_{BA \uparrow \uparrow }(\tau-\tau') G^{dd}_{BA 22}(\tau-\tau') 
% \nonumber \\ & \qquad \qquad
+ {\mathfrak F}_{BA \downarrow \uparrow }(\tau-\tau') G^{bd}_{BA 22}(\tau-\tau') 
- {\mathfrak F}_{BA \downarrow \uparrow }(\tau-\tau') F^{*dd}_{BA 12}(\tau-\tau') 
\nonumber \\ & \qquad \qquad
+ {\mathfrak F}_{BA \downarrow \uparrow }(\tau-\tau') F^{bb}_{BA 21}(\tau-\tau') 
+ {\mathfrak G}_{BA \downarrow \downarrow}(\tau-\tau') F^{*bd}_{BA 12}(\tau-\tau') 
\nonumber \\ & \qquad \qquad
+ {\mathfrak G}_{BA \downarrow \downarrow}(\tau-\tau') F^{bd}_{BA 21}(\tau-\tau') 
- {\mathfrak G}_{BA \downarrow \downarrow}(\tau-\tau') G^{bb}_{BA 22}(\tau-\tau') 
\nonumber \\ & \qquad \qquad
- {\mathfrak G}_{BA \downarrow \downarrow}(\tau-\tau') G^{*dd}_{BA 11}(\tau-\tau')  
- {\mathfrak F}_{BA \downarrow \uparrow }(\tau-\tau') G^{*db}_{BA 11}(\tau-\tau') 
% \nonumber \\ & \qquad \qquad
% + {\mathfrak F}^{*}_{BA \uparrow \downarrow}(\tau-\tau') G^{bd}_{BA 11}(\tau-\tau')  
% + {\mathfrak G}^{*}_{BA \uparrow \uparrow }(\tau-\tau') G^{*bb}_{BA 11}(\tau-\tau')  
\Big)
\nonumber \\ & \qquad
+ F^{*f}_{AA \downarrow \uparrow }(\tau-\tau_{2}) \Big( 
% {  \mathfrak F}^{*}_{AA \uparrow \downarrow }(\tau-\tau') B^{bd}_{AA 12}(\tau-\tau') 
% - {\mathfrak F}^{*}_{AA \uparrow \downarrow }(\tau-\tau') F^{dd}_{AA 11}(\tau-\tau') 
% \nonumber \\ & \qquad \qquad
% - {\mathfrak F}^{*}_{AA \uparrow \downarrow }(\tau-\tau') F^{*bb}_{AA 22}(\tau-\tau') 
% + {\mathfrak G}^{*}_{AA \uparrow \uparrow }(\tau-\tau') F^{*bd}_{AA 22}(\tau-\tau') 
% \nonumber \\ & \qquad \qquad
% - {\mathfrak G}^{*}_{AA \uparrow \uparrow }(\tau-\tau') F^{bd}_{AA 11}(\tau-\tau') 
% - {\mathfrak G}^{*}_{AA \uparrow \uparrow }(\tau-\tau') G^{dd}_{AA 12}(\tau-\tau') 
% \nonumber \\ & \qquad \qquad
+ {\mathfrak F}_{AA \uparrow \downarrow }(\tau-\tau') B^{bd}_{AA 12}(\tau-\tau') 
+ {\mathfrak F}_{AA \uparrow \downarrow }(\tau-\tau') F^{*dd}_{AA 22}(\tau-\tau') 
\nonumber \\ & \qquad \qquad
+ {\mathfrak F}_{AA \uparrow \downarrow }(\tau-\tau') F^{bb}_{AA 11}(\tau-\tau') 
- {\mathfrak G}_{AA \downarrow \downarrow }(\tau-\tau') F^{*bd}_{AA 22}(\tau-\tau') 
\nonumber \\ & \qquad \qquad
+ {\mathfrak G}_{AA \downarrow \downarrow }(\tau-\tau') F^{bd}_{AA 11}(\tau-\tau') 
- {\mathfrak G}_{AA \downarrow \downarrow }(\tau-\tau') G^{bb}_{AA 12}(\tau-\tau') 
% \nonumber \\ & \qquad \qquad
% + {\mathfrak F}^{*}_{AA \uparrow \downarrow }(\tau-\tau') G^{*bd}_{AA 21}(\tau-\tau')  
% + {\mathfrak G}^{*}_{AA \uparrow \uparrow }(\tau-\tau') G^{*bb}_{AA 21}(\tau-\tau')  
\nonumber \\ & \qquad \qquad
+ {\mathfrak G}_{AA \downarrow \downarrow }(\tau-\tau') G^{*dd}_{AA 21}(\tau-\tau')  
+ {\mathfrak F}_{AA \uparrow \downarrow }(\tau-\tau') G^{*db}_{AA 21}(\tau-\tau')  
\Big)
\nonumber \\ & \qquad
+ G^{f}_{AA \uparrow \uparrow }(\tau-\tau_{2}) \Big(  
% - {\mathfrak F}^{*}_{AA \uparrow \downarrow }(\tau-\tau') G^{bd}_{AA 22}(\tau-\tau')
% + {\mathfrak F}^{*}_{AA \uparrow \downarrow }(\tau-\tau') F^{dd}_{AA 21}(\tau-\tau')
% \nonumber \\ & \qquad \qquad
% - {\mathfrak F}^{*}_{AA \uparrow \downarrow }(\tau-\tau') F^{*bb}_{AA 12}(\tau-\tau') 
% + {\mathfrak G}^{*}_{AA \uparrow \uparrow }(\tau-\tau')F^{*bd}_{AA 12}(\tau-\tau')
% \nonumber \\ & \qquad \qquad
% + {\mathfrak G}^{*}_{AA \uparrow \uparrow }(\tau-\tau') F^{bd}_{AA 21}(\tau-\tau') 
% + {\mathfrak G}^{*}_{AA \uparrow \uparrow }(\tau-\tau') G^{dd}_{AA 22}(\tau-\tau')
% \nonumber \\ & \qquad \qquad
- {\mathfrak F}_{AA \uparrow \downarrow }(\tau-\tau') G^{bd}_{AA 22}(\tau-\tau')
+ {\mathfrak F}_{AA \uparrow \downarrow }(\tau-\tau') F^{*dd}_{AA 12}(\tau-\tau') 
\nonumber \\ & \qquad \qquad
- {\mathfrak F}_{AA \uparrow \downarrow }(\tau-\tau') F^{bb}_{AA 21}(\tau-\tau')
- {\mathfrak G}_{AA \downarrow \downarrow }(\tau-\tau')F^{*bd}_{AA 12}(\tau-\tau')
\nonumber \\ & \qquad \qquad
- {\mathfrak G}_{AA \downarrow \downarrow }(\tau-\tau') F^{bd}_{AA 21}(\tau-\tau') 
+ {\mathfrak G}_{AA \downarrow \downarrow }(\tau-\tau') G^{bb}_{AA 22}(\tau-\tau')
% \nonumber \\ & \qquad \qquad
% + {\mathfrak F}^{*}_{AA \uparrow \downarrow }(\tau-\tau')  G^{*bd}_{AA 11}(\tau-\tau') 
% + {\mathfrak G}^{*}_{AA \uparrow \uparrow }(\tau-\tau') G^{*bb}_{AA 11}(\tau-\tau')
\nonumber \\ & \qquad \qquad
+ {\mathfrak G}_{AA \downarrow \downarrow }(\tau-\tau') G^{*dd}_{AA 11}(\tau-\tau')  
+ {\mathfrak F}_{AA \uparrow \downarrow }(\tau-\tau') G^{*db}_{AA 11}(\tau-\tau')  
\Big)     
\Big]=\delta(\tau-\tau_{2}).
  \end{align}

\subsection{Equation of motion for $f_{A\downarrow}$ particles}

%   \subsection{Equation of motion for $f_{\downarrow A}$ particle}

  \begin{align}
%   derivative
  & \partial_{ \tau} G^{f}_{AA \downarrow \downarrow }(\tau-\tau_{2}) 
  -\lambda_{A}^{-} F^{*f}_{AA \uparrow \downarrow }(\tau-\tau_{2})
  +\lambda_{A}^{z} G^{f}_{AA \downarrow \downarrow }(\tau-\tau_{2}) 
   -\frac{U}{2} n^{dd}_{AA 21}(\tau) F^{*f}_{AA \uparrow \downarrow }(\tau-\tau_{2}) 
   \nonumber \\ &
   -\frac{U}{2} \Big[
   + n^{dd}_{AA 22}(\tau) n^{F}_{AA \uparrow \downarrow }(\tau) F^{*f}_{AA \uparrow \downarrow }(\tau-\tau_{2})    
   + n^{dd}_{AA 22}(\tau) n_{AA \uparrow \uparrow }(\tau)  G^{f}_{AA \downarrow \downarrow }(\tau-\tau_{2})   
   \Big]
   \nonumber \\  &
   +\frac{v_{1}}{2} \Big[
   + F^{*f}_{BA \uparrow \downarrow }(\tau-\tau_{2}) \Big( 
   n^{bb}_{BA 21}(\tau) 
   - n^{Fbd}_{AB 11}(\tau) 
   + n^{*Fbd}_{AB 22}(\tau) 
   - n^{d}_{AB 21}(\tau) 
   + n^{bb}_{AB 21}(\tau) 
   - n^{Fbd}_{BA 11}(\tau) 
   + n^{*F db}_{BA 22}(\tau) 
   - n^{dd}_{BA 21}(\tau)
   \Big)
 \nonumber \\ & \qquad
+ G^{f}_{BA \downarrow \downarrow }(\tau-\tau_{2}) \Big( 
n^{bb}_{BA 11}(\tau) 
+ n^{Fbd}_{AB 12}(\tau)
+ n^{*Fdb}_{AB 21}(\tau) 
+ n^{dd}_{AB 22}(\tau) 
- n^{bb}_{AB 22}(\tau) 
+ n^{bd}_{BA 21}(\tau) 
+ F^{*db}_{BA 12}(\tau) 
- n^{dd}_{BA 11}(\tau) 
\Big)
  \Big]
  \nonumber \\ &
  %    derivative with t, integration over tau'
 + \frac{1}{2} \int {\rm d}{\tau'}\Big[
+ F^{*f}_{BA \uparrow \downarrow }(\tau'-\tau_{2}) \Big(  
% - {\mathfrak F}^{*}_{AB \uparrow \downarrow }(\tau-\tau') G^{bd}_{AB 12}(\tau-\tau')
% + {\mathfrak F}^{*}_{AB \uparrow \downarrow }(\tau-\tau') F^{dd}_{AB 11}(\tau-\tau')
% \nonumber \\ & \qquad \qquad
% + {\mathfrak F}^{*}_{AB \uparrow \downarrow }(\tau-\tau') F^{*bb}_{AB 22}(\tau-\tau')
% + {\mathfrak G}^{*}_{AB \uparrow \uparrow }(\tau-\tau') F^{*bd}_{AB 22}(\tau-\tau')
% \nonumber \\ & \qquad \qquad
% - {\mathfrak G}^{*}_{AB \uparrow \uparrow }(\tau-\tau') F^{bd}_{AB 11}(\tau-\tau')
% - {\mathfrak G}^{*}_{AB \uparrow \uparrow }(\tau-\tau') G^{dd}_{AB 12}(\tau-\tau')
% \nonumber \\ & \qquad \qquad
+ {\mathfrak F}_{AB \downarrow \uparrow }(\tau-\tau') G^{bd}_{AB 12}(\tau-\tau')
+ {\mathfrak F}_{AB \downarrow \uparrow }(\tau-\tau') F^{*dd}_{AB 22}(\tau-\tau')
\nonumber \\ & \qquad \qquad
+ {\mathfrak F}_{AB \downarrow \uparrow }(\tau-\tau') F^{bb}_{AB 11}(\tau-\tau')
+ {\mathfrak G}_{AB \downarrow \downarrow }(\tau-\tau')F^{*bd}_{AB 22}(\tau-\tau')
\nonumber \\ & \qquad \qquad
- {\mathfrak G}_{AB \downarrow \downarrow }(\tau-\tau')F^{bd}_{AB 11}(\tau-\tau')
+ {\mathfrak G}_{AB \downarrow \downarrow }(\tau-\tau')G^{bb}_{AB 12}(\tau-\tau')
% \nonumber \\ & \qquad \qquad
% + {\mathfrak G}^{*}_{AB \uparrow \uparrow }(\tau-\tau') G^{*bb}_{AB 21}(\tau-\tau')
% - {\mathfrak F}^{*}_{AB \uparrow \downarrow }(\tau-\tau') G^{*bd}_{AB 21}(\tau-\tau')
\nonumber \\ & \qquad \qquad
+ {\mathfrak F}_{AB \downarrow \uparrow }(\tau-\tau')G^{db}_{AB 21}(\tau-\tau')
- {\mathfrak G}_{AB \downarrow \downarrow }(\tau-\tau') G^{*dd}_{AB 21}(\tau-\tau') 
\Big)
\nonumber \\ & \qquad
+ G^{f}_{BA \downarrow \downarrow }(\tau'-\tau_{2}) \Big(  
% - {\mathfrak F}^{*}_{AB \uparrow \downarrow }(\tau-\tau')G^{bd}_{AB 11}(\tau-\tau')
% - {\mathfrak F}^{*}_{AB \uparrow \downarrow }(\tau-\tau') F^{dd}_{AB 12}(\tau-\tau')
% \nonumber \\ & \qquad \qquad
% + {\mathfrak F}^{*}_{AB \uparrow \downarrow }(\tau-\tau') F^{*bb}_{AB 21}(\tau-\tau')
% + {\mathfrak G}^{*}_{AB \uparrow \uparrow }(\tau-\tau') F^{*bd}_{AB 21}(\tau-\tau')
% \nonumber \\ & \qquad \qquad
% + {\mathfrak G}^{*}_{AB \uparrow \uparrow }(\tau-\tau') F^{bd}_{AB 12}(\tau-\tau')
% - {\mathfrak G}^{*}_{AB \uparrow \uparrow }(\tau-\tau')G^{dd}_{AB 11}(\tau-\tau')
% \nonumber \\ & \qquad \qquad
+ {\mathfrak F}_{AB \downarrow \uparrow }(\tau-\tau')G^{bd}_{AB 11}(\tau-\tau')
+ {\mathfrak F}_{AB \downarrow \uparrow }(\tau-\tau') F^{*dd}_{AB 21}(\tau-\tau')
\nonumber \\ & \qquad \qquad
- {\mathfrak F}_{AB \downarrow \uparrow }(\tau-\tau') F^{bb}_{AB 12}(\tau-\tau')
+ {\mathfrak G}_{AB \downarrow \downarrow }(\tau-\tau') F^{*bd}_{AB 21}(\tau-\tau')
\nonumber \\ & \qquad \qquad
+ {\mathfrak G}_{AB \downarrow \downarrow }(\tau-\tau') F^{bd}_{AB 12}(\tau-\tau')
+ {\mathfrak G}_{AB \downarrow \downarrow }(\tau-\tau') G^{bb}_{AB 11}(\tau-\tau')
% \nonumber \\ & \qquad \qquad
% - {\mathfrak G}^{*}_{AB \uparrow \uparrow }(\tau-\tau') G^{*bb}_{AB 22}(\tau-\tau')
% + {\mathfrak F}^{*}_{AB \uparrow \downarrow }(\tau-\tau') G^{*bd}_{AB 22}(\tau-\tau')
\nonumber \\ & \qquad \qquad
- {\mathfrak F}_{AB \downarrow \uparrow }(\tau-\tau') G^{*db}_{AB 22}(\tau-\tau')
+ {\mathfrak G}_{AB \downarrow \downarrow }(\tau-\tau') G^{*dd}_{AB 22}(\tau-\tau') 
\Big)
\nonumber \\ & \qquad
+ F^{*f}_{AA \uparrow \downarrow }(\tau'-\tau_{2}) \Big( 
% {\mathfrak F}^{*}_{AA \uparrow \downarrow }(\tau-\tau')G^{bd}_{AA 12}(\tau-\tau')
% - {\mathfrak F}^{*}_{AA \uparrow \downarrow }(\tau-\tau')F^{dd}_{AA 11}(\tau-\tau')
% \nonumber \\ & \qquad \qquad
% - {\mathfrak F}^{*}_{AA \uparrow \downarrow }(\tau-\tau')F^{*bb}_{AA 22}(\tau-\tau')
% + {\mathfrak G}^{*}_{AA \uparrow \uparrow }(\tau-\tau') F^{*bd}_{AA 22}(\tau-\tau')
% \nonumber \\ & \qquad \qquad
% - {\mathfrak G}^{*}_{AA \uparrow \uparrow }(\tau-\tau') F^{bd}_{AA 11}(\tau-\tau')
% - {\mathfrak G}^{*}_{AA \uparrow \uparrow }(\tau-\tau') G^{dd}_{AA 12}(\tau-\tau')
% \nonumber \\ & \qquad \qquad
+ {\mathfrak F}_{AA \downarrow \uparrow }(\tau-\tau') G^{bd}_{AA 12}(\tau-\tau')
+ {\mathfrak F}_{AA \downarrow \uparrow }(\tau-\tau') F^{*dd}_{AA 22}(\tau-\tau')
\nonumber \\ & \qquad \qquad
+ {\mathfrak F}_{AA \downarrow \uparrow }(\tau-\tau') F^{bb}_{AA 11}(\tau-\tau')
- {\mathfrak G}_{AA \downarrow \downarrow }(\tau-\tau') F^{*bd}_{AA 22}(\tau-\tau')
\nonumber \\ & \qquad \qquad
+ {\mathfrak G}_{AA \downarrow \downarrow }(\tau-\tau') F^{bd}_{AA 11}(\tau-\tau')
- {\mathfrak G}_{AA \downarrow \downarrow }(\tau-\tau') G^{bb}_{AA 12}(\tau-\tau')
\nonumber \\ & \qquad \qquad
% + {\mathfrak F}^{*}_{AA \uparrow \downarrow }(\tau-\tau')G^{*bd}_{AA 21}(\tau-\tau')
% + {\mathfrak G}^{*}_{AA \uparrow \uparrow }(\tau-\tau') G^{*bb}_{AA 21}(\tau-\tau')
% \nonumber \\ & \qquad \qquad
+ {\mathfrak G}_{AA \downarrow \downarrow }(\tau-\tau') G^{*dd}_{AA 21}(\tau-\tau')
+ {\mathfrak F}_{AA \downarrow \uparrow }(\tau-\tau') G^{*db}_{AA 21}(\tau-\tau')
\Big)
\nonumber \\ & \qquad
+ G^{f}_{AA \downarrow \downarrow }(\tau'-\tau_{2}) \Big(  
% - {\mathfrak F}^{*}_{AA \uparrow \downarrow }(\tau-\tau') G^{bd}_{AA 11}(\tau-\tau')
% - {\mathfrak F}^{*}_{AA \uparrow \downarrow }(\tau-\tau') F^{dd}_{AA 12}(\tau-\tau')
% \nonumber \\ & \qquad \qquad
% + {\mathfrak F}^{*}_{AA \uparrow \downarrow }(\tau-\tau') F^{*bb}_{AA 21}(\tau-\tau')
% - {\mathfrak G}^{*}_{AA \uparrow \uparrow }(\tau-\tau')F^{*bd}_{AA 21}(\tau-\tau')
% \nonumber \\ & \qquad \qquad
% - {\mathfrak G}^{*}_{AA \uparrow \uparrow }(\tau-\tau')F^{bd}_{AA 12}(\tau-\tau')
% + {\mathfrak G}^{*}_{AA \uparrow \uparrow }(\tau-\tau') G^{dd}_{AA 11}(\tau-\tau')
% \nonumber \\ & \qquad \qquad
- {\mathfrak F}_{AA \downarrow \uparrow }(\tau-\tau') G^{bd}_{AA 11}(\tau-\tau')
- {\mathfrak F}_{AA \downarrow \uparrow }(\tau-\tau') F^{*dd}_{AA 21}(\tau-\tau')
\nonumber \\ & \qquad \qquad
+ {\mathfrak F}_{AA \downarrow \uparrow }(\tau-\tau') F^{bb}_{AA 12}(\tau-\tau')
+ {\mathfrak G}_{AA \downarrow \downarrow }(\tau-\tau')F^{*bd}_{AA 21}(\tau-\tau')
\nonumber \\ & \qquad \qquad
+ {\mathfrak G}_{AA \downarrow \downarrow }(\tau-\tau')F^{bd}_{AA 12}(\tau-\tau')
+ {\mathfrak G}_{AA \downarrow \downarrow }(\tau-\tau')G^{bb}_{AA 11}(\tau-\tau')
% \nonumber \\ & \qquad \qquad
% + {\mathfrak F}^{*}_{AA \uparrow \downarrow }(\tau-\tau')G^{*bd}_{AA 22}(\tau-\tau')
% + {\mathfrak G}^{*}_{AA \uparrow \uparrow }(\tau-\tau') G^{*bb}_{AA 22}(\tau-\tau')
\nonumber \\ & \qquad \qquad
+ {\mathfrak G}_{AA \downarrow \downarrow }(\tau-\tau')G^{*dd}_{AA 22}(\tau-\tau')
+ {\mathfrak F}_{AA \downarrow \uparrow }(\tau-\tau')G^{*db}_{AA 22}(\tau-\tau')
\Big)
  \Big]
  \nonumber \\ &
  %    derivative with t', integration over tau
+\frac{1}{2} \int {\rm d}{\tau}\Big[
+ F^{*f}_{BA \uparrow \downarrow }(\tau-\tau_{2}) \Big(  
% - {\mathfrak F}^{*}_{BA \downarrow \uparrow }(\tau-\tau') G^{bd}_{BA 12}(\tau-\tau')
% + {\mathfrak F}^{*}_{BA \downarrow \uparrow }(\tau-\tau') F^{dd}_{BA 11}(\tau-\tau')
% \nonumber \\ & \qquad \qquad
% + {\mathfrak F}^{*}_{BA \downarrow \uparrow }(\tau-\tau') F^{*bb}_{BA 22}(\tau-\tau')
% + {\mathfrak G}^{*}_{BA \downarrow \downarrow }(\tau-\tau') F^{*bd}_{BA 22}(\tau-\tau')
% \nonumber \\ & \qquad \qquad
% - {\mathfrak G}^{*}_{BA \downarrow \downarrow }(\tau-\tau') F^{bd}_{BA 11}(\tau-\tau')
% - {\mathfrak G}^{*}_{BA \downarrow \downarrow }(\tau-\tau') G^{dd}_{BA 12}(\tau-\tau')
% \nonumber \\ & \qquad \qquad
+ {\mathfrak F}_{BA \uparrow \downarrow }(\tau-\tau') G^{bd}_{BA 12}(\tau-\tau')
+ {\mathfrak F}_{BA \uparrow \downarrow }(\tau-\tau') F^{*dd}_{BA 22}(\tau-\tau')
\nonumber \\ & \qquad \qquad
+ {\mathfrak F}_{BA \uparrow \downarrow }(\tau-\tau')F^{bb}_{BA 11}(\tau-\tau')
+ {\mathfrak G}_{BA \uparrow \uparrow }(\tau-\tau') F^{*bd}_{BA 22}(\tau-\tau')
\nonumber \\ & \qquad \qquad
- {\mathfrak G}_{BA \uparrow \uparrow }(\tau-\tau') F^{bd}_{BA 11}(\tau-\tau')
+ {\mathfrak G}_{BA \uparrow \uparrow }(\tau-\tau') G^{bb}_{BA 12}(\tau-\tau')
\nonumber \\ & \qquad \qquad
- {\mathfrak G}_{BA \uparrow \uparrow }(\tau-\tau') G^{*dd}_{BA 21}(\tau-\tau')
+ {\mathfrak F}_{BA \uparrow \downarrow }(\tau-\tau') G^{*db}_{BA 21}(\tau-\tau')
% \nonumber \\ & \qquad \qquad
% - {\mathfrak F}^{*}_{BA \downarrow \uparrow }(\tau-\tau') G^{*bd}_{BA 21}(\tau-\tau')
% + {\mathfrak G}^{*}_{BA \downarrow \downarrow }(\tau-\tau') G^{*bb}_{BA 21}(\tau-\tau')
\Big)
\nonumber \\ & \qquad
+ G^{f}_{BA \downarrow \downarrow }(\tau-\tau_{2}) \Big( 
% {\mathfrak F}^{*}_{BA \downarrow \uparrow }(\tau-\tau') G^{bd}_{BA 22}(\tau-\tau')
% - {\mathfrak F}^{*}_{BA \downarrow \uparrow }(\tau-\tau') F^{dd}_{BA 21}(\tau-\tau')
% \nonumber \\ & \qquad \qquad
% + {\mathfrak F}^{*}_{BA \downarrow \uparrow }(\tau-\tau') F^{*bb}_{BA 12}(\tau-\tau')
% + {\mathfrak G}^{*}_{BA \downarrow \downarrow }(\tau-\tau') F^{*bd}_{BA 12}(\tau-\tau')
% \nonumber \\ & \qquad \qquad
% + {\mathfrak G}^{*}_{BA \downarrow \downarrow }(\tau-\tau') F^{bd}_{BA 21}(\tau-\tau')
% + {\mathfrak G}^{*}_{BA \downarrow \downarrow }(\tau-\tau') G^{dd}_{BA 22}(\tau-\tau')
% \nonumber \\ & \qquad \qquad
- {\mathfrak F}_{BA \uparrow \downarrow }(\tau-\tau') G^{bd}_{BA 22}(\tau-\tau')
+ {\mathfrak F}_{BA \uparrow \downarrow }(\tau-\tau') F^{*dd}_{BA 12}(\tau-\tau')
\nonumber \\ & \qquad \qquad
- {\mathfrak F}_{BA \uparrow \downarrow }(\tau-\tau') F^{bb}_{BA 21}(\tau-\tau')
+ {\mathfrak G}_{BA \uparrow \uparrow }(\tau-\tau') F^{*bd}_{BA 12}(\tau-\tau')
\nonumber \\ & \qquad \qquad
+ {\mathfrak G}_{BA \uparrow \uparrow }(\tau-\tau') F^{bd}_{BA 21}(\tau-\tau')
- {\mathfrak G}_{BA \uparrow \uparrow }(\tau-\tau') G^{bb}_{BA 22}(\tau-\tau')
\nonumber \\ & \qquad \qquad
- {\mathfrak G}_{BA \uparrow \uparrow }(\tau-\tau') G^{*dd}_{BA 11}(\tau-\tau')
+ {\mathfrak F}_{BA \uparrow \downarrow }(\tau-\tau') G^{*db}_{BA 11}(\tau-\tau')
% \nonumber \\ & \qquad \qquad
% - {\mathfrak F}^{*}_{BA \downarrow \uparrow }(\tau-\tau') G^{*bd}_{BA 11}(\tau-\tau')
% + {\mathfrak G}^{*}_{BA \downarrow \downarrow }(\tau-\tau') G^{*bb}_{BA 11}(\tau-\tau')
\Big)
\nonumber \\ & \qquad
+ F^{*f}_{AA \uparrow \downarrow }(\tau-\tau_{2}) \Big( 
% {\mathfrak F}^{*}_{AA \downarrow \uparrow }(\tau-\tau') G^{bd}_{AA 12}(\tau-\tau')
% - {\mathfrak F}^{*}_{AA \downarrow \uparrow }(\tau-\tau') F^{dd}_{AA 11}(\tau-\tau')
% \nonumber \\ & \qquad \qquad
% - {\mathfrak F}^{*}_{AA \downarrow \uparrow }(\tau-\tau') F^{*bb}_{AA 22}(\tau-\tau')
% - {\mathfrak G}^{*}_{AA \downarrow \downarrow }(\tau-\tau') F^{*bd}_{AA 22}(\tau-\tau')
% \nonumber \\ & \qquad \qquad
% + {\mathfrak G}^{*}_{AA \downarrow \downarrow }(\tau-\tau') F^{bd}_{AA 11}(\tau-\tau')
% + {\mathfrak G}^{*}_{AA \downarrow \downarrow }(\tau-\tau') G^{dd}_{AA 12}(\tau-\tau')
% \nonumber \\ & \qquad \qquad
+ {\mathfrak F}_{AA \uparrow \downarrow  }(\tau-\tau') G^{bd}_{AA 12}(\tau-\tau')
+ {\mathfrak F}_{AA \uparrow \downarrow  }(\tau-\tau') F^{*dd}_{AA 22}(\tau-\tau')
\nonumber \\ & \qquad \qquad
+ {\mathfrak F}_{AA \uparrow \downarrow  }(\tau-\tau') F^{bb}_{AA 11}(\tau-\tau')
+ {\mathfrak G}_{AA \uparrow \uparrow }(\tau-\tau') F^{*bd}_{AA 22}(\tau-\tau')
\nonumber \\ & \qquad \qquad
- {\mathfrak G}_{AA \uparrow \uparrow }(\tau-\tau') F^{bd}_{AA 11}(\tau-\tau')
+ {\mathfrak G}_{AA \uparrow \uparrow }(\tau-\tau') G^{bb}_{AA 12}(\tau-\tau')
% \nonumber \\ & \qquad \qquad
% + {\mathfrak F}^{*}_{AA \downarrow \uparrow }(\tau-\tau') G^{*bd}_{AA 21}(\tau-\tau')
% - {\mathfrak G}^{*}_{AA \downarrow \downarrow }(\tau-\tau') G^{*bb}_{AA 21}(\tau-\tau')
\nonumber \\ & \qquad \qquad
- {\mathfrak G}_{AA \uparrow \uparrow }(\tau-\tau')G^{*dd}_{AA 21}(\tau-\tau')
+ {\mathfrak F}_{AA \uparrow \downarrow  }(\tau-\tau')G^{*db}_{AA 21}(\tau-\tau')
\Big)
\nonumber \\ & \qquad
+ G^{f}_{AA \downarrow \downarrow }(\tau-\tau_{2}) \Big( 
% {\mathfrak F}^{*}_{AA \downarrow \uparrow }(\tau-\tau') G^{bd}_{AA 22}(\tau-\tau')
% - {\mathfrak F}^{*}_{AA \downarrow \uparrow }(\tau-\tau')F^{dd}_{AA 21}(\tau-\tau')
% \nonumber \\ & \qquad \qquad
% + {\mathfrak F}^{*}_{AA \downarrow \uparrow }(\tau-\tau')F^{*bb}_{AA 12}(\tau-\tau')
% + {\mathfrak G}^{*}_{AA \downarrow \downarrow }(\tau-\tau') F^{*bd}_{AA 12}(\tau-\tau')
% \nonumber \\ & \qquad \qquad
% + {\mathfrak G}^{*}_{AA \downarrow \downarrow }(\tau-\tau') F^{bd}_{AA 21}(\tau-\tau')
% + {\mathfrak G}^{*}_{AA \downarrow \downarrow }(\tau-\tau')G^{dd}_{AA 22}(\tau-\tau')
% \nonumber \\ & \qquad \qquad
+ {\mathfrak F}_{AA \uparrow \downarrow  }(\tau-\tau') G^{bd}_{AA 22}(\tau-\tau')
- {\mathfrak F}_{AA \uparrow \downarrow  }(\tau-\tau') F^{*dd}_{AA 12}(\tau-\tau')
\nonumber \\ & \qquad \qquad
+ {\mathfrak F}_{AA \uparrow \downarrow  }(\tau-\tau') F^{bb}_{AA 21}(\tau-\tau')
- {\mathfrak G}_{AA \uparrow \uparrow }(\tau-\tau') F^{*bd}_{AA 12}(\tau-\tau')
\nonumber \\ & \qquad \qquad
- {\mathfrak G}_{AA \uparrow \uparrow }(\tau-\tau') F^{bd}_{AA 21}(\tau-\tau')
+ {\mathfrak G}_{AA \uparrow \uparrow }(\tau-\tau') G^{bb}_{AA 22}(\tau-\tau')
% \nonumber \\ & \qquad \qquad
% - {\mathfrak F}^{*}_{AA \downarrow \uparrow }(\tau-\tau') G^{*bd}_{AA 11}(\tau-\tau')
% + {\mathfrak G}^{*}_{AA \downarrow \downarrow }(\tau-\tau')G^{*bb}_{AA 11}(\tau-\tau')
\nonumber \\ & \qquad \qquad
+ {\mathfrak G}_{AA \uparrow \uparrow }(\tau-\tau')G^{*dd}_{AA 11}(\tau-\tau')
- {\mathfrak F}_{AA \uparrow \downarrow  }(\tau-\tau')G^{*db}_{AA 11}(\tau-\tau')
  \Big]= \delta(\tau-\tau_{2}).
  \end{align}

\subsection{Equation of motion for $f_{B\uparrow}$ particles}

%   \subsection{Equation of motion for $f_{\uparrow B}$ particle}

  \begin{align}
%   derivative
  & \partial_{ \tau} G^{f}_{BB \uparrow \uparrow }(\tau-\tau_{2}) 
  +\lambda_{B}^{-} F^{*f}_{BB \downarrow \uparrow}(\tau-\tau_{2})
  +\lambda_{B}^{z} G^{f}_{BB \uparrow \uparrow }(\tau-\tau_{2}) 
  +\frac{U}{2} n^{dd}_{BB 21}(\tau) F^{*f}_{BB \downarrow \uparrow}(\tau-\tau_{2}) 
  \nonumber \\ &
  -\frac{U}{2}\Big[
  + n^{dd}_{BB 22}(\tau) n^{F}_{BB \uparrow \downarrow}(\tau) F^{*f}_{BB \downarrow \uparrow}(\tau-\tau_{2}) 
  + n^{dd}_{BB 22}(\tau)n_{BB \downarrow \downarrow }(\tau) G^{f}_{BB \uparrow \uparrow }(\tau-\tau_{2}) 
  \Big]
  \nonumber \\ &
 + \frac{v_{1}}{2} \Big[
  + F^{*f}_{AB \downarrow \uparrow}(\tau-\tau_{2}) \Big( 
  n^{bb}_{BA 21}(\tau) 
  - n^{Fbd}_{AB 11}(\tau)
  + n^{*Fdb}_{AB 22}(\tau) 
  - n^{dd}_{AB 21}(\tau) 
  + n^{bb}_{AB 21}(\tau) 
  - n^{Fdb}_{BA 11}(\tau) 
  + n^{*Fdb}_{BA 22}(\tau) 
  - n^{dd}_{BA 21}(\tau) 
  \Big)
\nonumber \\ & \qquad
  + G^{f}_{AB \uparrow \uparrow }(\tau-\tau_{2}) \Big(  
  - n^{bb}_{BA 22}(\tau) 
  + n^{Fbd}_{AB 21}(\tau) 
  + n^{*Fdb}_{AB 12}(\tau) 
  - n^{dd}_{AB 11}(\tau)
  + n^{bb}_{AB 11}(\tau) 
  + n^{Fbd}_{BA 12}(\tau) 
  + n^{*Fdb}_{BA 21}(\tau) 
  + n^{dd}_{BA 22}(\tau) 
  \Big)
  \Big]
     \nonumber \\ &
%    derivative with t, integration over tau'
+ \frac{1}{2} \int {\rm d}{\tau'}\Big[
+ F^{*f}_{BB \downarrow \uparrow}(\tau'-\tau_{2}) \Big( 
% {\mathfrak F}^{*}_{BB \downarrow \uparrow}(\tau-\tau') G^{bd}_{BB 12}(\tau-\tau') 
% - {\mathfrak F}^{*}_{BB \downarrow \uparrow}(\tau-\tau') F^{dd}_{BB 11}(\tau-\tau') 
% \nonumber \\ & \qquad \qquad
% - {\mathfrak F}^{*}_{BB \downarrow \uparrow}(\tau-\tau') F^{*bb}_{BB 22}(\tau-\tau') 
% + {\mathfrak G}^{*}_{BB \downarrow \downarrow }(\tau-\tau')F^{*bd}_{BB 22}(\tau-\tau') 
% \nonumber \\ & \qquad \qquad
% - {\mathfrak G}^{*}_{BB \downarrow \downarrow }(\tau-\tau')F^{bd}_{BB 11}(\tau-\tau') 
% - {\mathfrak G}^{*}_{BB \downarrow \downarrow }(\tau-\tau')G^{dd}_{BB 12}(\tau-\tau') 
% \nonumber \\ & \qquad \qquad
+ {\mathfrak F}_{BB \uparrow \downarrow}(\tau-\tau') G^{bd}_{BB 12}(\tau-\tau') 
+ {\mathfrak F}_{BB \uparrow \downarrow}(\tau-\tau') F^{*dd}{BB 22}(\tau-\tau') 
\nonumber \\ & \qquad \qquad
+ {\mathfrak F}_{BB \uparrow \downarrow}(\tau-\tau')F^{bb}_{BB 11}(\tau-\tau')
- {\mathfrak G}_{BB \uparrow \uparrow }(\tau-\tau')F^{*bd}_{BB 22}(\tau-\tau') 
\nonumber \\ & \qquad \qquad
+ {\mathfrak G}_{BB \uparrow \uparrow }(\tau-\tau')F^{bd}_{BB 11}(\tau-\tau') 
- {\mathfrak G}_{BB \uparrow \uparrow }(\tau-\tau') G^{bb}_{BB 12}(\tau-\tau')
\nonumber \\ & \qquad \qquad
+ {\mathfrak F}_{BB \uparrow \downarrow}(\tau-\tau') G^{*db}_{BB 21}(\tau-\tau')
+ {\mathfrak G}_{BB \uparrow \uparrow }(\tau-\tau') G^{*dd}_{BB 21}(\tau-\tau') 
% \nonumber \\ & \qquad \qquad
% + {\mathfrak G}^{*}_{BB \downarrow \downarrow }(\tau-\tau') G^{*bb}_{BB 21}(\tau-\tau')
% + {\mathfrak F}^{*}_{BB \downarrow \uparrow}(\tau-\tau') G^{bd}_{BB 21}(\tau-\tau')
\Big)
\nonumber \\ & \qquad
+ G^{f}_{BB \uparrow \uparrow }(\tau'-\tau_{2}) \Big(  
% - {\mathfrak F}^{*}_{BB \downarrow \uparrow}(\tau-\tau') G^{bd}_{BB 11}(\tau-\tau')
% - {\mathfrak F}^{*}_{BB \downarrow \uparrow}(\tau-\tau')F^{dd}_{BB 12}(\tau-\tau') 
% \nonumber \\ & \qquad \qquad
% + {\mathfrak F}^{*}_{BB \downarrow \uparrow}(\tau-\tau')F^{*bb}_{BB 21}(\tau-\tau') 
% - {\mathfrak G}^{*}_{BB \downarrow \downarrow }(\tau-\tau')F^{*bd}_{BB 21}(\tau-\tau') 
% \nonumber \\ & \qquad \qquad
% - {\mathfrak G}^{*}_{BB \downarrow \downarrow }(\tau-\tau')F^{bd}_{BB 12}(\tau-\tau') 
% + {\mathfrak G}^{*}_{BB \downarrow \downarrow }(\tau-\tau')G^{dd}_{BB 11}(\tau-\tau') 
% \nonumber \\ & \qquad \qquad
- {\mathfrak F}_{BB \uparrow \downarrow}(\tau-\tau') G^{bd}_{BB 11}(\tau-\tau')
- {\mathfrak F}_{BB \uparrow \downarrow}(\tau-\tau') F^{*dd}_{BB 21}(\tau-\tau') 
\nonumber \\ & \qquad \qquad
+ {\mathfrak F}_{BB \uparrow \downarrow}(\tau-\tau')F^{bb}_{BB 12}(\tau-\tau') 
+ {\mathfrak G}_{BB \uparrow \uparrow }(\tau-\tau')F^{*bd}_{BB 21}(\tau-\tau') 
\nonumber \\ & \qquad \qquad
+ {\mathfrak G}_{BB \uparrow \uparrow }(\tau-\tau')F^{bd}_{BB 12}(\tau-\tau') 
+ {\mathfrak G}_{BB \uparrow \uparrow }(\tau-\tau')G^{bb}_{BB 11}(\tau-\tau') 
\nonumber \\ & \qquad \qquad
+ {\mathfrak F}_{BB \uparrow \downarrow}(\tau-\tau') G^{db}_{BB 22}(\tau-\tau') 
+ {\mathfrak G}_{BB \uparrow \uparrow }(\tau-\tau')  G^{*dd}_{BB 22}(\tau-\tau')
% \nonumber \\ & \qquad \qquad
% + {\mathfrak G}^{*}_{BB \downarrow \downarrow }(\tau-\tau') G^{*bb}_{BB 22}(\tau-\tau') 
% + {\mathfrak F}^{*}_{BB \downarrow \uparrow}(\tau-\tau') G^{*bd}_{BB 22}(\tau-\tau')
\Big)
\nonumber \\ & \qquad
+ F^{*f}_{AB \downarrow \uparrow}(\tau-\tau_{2}) \Big(  
% - {\mathfrak F}^{*}_{BA \downarrow \uparrow }(\tau-\tau')G^{bd}_{BA 12}(\tau-\tau') 
% + {\mathfrak F}^{*}_{BA \downarrow \uparrow }(\tau-\tau')F^{dd}_{BA 11}(\tau-\tau') 
% \nonumber \\ & \qquad \qquad
% + {\mathfrak F}^{*}_{BA \downarrow \uparrow }(\tau-\tau')F^{*bb}_{BA 22}(\tau-\tau') 
% + {\mathfrak G}^{*}_{BA \downarrow \uparrow}(\tau-\tau') F^{*bd}_{BA 22}(\tau-\tau') 
% \nonumber \\ & \qquad \qquad
% - {\mathfrak G}^{*}_{BA \downarrow \uparrow}(\tau-\tau') F^{bd}_{BA 11}(\tau-\tau')
% - {\mathfrak G}^{*}_{BA \downarrow \uparrow}(\tau-\tau') G^{dd}_{BA 12}(\tau-\tau')
% \nonumber \\ & \qquad \qquad
+ {\mathfrak F}_{BA \uparrow \downarrow }(\tau-\tau')G^{bd}_{BA 12}(\tau-\tau') 
+ {\mathfrak F}_{BA \uparrow \downarrow }(\tau-\tau') F^{*dd}_{BA 22}(\tau-\tau') 
\nonumber \\ & \qquad \qquad
+ {\mathfrak F}_{BA \uparrow \downarrow }(\tau-\tau') F^{bb}_{BA 11}(\tau-\tau') 
+ {\mathfrak G}_{BA \uparrow \uparrow }(\tau-\tau') F^{*bd}_{BA 22}(\tau-\tau') 
\nonumber \\ & \qquad \qquad
- {\mathfrak G}_{BA \uparrow \uparrow }(\tau-\tau') F^{bd}_{BA 11}(\tau-\tau')
+ {\mathfrak G}_{BA \uparrow \uparrow }(\tau-\tau') G^{bb}_{BA 12}(\tau-\tau') 
\nonumber \\ & \qquad \qquad
- {\mathfrak G}_{BA \uparrow \uparrow }(\tau-\tau') G^{*dd}_{BA 21}(\tau-\tau') 
+ {\mathfrak F}_{BA \uparrow \downarrow }(\tau-\tau') G^{*db}_{BA 21}(\tau-\tau')
% \nonumber \\ & \qquad \qquad
% - {\mathfrak F}^{*}_{BA \downarrow \uparrow }(\tau-\tau') G^{*bd}_{BA 21}(\tau-\tau')
% + {\mathfrak G}^{*}_{BA \downarrow \uparrow}(\tau-\tau') G^{*bb}_{BA 21}(\tau-\tau')
\Big)
\nonumber \\ & \qquad
+ G^{f}_{AB \uparrow \uparrow }(\tau'-\tau_{2}) \Big(  
% - {\mathfrak F}^{*}_{BA \downarrow \uparrow }(\tau-\tau')G^{bd}_{BA 11}(\tau-\tau') 
% - {\mathfrak F}^{*}_{BA \downarrow \uparrow }(\tau-\tau')*F^{dd}_{BA 12}(\tau-\tau') 
% \nonumber \\ & \qquad \qquad
% + {\mathfrak F}^{*}_{BA \downarrow \uparrow }(\tau-\tau')F^{*bb}_{BA 21}(\tau-\tau') 
% + {\mathfrak G}^{*}_{BA \downarrow \uparrow}(\tau-\tau')F^{*bd}_{BA 21}(\tau-\tau') 
% \nonumber \\ & \qquad \qquad
% + {\mathfrak G}^{*}_{BA \downarrow \uparrow}(\tau-\tau') F^{bd}_{BA 12}(\tau-\tau') 
% - {\mathfrak G}^{*}_{BA \downarrow \uparrow}(\tau-\tau')G^{dd}_{BA 11}(\tau-\tau') 
% \nonumber \\ & \qquad \qquad
+ {\mathfrak F}_{BA \uparrow \downarrow }(\tau-\tau')G^{bd}_{BA 11}(\tau-\tau') 
+ {\mathfrak F}_{BA \uparrow \downarrow }(\tau-\tau')F^{*dd}_{BA 21}(\tau-\tau') 
\nonumber \\ & \qquad \qquad
- {\mathfrak F}_{BA \uparrow \downarrow }(\tau-\tau')F^{bb}_{BA 12}(\tau-\tau') 
+ {\mathfrak G}_{BA \uparrow \uparrow }(\tau-\tau')F^{*bd}_{BA 21}(\tau-\tau') 
\nonumber \\ & \qquad \qquad
+ {\mathfrak G}_{BA \uparrow \uparrow }(\tau-\tau') F^{bd}_{BA 12}(\tau-\tau') 
+ {\mathfrak G}_{BA \uparrow \uparrow }(\tau-\tau')G^{bb}_{BA 11}(\tau-\tau') 
\nonumber \\ & \qquad \qquad
+ {\mathfrak G}_{BA \uparrow \uparrow }(\tau-\tau') G^{*dd}_{BA 22}(\tau-\tau')
- {\mathfrak F}_{BA \uparrow \downarrow }(\tau-\tau') G^{*db}_{BA 22}(\tau-\tau')
% \nonumber \\ & \qquad \qquad
% + {\mathfrak F}^{*}_{BA \downarrow \uparrow }(\tau-\tau')G^{*bd}_{BA 22}(\tau-\tau') 
% - {\mathfrak G}^{*}_{BA \downarrow \uparrow}(\tau-\tau') G^{*bb}_{BA 22}(\tau-\tau')
\Big)
\Big]
   \nonumber \\ &
%    derivative with t', integration over tau
+ \frac{1}{2} \int {\rm d}{\tau}\Big[
+ F^{*f}_{BB \downarrow \uparrow}(\tau-\tau_{2}) \Big( 
% {\mathfrak F}^{*}_{BB \uparrow \downarrow}(\tau-\tau') G^{bd}_{BB 12}(\tau-\tau') 
% - {\mathfrak F}^{*}_{BB \uparrow \downarrow}(\tau-\tau') F^{dd}_{BB 11}(\tau-\tau') 
% \nonumber \\ & \qquad \qquad
% - {\mathfrak F}^{*}_{BB \uparrow \downarrow}(\tau-\tau') F^{*bb}_{BB 22}(\tau-\tau') 
% - {\mathfrak G}^{*}_{BB \uparrow \uparrow }(\tau-\tau')F^{*bd}_{BB 22}(\tau-\tau') 
% \nonumber \\ & \qquad \qquad
% + {\mathfrak G}^{*}_{BB \uparrow \uparrow }(\tau-\tau')F^{bd}_{BB 11}(\tau-\tau') 
% + {\mathfrak G}^{*}_{BB \uparrow \uparrow }(\tau-\tau')G^{dd}_{BB 12}(\tau-\tau') 
% \nonumber \\ & \qquad \qquad
+ {\mathfrak F}_{BB \downarrow \uparrow}(\tau-\tau') G^{bd}_{BB 12}(\tau-\tau') 
+ {\mathfrak F}_{BB \downarrow \uparrow}(\tau-\tau') F^{*dd}{BB 22}(\tau-\tau') 
\nonumber \\ & \qquad \qquad
+ {\mathfrak F}_{BB \downarrow \uparrow}(\tau-\tau')F^{bb}_{BB 11}(\tau-\tau')
+ {\mathfrak G}_{BB \downarrow \downarrow }(\tau-\tau')F^{*bd}_{BB 22}(\tau-\tau') 
\nonumber \\ & \qquad \qquad
- {\mathfrak G}_{BB \downarrow \downarrow }(\tau-\tau')F^{bd}_{BB 11}(\tau-\tau') 
+ {\mathfrak G}_{BB \downarrow \downarrow }(\tau-\tau') G^{bb}_{BB 12}(\tau-\tau')
\nonumber \\ & \qquad \qquad
+ {\mathfrak F}_{BB \downarrow \uparrow}(\tau-\tau') G^{*db}_{BB 21}(\tau-\tau') 
- {\mathfrak G}_{BB \downarrow \downarrow }(\tau-\tau') G^{*dd}_{BB 21}(\tau-\tau')
% \nonumber \\ & \qquad \qquad
% - {\mathfrak G}^{*}_{BB \uparrow \uparrow }(\tau-\tau') G^{*bb}_{BB 21}(\tau-\tau')
% + {\mathfrak F}^{*}_{BB \uparrow \downarrow}(\tau-\tau') G^{bd}_{BB 21}(\tau-\tau')
\Big)
\nonumber \\ & \qquad
+ G^{f}_{BB \uparrow \uparrow }(\tau-\tau_{2}) \Big( 
% {\mathfrak F}^{*}_{BB \uparrow \downarrow}(\tau-\tau')G^{bd}_{BB 22}(\tau-\tau') 
% - {\mathfrak F}^{*}_{BB \uparrow \downarrow}(\tau-\tau') F^{dd}_{BB 21}(\tau-\tau') 
% \nonumber \\ & \qquad \qquad
% + {\mathfrak F}^{*}_{BB \uparrow \downarrow}(\tau-\tau')F^{*bb}_{BB 12}(\tau-\tau') 
% + {\mathfrak G}^{*}_{BB \uparrow \uparrow }(\tau-\tau') F^{*bd}_{BB 12}(\tau-\tau') 
% \nonumber \\ & \qquad \qquad
% + {\mathfrak G}^{*}_{BB \uparrow \uparrow }(\tau-\tau')F^{bd}_{BB 21}(\tau-\tau') 
% + {\mathfrak G}^{*}_{BB \uparrow \uparrow }(\tau-\tau') G^{dd}_{BB 22}(\tau-\tau')
% \nonumber \\ & \qquad \qquad
+ {\mathfrak F}_{BB \downarrow \uparrow}(\tau-\tau')G^{bd}_{BB 22}(\tau-\tau') 
- {\mathfrak F}_{BB \downarrow \uparrow}(\tau-\tau') F^{*dd}_{BB 12}(\tau-\tau') 
\nonumber \\ & \qquad \qquad
+ {\mathfrak F}_{BB \downarrow \uparrow}(\tau-\tau')F^{bb}_{BB 21}(\tau-\tau') 
- {\mathfrak G}_{BB \downarrow \downarrow }(\tau-\tau') F^{*bd}_{BB 12}(\tau-\tau') 
\nonumber \\ & \qquad \qquad
- {\mathfrak G}_{BB \downarrow \downarrow }(\tau-\tau')F^{bd}_{BB 21}(\tau-\tau') 
+ {\mathfrak G}_{BB \downarrow \downarrow }(\tau-\tau')G^{bb}_{BB 22}(\tau-\tau') 
\nonumber \\ & \qquad \qquad
- {\mathfrak F}_{BB \downarrow \uparrow}(\tau-\tau') G^{*db}_{BB 11}(\tau-\tau')
+ {\mathfrak G}_{BB \downarrow \downarrow }(\tau-\tau') G^{*dd}_{BB 11}(\tau-\tau')
% \nonumber \\ & \qquad \qquad
% + {\mathfrak G}^{*}_{BB \uparrow \uparrow }(\tau-\tau') G^{*bb}_{BB 11}(\tau-\tau')
% - {\mathfrak F}^{*}_{BB \uparrow \downarrow}(\tau-\tau') G^{*bd}_{BB 11}(\tau-\tau')
\Big)
\nonumber \\ & \qquad
+ F^{*f}_{AB \downarrow \uparrow}(\tau-\tau_{2}) \Big(  
% - {\mathfrak F}^{*}_{AB \uparrow \downarrow }(\tau-\tau') G^{bd}_{AB 12}(\tau-\tau') 
% + {\mathfrak F}^{*}_{AB \uparrow \downarrow }(\tau-\tau') F^{dd}_{AB 11}(\tau-\tau') 
% \nonumber \\ & \qquad \qquad
% + {\mathfrak F}^{*}_{AB \uparrow \downarrow }(\tau-\tau')F^{*bb}_{AB 22}(\tau-\tau') 
% + {\mathfrak G}^{*}_{AB \uparrow \uparrow }(\tau-\tau') F^{*bd}_{AB 22}(\tau-\tau') 
% \nonumber \\ & \qquad \qquad
% - {\mathfrak G}^{*}_{AB \uparrow \uparrow }(\tau-\tau') F^{bd}_{AB 11}(\tau-\tau')
% - {\mathfrak G}^{*}_{AB \uparrow \uparrow }(\tau-\tau') G^{dd}_{AB 12}(\tau-\tau') 
% \nonumber \\ & \qquad \qquad
+ {\mathfrak F}_{AB \downarrow \uparrow}(\tau-\tau') G^{bd}_{AB 12}(\tau-\tau') 
+ {\mathfrak F}_{AB \downarrow \uparrow}(\tau-\tau') F^{*dd}_{AB 22}(\tau-\tau')
\nonumber \\ & \qquad \qquad
+ {\mathfrak F}_{AB \downarrow \uparrow}(\tau-\tau') F^{bb}_{AB 11}(\tau-\tau') 
+ {\mathfrak G}_{AB \downarrow \downarrow }(\tau-\tau') F^{*bd}_{AB 22}(\tau-\tau') 
\nonumber \\ & \qquad \qquad
- {\mathfrak G}_{AB \downarrow \downarrow }(\tau-\tau') F^{bd}_{AB 11}(\tau-\tau')
+ {\mathfrak G}_{AB \downarrow \downarrow }(\tau-\tau') G^{bb}_{AB 12}(\tau-\tau') 
% \nonumber \\ & \qquad \qquad
% + {\mathfrak G}^{*}_{AB \uparrow \uparrow }(\tau-\tau') G^{*bb}_{AB 21}(\tau-\tau')
% - {\mathfrak F}^{*}_{AB \uparrow \downarrow }(\tau-\tau') G^{*bd}_{AB 21}(\tau-\tau') 
\nonumber \\ & \qquad \qquad
+ {\mathfrak F}_{AB \downarrow \uparrow}(\tau-\tau') G^{*db}_{AB 21}(\tau-\tau')
- {\mathfrak G}_{AB \downarrow \downarrow }(\tau-\tau') G^{*dd}_{AB 21}(\tau-\tau')
\Big)
\nonumber \\ & \qquad
+ G^{f}_{AB \uparrow \uparrow }(\tau-\tau_{2}) \Big( 
% {\mathfrak F}^{*}_{AB \uparrow \downarrow }(\tau-\tau') G^{bd}_{AB 22}(\tau-\tau') 
% - {\mathfrak F}^{*}_{AB \uparrow \downarrow }(\tau-\tau')F^{dd}_{AB 21}(\tau-\tau') 
% \nonumber \\ & \qquad \qquad
% + {\mathfrak F}^{*}_{AB \uparrow \downarrow }(\tau-\tau') F^{*bb}{AB 12}(\tau-\tau') 
% + {\mathfrak G}^{*}_{AB \uparrow \uparrow }(\tau-\tau') F^{*bd}_{AB 12}(\tau-\tau') 
% \nonumber \\ & \qquad \qquad
% + {\mathfrak G}^{*}_{AB \uparrow \uparrow }(\tau-\tau') F^{bd}_{AB 21}(\tau-\tau') 
% + {\mathfrak G}^{*}_{AB \uparrow \uparrow }(\tau-\tau') G^{dd}_{AB 22}(\tau-\tau') 
% \nonumber \\ & \qquad \qquad
- {\mathfrak F}_{AB \downarrow \uparrow}(\tau-\tau') G^{bd}_{AB 22}(\tau-\tau') 
+ {\mathfrak F}_{AB \downarrow \uparrow}(\tau-\tau') F^{*dd}_{AB 12}(\tau-\tau') 
\nonumber \\ & \qquad \qquad
- {\mathfrak F}_{AB \downarrow \uparrow}(\tau-\tau')F^{bb}_{AB 21}(\tau-\tau') 
+ {\mathfrak G}_{AB \downarrow \downarrow }(\tau-\tau') F^{*bd}_{AB 12}(\tau-\tau') 
\nonumber \\ & \qquad \qquad
+ {\mathfrak G}_{AB \downarrow \downarrow }(\tau-\tau') F^{bd}_{AB 21}(\tau-\tau') 
- {\mathfrak G}_{AB \downarrow \downarrow }(\tau-\tau') G^{bb}_{AB 22}(\tau-\tau') 
% \nonumber \\ & \qquad \qquad
% + {\mathfrak G}^{*}_{AB \uparrow \uparrow }(\tau-\tau') G^{*bb}_{AB 11}(\tau-\tau')
% - {\mathfrak F}^{*}_{AB \uparrow \downarrow }(\tau-\tau') G^{*bd}_{AB 11}(\tau-\tau')
\nonumber \\ & \qquad \qquad
+ {\mathfrak F}_{AB \downarrow \uparrow}(\tau-\tau') G^{*db}_{AB 11}(\tau-\tau')
- {\mathfrak G}_{AB \downarrow \downarrow }(\tau-\tau') G^{dd}_{AB 11}(\tau-\tau') 
\Big)
\Big]=\delta(\tau-\tau_{2}).
  \end{align}

\subsection{Equation of motion for $f_{B\downarrow}$ particles}

%   \subsection{Equation of motion for $f_{\downarrow B}$ particle}

  \begin{align}
%   derivative
  & \partial_{ \tau} G^{f}_{BB \downarrow \downarrow }(\tau-\tau_{2})
  +\lambda_{B}^{-} F^{*f}_{BB \uparrow \downarrow }(\tau'-\tau_{2}) 
  +\lambda_{B}^{z} G^{f}_{BB \downarrow \downarrow }(\tau-\tau_{2}) 
  +\frac{U}{2} n^{dd}_{BB 21}(\tau) F^{*f}_{BB \uparrow \downarrow }(\tau'-\tau_{2}) 
  \nonumber \\ &
  \frac{U}{2} \Big[
  n^{dd}_{BB 22}(\tau)n^{F}_{BB \uparrow \downarrow }(\tau) F^{*f}_{BB \uparrow \downarrow }(\tau'-\tau_{2}) 
  +n^{dd}_{BB 22}(\tau)n_{BB \uparrow \uparrow }(\tau) G^{f}_{BB \downarrow \downarrow }(\tau-\tau_{2}) 
  \Big]
  \nonumber \\ &
  +\frac{v_{1}}{2}\Big[
  + F^{*f}_{AB \uparrow \downarrow }(\tau-\tau_{2}) \Big(  
  - n^{bb}_{BA 21}(\tau)
  + n^{F bd}_{AB 11}(\tau)
  - n^{*Fdb}_{AB 22}(\tau) 
  + n^{dd}_{AB 21}(\tau) 
  - n^{bb}_{AB 21}(\tau) 
  + n^{Fbd}_{BA 11}(\tau) 
  - n^{*Fdb}_{BA 22}(\tau) 
  + n^{dd}_{BA 21}(\tau) 
  \Big)
 \nonumber \\ & \qquad
+ G^{f}_{AB \downarrow \downarrow }(\tau-\tau_{2}) \Big(  
- n^{bb}_{BA 22}(\tau) 
+ n^{F bd}_{AB 21}(\tau)
+ n^{*Fdb}_{AB 12}(\tau) 
- n^{dd}_{AB 11}(\tau) 
+ n^{bb}_{AB 11}(\tau) 
+ n^{bd}_{BA 12}(\tau) 
+ n^{*Fdb}_{BA 21}(\tau) 
+ n^{dd}_{BA 22}(\tau) 
\Big)
  \Big]
 \nonumber \\ &
%    derivative with t, integration over tau'
+\frac{1}{2} \int {\rm d}{\tau'}\Big[
+ F^{*f}_{AB \uparrow \downarrow }(\tau'-\tau_{2}) \Big(  
% - {\mathfrak F}^{*}_{BA \uparrow \downarrow }(\tau-\tau')G^{bd}_{BA 12}(\tau-\tau') 
% + {\mathfrak F}^{*}_{BA \uparrow \downarrow }(\tau-\tau')F^{dd}_{BA 11}(\tau-\tau') 
% \nonumber \\ & \qquad \qquad
% + {\mathfrak F}^{*}_{BA \uparrow \downarrow }(\tau-\tau')F^{*bb}_{BA 22}(\tau-\tau') 
% - {\mathfrak G}^{*}_{BA \uparrow \uparrow }(\tau-\tau')F^{*bd}_{BA 22}(\tau-\tau') 
% \nonumber \\ & \qquad \qquad
% + {\mathfrak G}^{*}_{BA \uparrow \uparrow }(\tau-\tau')F^{bd}_{BA 11}(\tau-\tau') 
% + {\mathfrak G}^{*}_{BA \uparrow \uparrow }(\tau-\tau')G^{dd}_{BA 12}(\tau-\tau') 
% \nonumber \\ & \qquad \qquad
+ {\mathfrak F}_{BA \downarrow \uparrow }(\tau-\tau')G^{bd}_{BA 12}(\tau-\tau') 
+ {\mathfrak F}_{BA \downarrow \uparrow }(\tau-\tau')F^{*dd}_{BA 22}(\tau-\tau')
\nonumber \\ & \qquad \qquad
+ {\mathfrak F}_{BA \downarrow \uparrow }(\tau-\tau')F^{bb}_{BA 11}(\tau-\tau') 
- {\mathfrak G}_{BA \downarrow \downarrow }(\tau-\tau')F^{*bd}_{BA 22}(\tau-\tau') 
\nonumber \\ & \qquad \qquad
+ {\mathfrak G}_{BA \downarrow \downarrow }(\tau-\tau')F^{bd}_{BA 11}(\tau-\tau') 
- {\mathfrak G}_{BA \downarrow \downarrow }(\tau-\tau')G^{bb}_{BA 12}(\tau-\tau') 
\nonumber \\ & \qquad \qquad
+ {\mathfrak G}_{BA \downarrow \downarrow }(\tau-\tau') G^{*dd}_{BA 21}(\tau-\tau')
+ {\mathfrak F}_{BA \downarrow \uparrow }(\tau-\tau')G^{*db}_{BA 21}(\tau-\tau')
% \nonumber \\ & \qquad \qquad
% - {\mathfrak F}^{*}_{BA \uparrow \downarrow }(\tau-\tau') G^{*bd}_{BA 21}(\tau-\tau')
% - {\mathfrak G}^{*}_{BA \uparrow \uparrow }(\tau-\tau') G^{*bb}_{BA 21}(\tau-\tau')
\Big)
\nonumber \\ & \qquad
+ G^{f}_{AB \downarrow \downarrow }(\tau'-\tau_{2}) \Big(
% {\mathfrak F}^{*}_{BA \uparrow \downarrow }(\tau-\tau')G^{bd}_{BA 11}(\tau-\tau')
% + {\mathfrak F}^{*}_{BA \uparrow \downarrow }(\tau-\tau')F^{dd}_{BA 12}(\tau-\tau') 
% \nonumber \\ & \qquad \qquad
% - {\mathfrak F}^{*}_{BA \uparrow \downarrow }(\tau-\tau')F^{*bb}_{BA 21}(\tau-\tau') 
% + {\mathfrak G}^{*}_{BA \uparrow \uparrow }(\tau-\tau')F^{*bd}_{BA 21}(\tau-\tau') 
% \nonumber \\ & \qquad \qquad
% + {\mathfrak G}^{*}_{BA \uparrow \uparrow }(\tau-\tau')F^{bd}_{BA 12}(\tau-\tau') 
% - {\mathfrak G}^{*}_{BA \uparrow \uparrow }(\tau-\tau')G^{dd}_{BA 11}(\tau-\tau') 
% \nonumber \\ & \qquad \qquad
- {\mathfrak F}_{BA \downarrow \uparrow }(\tau-\tau')G^{bd}_{BA 11}(\tau-\tau') 
- {\mathfrak F}_{BA \downarrow \uparrow }(\tau-\tau')F^{*dd}_{BA 21}(\tau-\tau') 
\nonumber \\ & \qquad \qquad
+ {\mathfrak F}_{BA \downarrow \uparrow }(\tau-\tau')F^{bb}_{BA 12}(\tau-\tau') 
+ {\mathfrak G}_{BA \downarrow \downarrow }(\tau-\tau')F^{*bd}_{BA 21}(\tau-\tau') 
\nonumber \\ & \qquad \qquad
+ {\mathfrak G}_{BA \downarrow \downarrow }(\tau-\tau')F^{bd}_{BA 12}(\tau-\tau') 
+ {\mathfrak G}_{BA \downarrow \downarrow }(\tau-\tau')G^{bb}_{BA 11}(\tau-\tau')
\nonumber \\ & \qquad \qquad
+ {\mathfrak G}_{BA \downarrow \downarrow }(\tau-\tau') G^{*dd}_{BA 22}(\tau-\tau')
+ {\mathfrak F}_{BA \downarrow \uparrow }(\tau-\tau')G^{*db}_{BA 22}(\tau-\tau')
% \nonumber \\ & \qquad \qquad
% - {\mathfrak F}^{*}_{BA \uparrow \downarrow }(\tau-\tau') G^{*bd}_{BA 22}(\tau-\tau')
% - {\mathfrak G}^{*}_{BA \uparrow \uparrow }(\tau-\tau') G^{*bb}_{BA 22}(\tau-\tau')
\Big)
\nonumber \\ & \qquad
+ F^{*f}_{BB \uparrow \downarrow }(\tau'-\tau_{2}) \Big( 
% {\mathfrak F}^{*}_{BB \uparrow \downarrow }(\tau-\tau')G^{bd}_{BB 12}(\tau-\tau') 
% - {\mathfrak F}^{*}_{BB \uparrow \downarrow }(\tau-\tau')F^{dd}_{BB 11}(\tau-\tau')
% \nonumber \\ & \qquad \qquad
% - {\mathfrak F}^{*}_{BB \uparrow \downarrow }(\tau-\tau')F^{*bb}_{BB 22}(\tau-\tau')
% - {\mathfrak G}^{*}_{BB \uparrow \uparrow }(\tau-\tau')F^{*bd}_{BB 22}(\tau-\tau') 
% \nonumber \\ & \qquad \qquad
% + {\mathfrak G}^{*}_{BB \uparrow \uparrow }(\tau-\tau')F^{bd}_{BB 11}(\tau-\tau') 
% + {\mathfrak G}^{*}_{BB \uparrow \uparrow }(\tau-\tau')G^{dd}_{BB 12}(\tau-\tau') 
% \nonumber \\ & \qquad \qquad
+ {\mathfrak F}_{BB \downarrow \uparrow }(\tau-\tau')G^{bd}_{BB 12}(\tau-\tau') 
+ {\mathfrak F}_{BB \downarrow \uparrow }(\tau-\tau')F^{*dd}_{BB 22}(\tau-\tau')
\nonumber \\ & \qquad \qquad
+ {\mathfrak F}_{BB \downarrow \uparrow }(\tau-\tau')F^{bb}_{BB 11}(\tau-\tau') 
+ {\mathfrak G}_{BB \downarrow \downarrow }(\tau-\tau')F^{*bd}_{BB 22}(\tau-\tau') 
\nonumber \\ & \qquad \qquad
- {\mathfrak G}_{BB \downarrow \downarrow }(\tau-\tau')F^{bd}_{BB 11}(\tau-\tau') 
+ {\mathfrak G}_{BB \downarrow \downarrow }(\tau-\tau')G^{bb}_{BB 12}(\tau-\tau') 
\nonumber \\ & \qquad \qquad
+ {\mathfrak F}_{BB \downarrow \uparrow }(\tau-\tau') G^{*db}_{BB 21}(\tau-\tau')
- {\mathfrak G}_{BB \downarrow \downarrow }(\tau-\tau') G^{dd}_{BB 21}(\tau-\tau')
% \nonumber \\ & \qquad \qquad
% - {\mathfrak G}^{*}_{BB \uparrow \uparrow }(\tau-\tau') G^{*bb}_{BB 21}(\tau-\tau') 
% + {\mathfrak F}^{*}_{BB \uparrow \downarrow }(\tau-\tau') G^{*bd}_{BB 21}(\tau-\tau')
\Big)
\nonumber \\ & \qquad
+ G^{f}_{BB \downarrow \downarrow }(\tau-\tau_{2}) \Big( 
% {\mathfrak F}^{*}_{BB \uparrow \downarrow }(\tau-\tau')G^{bd}_{BB 11}(\tau-\tau') 
% + {\mathfrak F}^{*}_{BB \uparrow \downarrow }(\tau-\tau')F^{dd}_{BB 12}(\tau-\tau') 
% \nonumber \\ & \qquad \qquad
% - {\mathfrak F}^{*}_{BB \uparrow \downarrow }(\tau-\tau')F^{*bb}_{BB 21}(\tau-\tau') 
% - {\mathfrak G}^{*}_{BB \uparrow \uparrow }(\tau-\tau')F^{*bd}_{BB 21}(\tau-\tau') 
% \nonumber \\ & \qquad \qquad
% - {\mathfrak G}^{*}_{BB \uparrow \uparrow }(\tau-\tau')F^{bd}_{BB 12}(\tau-\tau') 
% + {\mathfrak G}^{*}_{BB \uparrow \uparrow }(\tau-\tau')G^{dd}_{BB 11}(\tau-\tau') 
% \nonumber \\ & \qquad \qquad
+ {\mathfrak F}_{BB \downarrow \uparrow }(\tau-\tau')G^{bd}_{BB 11}(\tau-\tau') 
+ {\mathfrak F}_{BB \downarrow \uparrow }(\tau-\tau')F^{*dd}_{BB 21}(\tau-\tau') 
\nonumber \\ & \qquad \qquad
- {\mathfrak F}_{BB \downarrow \uparrow }(\tau-\tau')F^{bb}_{BB 12}(\tau-\tau') 
+ {\mathfrak G}_{BB \downarrow \downarrow }(\tau-\tau')F^{*bd}_{BB 21}(\tau-\tau') 
\nonumber \\ & \qquad \qquad
+ {\mathfrak G}_{BB \downarrow \downarrow }(\tau-\tau')F^{bd}_{BB 12}(\tau-\tau') 
+ {\mathfrak G}_{BB \downarrow \downarrow }(\tau-\tau')G^{bb}_{BB 11}(\tau-\tau') 
\nonumber \\ & \qquad \qquad
- {\mathfrak F}_{BB \downarrow \uparrow }(\tau-\tau') G^{*db}_{BB 22}(\tau-\tau')
+ {\mathfrak G}_{BB \downarrow \downarrow }(\tau-\tau') G^{*dd}_{BB 22}(\tau-\tau')
% \nonumber \\ & \qquad \qquad
% + {\mathfrak G}^{*}_{BB \uparrow \uparrow }(\tau-\tau') G^{*bb}_{BB 22}(\tau-\tau')
% - {\mathfrak F}^{*}_{BB \uparrow \downarrow }(\tau-\tau') G^{bd}_{BB 22}(\tau-\tau')
\Big)
\Big]
  \nonumber \\ &
%    derivative with t', integration over tau
+ \frac{1}{2} \int {\rm d}{\tau}\Big[
+ F^{*f}_{BB \uparrow \downarrow }(\tau'-\tau_{2}) \Big( 
% {\mathfrak F}^{*}_{BB \downarrow \uparrow }(\tau-\tau')G^{bd}_{BB 12}(\tau-\tau') 
% - {\mathfrak F}^{*}_{BB \downarrow \uparrow }(\tau-\tau')F^{dd}_{BB 11}(\tau-\tau')
% \nonumber \\ & \qquad \qquad
% - {\mathfrak F}^{*}_{BB \downarrow \uparrow }(\tau-\tau')F^{*bb}_{BB 22}(\tau-\tau')
% + {\mathfrak G}^{*}_{BB \downarrow \downarrow }(\tau-\tau')F^{*bd}_{BB 22}(\tau-\tau') 
% \nonumber \\ & \qquad \qquad
% - {\mathfrak G}^{*}_{BB \downarrow \downarrow }(\tau-\tau')F^{bd}_{BB 11}(\tau-\tau') 
% - {\mathfrak G}^{*}_{BB \downarrow \downarrow }(\tau-\tau')G^{dd}_{BB 12}(\tau-\tau')
% \nonumber \\ & \qquad \qquad
+ {\mathfrak F}_{BB \uparrow \downarrow }(\tau-\tau')G^{bd}_{BB 12}(\tau-\tau') 
+ {\mathfrak F}_{BB \uparrow \downarrow }(\tau-\tau')F^{*dd}_{BB 22}(\tau-\tau')
\nonumber \\ & \qquad \qquad
+ {\mathfrak F}_{BB \uparrow \downarrow }(\tau-\tau')F^{bb}_{BB 11}(\tau-\tau') 
- {\mathfrak G}_{BB \uparrow \uparrow }(\tau-\tau')F^{*bd}_{BB 22}(\tau-\tau') 
\nonumber \\ & \qquad \qquad
+ {\mathfrak G}_{BB \uparrow \uparrow }(\tau-\tau')F^{bd}_{BB 11}(\tau-\tau') 
- {\mathfrak G}_{BB \uparrow \uparrow }(\tau-\tau')G^{bb}_{BB 12}(\tau-\tau') 
\nonumber \\ & \qquad \qquad
+ {\mathfrak F}_{BB \uparrow \downarrow }(\tau-\tau') G^{*db}_{BB 21}(\tau-\tau')
+ {\mathfrak G}_{BB \uparrow \uparrow }(\tau-\tau') G^{dd}_{BB 21}(\tau-\tau')
% \nonumber \\ & \qquad \qquad
% + {\mathfrak G}^{*}_{BB \downarrow \downarrow }(\tau-\tau') G^{*bb}_{BB 21}(\tau-\tau')
% + {\mathfrak F}^{*}_{BB \downarrow \uparrow }(\tau-\tau') G^{*bd}_{BB 21}(\tau-\tau')
\Big)
\nonumber \\ & \qquad
+ G^{f}_{BB \downarrow \downarrow }(\tau-\tau_{2}) \Big(  
% - {\mathfrak F}^{*}_{BB \downarrow \uparrow }(\tau-\tau')G^{bd}_{BB 22}(\tau-\tau') 
% + {\mathfrak F}^{*}_{BB \downarrow \uparrow }(\tau-\tau')F^{dd}_{BB 21}(\tau-\tau') 
% \nonumber \\ & \qquad \qquad
% - {\mathfrak F}^{*}_{BB \downarrow \uparrow }(\tau-\tau')F^{*bb}_{BB 12}(\tau-\tau') 
% + {\mathfrak G}^{*}_{BB \downarrow \downarrow }(\tau-\tau')F^{*bd}_{BB 12}(\tau-\tau')
% \nonumber \\ & \qquad \qquad
% + {\mathfrak G}^{*}_{BB \downarrow \downarrow }(\tau-\tau')F^{bd}_{BB 21}(\tau-\tau') 
% + {\mathfrak G}^{*}_{BB \downarrow \downarrow }(\tau-\tau')G^{dd}_{BB 22}(\tau-\tau') 
% \nonumber \\ & \qquad \qquad
- {\mathfrak F}_{BB \uparrow \downarrow }(\tau-\tau')G^{bd}_{BB 22}(\tau-\tau') 
+ {\mathfrak F}_{BB \uparrow \downarrow }(\tau-\tau')F^{*dd}_{BB 12}(\tau-\tau') 
\nonumber \\ & \qquad \qquad
- {\mathfrak F}_{BB \uparrow \downarrow }(\tau-\tau')F^{bb}_{BB 21}(\tau-\tau') 
- {\mathfrak G}_{BB \uparrow \uparrow }(\tau-\tau')F^{*bd}_{BB 12}(\tau-\tau') 
\nonumber \\ & \qquad \qquad
- {\mathfrak G}_{BB \uparrow \uparrow }(\tau-\tau')F^{bd}_{BB 21}(\tau-\tau') 
+ {\mathfrak G}_{BB \uparrow \uparrow }(\tau-\tau')G^{bb}_{BB 22}(\tau-\tau') 
\nonumber \\ & \qquad \qquad
+ {\mathfrak F}_{BB \uparrow \downarrow }(\tau-\tau') G^{*db}_{BB 11}(\tau-\tau')
+ {\mathfrak G}_{BB \uparrow \uparrow }(\tau-\tau') G^{*dd}_{BB 11}(\tau-\tau')
% \nonumber \\ & \qquad \qquad
% + {\mathfrak G}^{*}_{BB \downarrow \downarrow }(\tau-\tau') G^{*bb}_{BB 11}(\tau-\tau')
% + {\mathfrak F}^{*}_{BB \downarrow \uparrow }(\tau-\tau') G^{*bd}_{BB 11}(\tau-\tau')
\Big)
\nonumber \\ & \qquad
+ F^{*f}_{AB \uparrow \downarrow }(\tau-\tau_{2}) \Big(  
% - {\mathfrak F}^{*}_{AB \downarrow \uparrow }(\tau-\tau')G^{bd}_{AB 12}(\tau-\tau') 
% + {\mathfrak F}^{*}_{AB \downarrow \uparrow }(\tau-\tau')F^{dd}_{AB 11}(\tau-\tau') 
% \nonumber \\ & \qquad \qquad
% + {\mathfrak F}^{*}_{AB \downarrow \uparrow }(\tau-\tau')F^{*bb}_{AB 22}(\tau-\tau') 
% - {\mathfrak G}^{*}_{AB \downarrow \downarrow }(\tau-\tau')F^{*bd}_{AB 22}(\tau-\tau')
% \nonumber \\ & \qquad \qquad
% + {\mathfrak G}^{*}_{AB \downarrow \downarrow }(\tau-\tau')F^{bd}_{AB 11}(\tau-\tau') 
% + {\mathfrak G}^{*}_{AB \downarrow \downarrow }(\tau-\tau')G^{dd}_{AB 12}(\tau-\tau') 
% \nonumber \\ & \qquad \qquad
+ {\mathfrak F}_{AB \uparrow \downarrow }(\tau-\tau')G^{bd}_{AB 12}(\tau-\tau') 
+ {\mathfrak F}_{AB \uparrow \downarrow }(\tau-\tau')F^{*dd}_{AB 22}(\tau-\tau') 
\nonumber \\ & \qquad \qquad
+ {\mathfrak F}_{AB \uparrow \downarrow }(\tau-\tau')F^{bb}_{AB 11}(\tau-\tau') 
- {\mathfrak G}_{AB \uparrow \uparrow }(\tau-\tau')F^{*bd}_{AB 22}(\tau-\tau')
\nonumber \\ & \qquad \qquad
+ {\mathfrak G}_{AB \uparrow \uparrow }(\tau-\tau')F^{bd}_{AB 11}(\tau-\tau') 
- {\mathfrak G}_{AB \uparrow \uparrow }(\tau-\tau')G^{bb}_{AB 12}(\tau-\tau') 
% \nonumber \\ & \qquad \qquad
% - {\mathfrak G}^{*}_{AB \downarrow \downarrow }(\tau-\tau') G^{*bb}_{AB 21}(\tau-\tau')
% - {\mathfrak F}^{*}_{AB \downarrow \uparrow }(\tau-\tau') G^{*bd}_{AB 21}(\tau-\tau')
\nonumber \\ & \qquad \qquad
+ {\mathfrak F}_{AB \uparrow \downarrow }(\tau-\tau') G^{*db}_{AB 21}(\tau-\tau')
+ {\mathfrak G}_{AB \uparrow \uparrow }(\tau-\tau') G^{*dd}_{AB 21}(\tau-\tau')
\Big)
\nonumber \\ & \qquad
+ G^{f}_{AB \downarrow \downarrow }(\tau-\tau_{2}) \Big(  
% - {\mathfrak F}^{*}_{AB \downarrow \uparrow }(\tau-\tau')G^{bd}_{AB 22}(\tau-\tau') 
% + {\mathfrak F}^{*}_{AB \downarrow \uparrow }(\tau-\tau')F^{dd}_{AB 21}(\tau-\tau') 
% \nonumber \\ & \qquad \qquad
% - {\mathfrak F}^{*}_{AB \downarrow \uparrow }(\tau-\tau')F^{*bb}_{AB 12}(\tau-\tau') 
% + {\mathfrak G}^{*}_{AB \downarrow \downarrow }(\tau-\tau')F^{*bd}_{AB 12}(\tau-\tau')
% \nonumber \\ & \qquad \qquad
% + {\mathfrak G}^{*}_{AB \downarrow \downarrow }(\tau-\tau')F^{bd}_{AB 21}(\tau-\tau') 
% + {\mathfrak G}^{*}_{AB \downarrow \downarrow }(\tau-\tau')G^{dd}_{AB 22}(\tau-\tau') 
% \nonumber \\ & \qquad \qquad
+ {\mathfrak F}_{AB \uparrow \downarrow }(\tau-\tau')G^{bd}_{AB 22}(\tau-\tau') 
- {\mathfrak F}_{AB \uparrow \downarrow }(\tau-\tau')F^{*dd}_{AB 12}(\tau-\tau')
\nonumber \\ & \qquad \qquad
+ {\mathfrak F}_{AB \uparrow \downarrow }(\tau-\tau')F^{bb}_{AB 21}(\tau-\tau') 
+ {\mathfrak G}_{AB \uparrow \uparrow }(\tau-\tau')F^{*bd}_{AB 12}(\tau-\tau') 
\nonumber \\ & \qquad \qquad
+ {\mathfrak G}_{AB \uparrow \uparrow }(\tau-\tau')F^{bd}_{AB 21}(\tau-\tau') 
- {\mathfrak G}_{AB \uparrow \uparrow }(\tau-\tau')G^{bb}_{AB 22}(\tau-\tau') 
% \nonumber \\ & \qquad \qquad
% + {\mathfrak G}^{*}_{AB \downarrow \downarrow }(\tau-\tau') G^{*bb}_{AB 11}(\tau-\tau')
% + {\mathfrak F}^{*}_{AB \downarrow \uparrow }(\tau-\tau') G^{*bd}_{AB 11}(\tau-\tau')
\nonumber \\ & \qquad \qquad
- {\mathfrak F}_{AB \uparrow \downarrow }(\tau-\tau') G^{*db}_{AB 11}(\tau-\tau')
- {\mathfrak G}_{AB \uparrow \uparrow }(\tau-\tau') G^{*dd}_{AB 11}(\tau-\tau')
  \Big)
\Big]=\delta(\tau-\tau_{2}).
  \end{align}

\subsection{Equation of motion for $b_{A 1}$ particles }

%   \subsection{Equation of motion for $b_{1 A}$ particle}

 \begin{align}
%   derivative
  & \partial_{ \tau} G^{bb}_{AA 11}(\tau-\tau_{2}) 
  + \lambda_{A}^{-} G^{bb}_{AA 21}(\tau-\tau_{2}) 
  + \lambda_{A}^{z} G^{bb}_{AA 11}(\tau-\tau_{2}) 
  - \Big(\mu-\frac{U}{2} \Big) G^{bb}_{AA 11}(\tau-\tau_{2}) 
  \nonumber \\
  & +\frac{v_{1}}{2} \Big[
  + F^{*db}_{BA 21}(\tau-\tau2)  \Big( 
  n^{f}_{BA \uparrow \uparrow }(\tau) 
  + n^{f}_{BA \downarrow \downarrow }(\tau)
  \Big)
   + F^{*db}_{BA 11}(\tau-\tau_{2})  \Big(
   - n^{Ff}_{BA \uparrow \downarrow }(\tau) 
   + n^{Ff}_{BA \downarrow \uparrow  }(\tau)
   \Big)
   \nonumber \\ & \qquad
   + G^{bb}_{BA 21}(\tau-\tau_{2})  \Big( 
   n^{Ff}_{BA \uparrow \downarrow }(\tau) 
   - n^{Ff}_{BA \downarrow \uparrow  }(\tau) 
   \Big)
   + G^{bb}_{BA 11}(\tau-\tau_{2})  \Big( 
   n^{f}_{BA \uparrow \uparrow }(\tau) 
   + n^{f}_{BA \downarrow \downarrow }(\tau)
   \Big)
     \Big]
\nonumber \\ &
%      derivative with tp
+\frac{1}{2} \int {\rm d}{\tau}\Big[
+ F^{*db}_{BA 21}(\tau-\tau_{2})  \Big(  
-G^{f}_{AB \uparrow \uparrow }(\tau'-\tau) {\mathfrak G}_{BA \uparrow \uparrow }(\tau-\tau')
-G^{f}_{AB \downarrow \downarrow }(\tau'-\tau) {\mathfrak G}_{BA \downarrow \downarrow }(\tau-\tau')
\Big)
\nonumber \\ & \qquad
+ F^{*db}_{BA 11}(\tau-\tau_{2})  \Big( 
+ F^{f}_{AB \downarrow \uparrow }(\tau'-\tau) {\mathfrak G}_{BA \downarrow \downarrow }(\tau-\tau')
-F^{f}_{AB \uparrow \downarrow }(\tau'-\tau) {\mathfrak G}_{BA \uparrow \uparrow }(\tau-\tau')
\Big)
\nonumber \\ & \qquad
+ G^{db}_{BA 21}(\tau-\tau_{2})  \Big(  
-F^{f}_{AB \downarrow \uparrow }(\tau'-\tau){\mathfrak F}_{BA \uparrow \downarrow }(\tau-\tau')
-F^{f}_{AB \uparrow \downarrow }(\tau'-\tau) {\mathfrak F}_{BA \downarrow \uparrow }(\tau-\tau')
\Big)
\nonumber \\ & \qquad
+ G^{db}_{BA 11}(\tau-\tau_{2})  \Big( 
  + G^{f}_{AB \uparrow \uparrow }(\tau'-\tau) {\mathfrak F}_{BA \downarrow \uparrow }(\tau-\tau')
-G^{f}_{AB \downarrow \downarrow }(\tau'-\tau){\mathfrak F}_{BA \uparrow \downarrow }(\tau-\tau')
\Big)
\nonumber \\ & \qquad
+ F^{*db}_{AA 21}(\tau-\tau_{2})  \Big(  
-G^{f}_{AA \uparrow \uparrow }(\tau'-\tau) {\mathfrak G}_{AA \uparrow \uparrow }(\tau-\tau')
-G^{f}_{AA \downarrow \downarrow }(\tau'-\tau) {\mathfrak G}_{AA \downarrow \downarrow }(\tau-\tau')
\Big)
\nonumber \\ & \qquad
+ F^{*db}_{AA 11}(\tau-\tau_{2})  \Big(  
-F^{f}_{AA \downarrow \uparrow }(\tau'-\tau) {\mathfrak G}_{AA \downarrow \downarrow }(\tau-\tau')
+ F^{f}_{AA \uparrow \downarrow }(\tau'-\tau) {\mathfrak G}_{AA \uparrow \uparrow }(\tau-\tau')
\Big)
\nonumber \\ & \qquad
+ G^{db}_{AA 21}(\tau-\tau_{2})  \Big(  
-F^{f}_{AA \downarrow \uparrow }(\tau'-\tau) {\mathfrak F}_{AA \uparrow \downarrow }(\tau-\tau')
-F^{f}_{AA \uparrow \downarrow }(\tau'-\tau) {\mathfrak F}_{AA \downarrow \uparrow }(\tau-\tau')
\Big)
\nonumber \\ & \qquad
+ G^{db}_{AA 11}(\tau-\tau_{2})  \Big(  
-G^{f}_{AA \uparrow \uparrow }(\tau'-\tau) {\mathfrak F}_{AA \downarrow \uparrow }(\tau-\tau')
+ G^{f}_{AA \downarrow \downarrow }(\tau'-\tau) {\mathfrak F}_{AA \uparrow \downarrow }(\tau-\tau')
\Big)
\nonumber \\ & \qquad
+ F^{*bb}_{BA 21}(\tau-\tau_{2})  \Big(  
-G^{f}_{AB \uparrow \uparrow }(\tau'-\tau) {\mathfrak F}_{BA \downarrow \uparrow }(\tau-\tau')
+ G^{f}_{AB \downarrow \downarrow }(\tau'-\tau){\mathfrak F}_{BA \uparrow \downarrow }(\tau-\tau')
\Big)
\nonumber \\ & \qquad
+ F^{*bb}_{BA 11}(\tau-\tau_{2})  \Big(  
-F^{f}_{AB \downarrow \uparrow }(\tau'-\tau){\mathfrak F}_{BA \uparrow \downarrow }(\tau-\tau')
-F^{f}_{AB \uparrow \downarrow }(\tau'-\tau) {\mathfrak F}_{BA \downarrow \uparrow }(\tau-\tau')
\Big)
\nonumber \\ & \qquad
+ G^{bb}_{BA 21}(\tau-\tau_{2})  \Big(  
-F^{f}_{AB \downarrow \uparrow }(\tau'-\tau) {\mathfrak G}_{BA \downarrow \downarrow }(\tau-\tau')
+ F^{f}_{AB \uparrow \downarrow }(\tau'-\tau) {\mathfrak G}_{BA \uparrow \uparrow }(\tau-\tau')
\Big)
\nonumber \\ & \qquad
+ G^{bb}_{BA 11}(\tau-\tau_{2})  \Big(  
-G^{f}_{AB \uparrow \uparrow }(\tau'-\tau) {\mathfrak G}_{BA \uparrow \uparrow }(\tau-\tau')
-G^{f}_{AB \downarrow \downarrow }(\tau'-\tau) {\mathfrak G}_{BA \downarrow \downarrow }(\tau-\tau')
\Big)
\nonumber \\ & \qquad
+ F^{*bb}_{AA 21}(\tau-\tau_{2})  \Big( 
 + G^{f}_{AA \uparrow \uparrow }(\tau'-\tau) {\mathfrak F}_{AA \downarrow \uparrow }(\tau-\tau') 
-G^{f}_{AA \downarrow \downarrow }(\tau'-\tau) {\mathfrak F}_{AA \uparrow \downarrow }(\tau-\tau')
\Big)
\nonumber \\ & \qquad
+ F^{*bb}_{AA 11}(\tau-\tau_{2})  \Big(  
-F^{f}_{AA \downarrow \uparrow }(\tau'-\tau) {\mathfrak F}_{AA \uparrow \downarrow }(\tau-\tau')
-F^{f}_{AA \uparrow \downarrow }(\tau'-\tau) {\mathfrak F}_{AA \downarrow \uparrow }(\tau-\tau')
\Big)
\nonumber \\ & \qquad
+ G^{bb}_{AA 21}(\tau-\tau_{2})  \Big( 
+ F^{f}_{AA \downarrow \uparrow }(\tau'-\tau) {\mathfrak G}_{AA \downarrow \downarrow }(\tau-\tau')
-F^{f}_{AA \uparrow \downarrow }(\tau'-\tau) {\mathfrak G}_{AA \uparrow \uparrow }(\tau-\tau')
\Big)
\nonumber \\ & \qquad
+ G^{bb}_{AA 11}(\tau-\tau_{2})  \Big(  
-G^{f}_{AA \uparrow \uparrow }(\tau'-\tau) {\mathfrak G}_{AA \uparrow \uparrow }(\tau-\tau')
-G^{f}_{AA \downarrow \downarrow }(\tau'-\tau) {\mathfrak G}_{AA \downarrow \downarrow }(\tau-\tau')
\Big)
\Big]=\delta(\tau-\tau_{2}).
 \end{align}

\subsection{Equation of motion for $b_{2A}$ particles }

%   \subsection{Equation of motion for $b_{2 A}$ particle}

 \begin{align}
%   derivative
  & \partial_{ \tau} G^{bb}_{AA 22}(\tau-\tau_{2}) 
  +\lambda^{+}_{A} G^{bb}_{AB 12}(\tau-\tau_{2}) 
  -\lambda_{A}^{z} G^{bb}_{AA 22}(\tau-\tau_{2}) 
  - \big( \mu -\frac{U}{2} \big) G^{bb}_{AA 22}(\tau-\tau_{2}) 
  \nonumber \\
  & +\frac{v_{1}}{2} \Big[
  + F^{*db}_{BA 22}(\tau-\tau_{2}) * ( 
  n^{*F}_{BA \uparrow \downarrow }(\tau) 
  - n^{*Ff}_{BA \downarrow \uparrow }(\tau) 
  )
  + F^{*db}_{BA 12}(\tau-\tau_{2}) * (
  -n^{f}_{AB \uparrow \uparrow }(\tau) 
  + -n^{f}_{AB \downarrow \downarrow }(\tau) 
  )
  \nonumber \\ & \qquad 
  + G^{bb}_{BA 22}(\tau-\tau_{2}) * (  
  - -n^{f}_{AB \uparrow \uparrow }(\tau) 
  - -n^{f}_{AB \downarrow \downarrow }(\tau) 
  )
  + G^{bb}_{BA 12}(\tau-\tau_{2}) * ( 
  n^{*Ff}_{BA \uparrow \downarrow }(\tau) 
  - n^{*Ff}_{BA \downarrow \uparrow }(\tau) 
  )
  \Big]
   \nonumber \\ &
%    derivative with respect to t', integration on t
+\frac{1}{2} \int {\rm d}{\tau} \Big[
+ F^{*db}_{BA 22}(\tau-\tau_{2}) \Big(  
- F^{*f}_{AB \downarrow \uparrow }(\tau'-\tau) {\mathfrak G}_{BA \uparrow \uparrow }(\tau-\tau') 
+ F^{*f}_{AB \uparrow \downarrow}(\tau'-\tau) {\mathfrak G}_{BA \downarrow \downarrow }(\tau-\tau') 
\Big)
  \nonumber \\ & \qquad
+ F^{*db}_{BA 12}(\tau-\tau_{2}) \Big(  
- G^{*f}_{AB \uparrow \uparrow }(\tau'-\tau) {\mathfrak G}_{BA \downarrow \downarrow }(\tau-\tau') 
- G^{*f}_{AB \downarrow \downarrow }(\tau'-\tau) {\mathfrak G}_{BA \uparrow \uparrow }(\tau-\tau') 
\Big)
  \nonumber \\ & \qquad
+ G^{db}_{BA 22}(\tau-\tau_{2}) \Big( 
G^{*f}_{AB \uparrow \uparrow }(\tau'-\tau) {\mathfrak F}_{BA \uparrow \downarrow }(\tau-\tau') 
- G^{*f}_{AB \downarrow \downarrow }(\tau'-\tau) {\mathfrak F}_{BA \downarrow \uparrow }(\tau-\tau')
\Big)
  \nonumber \\ & \qquad
+ G^{db}_{BA 12}(\tau-\tau_{2}) \Big( 
F^{*f}_{AB \downarrow \uparrow }(\tau'-\tau) {\mathfrak F}_{BA \downarrow \uparrow }(\tau-\tau')
+ F^{*f}_{AB \uparrow \downarrow}(\tau'-\tau) {\mathfrak F}_{BA \uparrow \downarrow }(\tau-\tau') 
\Big)
  \nonumber \\ & \qquad
 + F^{*db}_{AA 22}(\tau-\tau_{2}) \Big(  
 - F^{*f}_{AA \downarrow \uparrow }(\tau'-\tau) {\mathfrak G}_{AA \uparrow \uparrow }(\tau-\tau') 
 + F^{*f}_{AA \uparrow \downarrow}(\tau'-\tau) {\mathfrak G}_{AA \downarrow \downarrow }(\tau-\tau')
 \Big)
  \nonumber \\ & \qquad
 + F^{*db}_{AA 12}(\tau-\tau_{2}) \Big( 
 G^{*f}_{AA \uparrow \uparrow }(\tau'-\tau) {\mathfrak G}_{AA \downarrow \downarrow }(\tau-\tau')
 + G^{*f}_{AA \downarrow \downarrow }(\tau'-\tau) {\mathfrak G}_{AA \uparrow \uparrow }(\tau-\tau') 
 \Big)
  \nonumber \\ & \qquad
 + G^{db}_{AA 22}(\tau-\tau_{2}) \Big( 
 G^{*f}_{AA \uparrow \uparrow }(\tau'-\tau) {\mathfrak F}_{AA \uparrow \downarrow}(\tau-\tau') 
 - G^{*f}_{AA \downarrow \downarrow }(\tau'-\tau) {\mathfrak F}_{AA \downarrow \uparrow }(\tau-\tau')
 \Big)
  \nonumber \\ & \qquad
 + G^{db}_{AA 12}(\tau-\tau_{2}) \Big(  
 - F^{*f}_{AA \downarrow \uparrow }(\tau'-\tau) {\mathfrak F}_{AA \downarrow \uparrow }(\tau-\tau')
 - F^{*f}_{AA \uparrow \downarrow}(\tau'-\tau) {\mathfrak F}_{AA \uparrow \downarrow}(\tau-\tau') 
 \Big)
  \nonumber \\ & \qquad
 + F^{*bb}_{BA 22}(\tau-\tau_{2}) \Big(  
 - F^{*f}_{AB \downarrow \uparrow }(\tau'-\tau) {\mathfrak F}_{BA \downarrow \uparrow }(\tau-\tau')
 - F^{*f}_{AB \uparrow \downarrow}(\tau'-\tau) {\mathfrak F}_{BA \uparrow \downarrow }(\tau-\tau') 
 \Big)
  \nonumber \\ & \qquad
 + F^{*bb}_{BA 12}(\tau-\tau_{2}) \Big( 
 G^{*f}_{AB \uparrow \uparrow }(\tau'-\tau) {\mathfrak F}_{BA \uparrow \downarrow }(\tau-\tau') 
 - G^{*f}_{AB \downarrow \downarrow }(\tau'-\tau) {\mathfrak F}_{BA \downarrow \uparrow }(\tau-\tau')
 \Big)
  \nonumber \\ & \qquad
 + G^{bb}_{BA 22}(\tau-\tau_{2}) \Big( 
 G^{*f}_{AB \uparrow \uparrow }(\tau'-\tau) {\mathfrak G}_{BA \downarrow \downarrow }(\tau-\tau') 
 + G^{*f}_{AB \downarrow \downarrow }(\tau'-\tau) {\mathfrak G}_{BA \uparrow \uparrow }(\tau-\tau') 
 \Big)
  \nonumber \\ & \qquad
 + G^{bb}_{BA 12}(\tau-\tau_{2}) \Big(  
 - F^{*f}_{AB \downarrow \uparrow }(\tau'-\tau) {\mathfrak G}_{BA \uparrow \uparrow }(\tau-\tau') 
 + F^{*f}_{AB \uparrow \downarrow}(\tau'-\tau) {\mathfrak G}_{BA \downarrow \downarrow }(\tau-\tau') 
 \Big)
  \nonumber \\ & \qquad
 + F^{*bb}_{AA 22}(\tau-\tau_{2}) \Big( 
 F^{*f}_{AA \downarrow \uparrow }(\tau'-\tau) {\mathfrak F}_{AA \downarrow \uparrow }(\tau-\tau')
 + F^{*f}_{AA \uparrow \downarrow}(\tau'-\tau) {\mathfrak F}_{AA \uparrow \downarrow}(\tau-\tau') 
 \Big)
  \nonumber \\ & \qquad
 + F^{*bb}_{AA 12}(\tau-\tau_{2}) \Big( 
 G^{*f}_{AA \uparrow \uparrow }(\tau'-\tau) {\mathfrak F}_{AA \uparrow \downarrow}(\tau-\tau') 
 - G^{*f}_{AA \downarrow \downarrow }(\tau'-\tau) {\mathfrak F}_{AA \downarrow \uparrow }(\tau-\tau')
 \Big)
  \nonumber \\ & \qquad
 + G^{bb}_{AA 22}(\tau-\tau_{2}) \Big(  
 - G^{*f}_{AA \uparrow \uparrow }(\tau'-\tau) {\mathfrak G}_{AA \downarrow \downarrow }(\tau-\tau')
 - G^{*f}_{AA \downarrow \downarrow }(\tau'-\tau) {\mathfrak G}_{AA \uparrow \uparrow }(\tau-\tau') 
 \Big)
  \nonumber \\ & \qquad
 + G^{bb}_{AA 12}(\tau-\tau_{2}) \Big(  
 - F^{*f}_{AA \downarrow \uparrow }(\tau'-\tau) {\mathfrak G}_{AA \uparrow \uparrow }(\tau-\tau') 
 + F^{*f}_{AA \uparrow \downarrow}(\tau'-\tau) {\mathfrak G}_{AA \downarrow \downarrow }(\tau-\tau')
 \Big)
 \Big]=\delta(\tau,\tau_{2}).
 \end{align}

\subsection{ Equation of motion for $b_{1B}$ particles }

%   \subsection{Equation of motion for $b_{1 B}$ particle}

  \begin{align}
%   derivative
  & \partial_{ \tau} G^{bb}_{BB 11}(\tau-\tau_{2})
  +\lambda_{B}^{-} G^{bb}_{BB 21}(\tau-\tau_{2}) 
  +\lambda_{B}^{z} G^{bb}_{BB 11}(\tau-\tau_{2}) 
  - (\mu -\frac{U}{1}) G^{bb}_{BB 11}(\tau-\tau_{2}) 
  \nonumber \\ &
  +\frac{v_{1}}{2}\Big[
  + F^{*db}_{AB 21}(\tau-\tau_{2}) \Big( 
  n^{Ff}_{AB \uparrow \uparrow }(\tau) 
  + n^{f}_{AB \downarrow \downarrow }(\tau) 
  \Big)
  + F^{*db}_{AB 11}(\tau-\tau_{2}) \Big(
  n^{Ff}_{AB \uparrow \downarrow }(\tau) 
  - n^{Ff}_{AB \downarrow \uparrow }(\tau)
    \Big)
    \nonumber \\ & \qquad
  + G^{bb}_{AB 21}(\tau-\tau_{2}) \Big(  
  - n^{Ff}_{AB \uparrow \downarrow }(\tau) 
  + n^{Ff}_{AB \downarrow \uparrow }(\tau)
    \Big)
  + G^{bb}_{AB 11}(\tau-\tau_{2}) \Big(
  n^{Ff}_{AB \uparrow \uparrow }(\tau) 
  + n^{f}_{AB \downarrow \downarrow }(\tau) 
  \Big)
  \Big]
  \nonumber \\ &
+\frac{1}{2} \int {\rm d}{\tau} \Big[
+ F^{*db}_{BB 21}(\tau-\tau_{2})  \Big(  
- G^{f}_{BB \uparrow \uparrow }(\tau'-\tau) {\mathfrak G}_{BB \uparrow \uparrow }(\tau -\tau') 
- G^{f}_{BB \downarrow \downarrow }(\tau'-\tau) {\mathfrak G}_{BB \downarrow \downarrow }(\tau-\tau')
\Big)
\nonumber \\ &\qquad 
+ F^{*db}_{BB 11}(\tau-\tau_{2})  \Big( 
F^{f}_{BB \downarrow \uparrow }(\tau'-\tau) {\mathfrak G}_{BB \downarrow \downarrow }(\tau-\tau')
- F^{f}_{BB \uparrow \downarrow }(\tau'-\tau) {\mathfrak G}_{BB \uparrow \uparrow }(\tau -\tau') 
\Big)
\nonumber \\ &\qquad 
+ G^{db}_{BB 21}(\tau-\tau_{2})  \Big(  
- F^{f}_{BB \downarrow \uparrow }(\tau'-\tau) {\mathfrak F}_{BB \uparrow \downarrow }(\tau-\tau')
- F^{f}_{BB \uparrow \downarrow }(\tau'-\tau) {\mathfrak F}_{BB \downarrow \uparrow }(\tau-\tau') 
\Big)
\nonumber \\ &\qquad 
+ G^{db}_{BB 11}(\tau-\tau_{2})  \Big( 
G^{f}_{BB \uparrow \uparrow }(\tau'-\tau) {\mathfrak F}_{BB \downarrow \uparrow }(\tau-\tau') 
- G^{f}_{BB \downarrow \downarrow }(\tau'-\tau) {\mathfrak F}_{BB \uparrow \downarrow }(\tau-\tau')
\Big)
\nonumber \\ &\qquad 
+ F^{*db}_{AB 21}(\tau-\tau_{2})  \Big(  
- G^{f}_{BA \uparrow \uparrow }(\tau'-\tau) {\mathfrak G}_{AB \uparrow \uparrow }(\tau-\tau')
- G^{f}_{BA \downarrow \downarrow }(\tau'-\tau) {\mathfrak G}_{AB \downarrow \downarrow }(\tau-\tau') 
\Big)
\nonumber \\ &\qquad 
+ F^{*db}_{AB 11}(\tau-\tau_{2})  \Big(  
- F^{f}_{BA \downarrow \uparrow }(\tau'-\tau) {\mathfrak G}_{AB \downarrow \downarrow }(\tau-\tau') 
+ F^{f}_{BA \uparrow \downarrow }(\tau'-\tau) {\mathfrak G}_{AB \uparrow \uparrow }(\tau-\tau')
\Big)
\nonumber \\ &\qquad 
+ G^{db}_{AB 21}(\tau-\tau_{2})  \Big(  
- F^{f}_{BA \downarrow \uparrow }(\tau'-\tau) {\mathfrak F}_{AB \uparrow \downarrow }(\tau-\tau')
- F^{f}_{BA \uparrow \downarrow }(\tau'-\tau) {\mathfrak F}_{AB \downarrow \uparrow }(\tau-\tau') 
\Big)
\nonumber \\ &\qquad 
+ G^{db}_{AB 11}(\tau-\tau_{2})  \Big(  
- G^{f}_{BA \uparrow \uparrow }(\tau'-\tau) {\mathfrak F}_{AB \downarrow \uparrow }(\tau-\tau') 
+ G^{f}_{BA \downarrow \downarrow }(\tau'-\tau) {\mathfrak F}_{AB \uparrow \downarrow }(\tau-\tau')
\Big)
\nonumber \\ &\qquad 
+ F^{*bb}_{BB 21}(\tau-\tau_{2})  \Big(  
- G^{f}_{BB \uparrow \uparrow }(\tau'-\tau) {\mathfrak F}_{BB \downarrow \uparrow }(\tau-\tau') 
+ G^{f}_{BB \downarrow \downarrow }(\tau'-\tau) {\mathfrak F}_{BB \uparrow \downarrow }(\tau-\tau')
\Big)
\nonumber \\ &\qquad 
+ F^{*bb}_{BB 11}(\tau-\tau_{2})  \Big(  
- F^{f}_{BB \downarrow \uparrow }(\tau'-\tau) {\mathfrak F}_{BB \uparrow \downarrow }(\tau-\tau')
- F^{f}_{BB \uparrow \downarrow }(\tau'-\tau) {\mathfrak F}_{BB \downarrow \uparrow }(\tau-\tau') 
\Big)
\nonumber \\ &\qquad 
+ G^{bb}_{BB 21}(\tau-\tau_{2})  \Big(  
- F^{f}_{BB \downarrow \uparrow }(\tau'-\tau) {\mathfrak G}_{BB \downarrow \downarrow }(\tau-\tau')
+ F^{f}_{BB \uparrow \downarrow }(\tau'-\tau) {\mathfrak G}_{BB \uparrow \uparrow }(\tau -\tau') 
\Big)
\nonumber \\ &\qquad 
+ G^{bb}_{BB 11}(\tau-\tau_{2})  \Big(  
- G^{f}_{BB \uparrow \uparrow }(\tau'-\tau) {\mathfrak G}_{BB \uparrow \uparrow }(\tau -\tau') 
- G^{f}_{BB \downarrow \downarrow }(\tau'-\tau) {\mathfrak G}_{BB \downarrow \downarrow }(\tau-\tau')
\Big)
\nonumber \\ &\qquad 
+ F^{*bb}_{AB 21}(\tau-\tau_{2})  \Big( 
G^{f}_{BA \uparrow \uparrow }(\tau'-\tau) {\mathfrak F}_{AB \downarrow \uparrow }(\tau-\tau') 
- G^{f}_{BA \downarrow \downarrow }(\tau'-\tau) {\mathfrak F}_{AB \uparrow \downarrow }(\tau-\tau')
\Big)
\nonumber \\ &\qquad 
+ F^{*bb}_{AB 11}(\tau-\tau_{2})  \Big(  
- F^{f}_{BA \downarrow \uparrow }(\tau'-\tau) {\mathfrak F}_{AB \uparrow \downarrow }(\tau-\tau')
- F^{f}_{BA \uparrow \downarrow }(\tau'-\tau) {\mathfrak F}_{AB \downarrow \uparrow }(\tau-\tau') 
\Big)
\nonumber \\ &\qquad 
+ G^{bb}_{AB 21}(\tau-\tau_{2})  \Big( 
F^{f}_{BA \downarrow \uparrow }(\tau'-\tau) {\mathfrak G}_{AB \downarrow \downarrow }(\tau-\tau') 
- F^{f}_{BA \uparrow \downarrow }(\tau'-\tau) {\mathfrak G}_{AB \uparrow \uparrow }(\tau-\tau')
\Big)
\nonumber \\ &\qquad 
+ G^{bb}_{AB 11}(\tau-\tau_{2})  \Big(  
- G^{f}_{BA \uparrow \uparrow }(\tau'-\tau) {\mathfrak G}_{AB \uparrow \uparrow }(\tau-\tau')
- G^{f}_{BA \downarrow \downarrow }(\tau'-\tau) {\mathfrak G}_{AB \downarrow \downarrow }(\tau-\tau') 
\Big)
\Big]=\delta(\tau-\tau_{2}).
  \end{align}

\subsection{ Equation of motion for $b_{2B}$ particles }

%   \subsection{Equation of motion for $b_{2 B}$ particle}

  \begin{align}
%   derivative
  & \partial_{ \tau} G^{bb}_{BB 22}(\tau-\tau_{2})
  + \lambda_{B}^{+} G^{bb}_{BB 12}(\tau-\tau_{2}) 
  - \lambda_{B}^{z} G^{bb}_{BB 22}(\tau-\tau_{2}) 
  - \big(\mu- \frac{U}{2} \big) G^{bb}_{BB 22}(\tau-\tau_{2}) 
  \nonumber \\ &
 + \frac{v_{1}}{2} \Big[
  + F^{*db}_{AB 22}(\tau-\tau_{2}) \Big(  
  - n^{*F}_{AB \uparrow \downarrow }(\tau) 
  + n^{*F}_{AB \downarrow \uparrow }(\tau) 
  \Big)
  + F^{*db}_{AB 12}(\tau-\tau_{2}) \Big( 
  -n^{*f}_{BA \uparrow \uparrow }(\tau) 
  + -n^{f}_{BA \downarrow \downarrow }(\tau) 
  \Big)
  \nonumber \\ & \qquad
  + G^{bb}_{AB 22}(\tau-\tau_{2}) \Big(  
  - -n^{*f}_{BA \uparrow \uparrow }(\tau) 
  - -n^{f}_{BA \downarrow \downarrow }(\tau) 
  \Big)
  + G^{bb}_{AB 12}(\tau-\tau_{2}) \Big(  
  - n^{*F}_{AB \uparrow \downarrow }(\tau) 
  + n^{*F}_{AB \downarrow \uparrow }(\tau) 
  \Big)
  \Big]
    \nonumber \\ &
%   derivative with respect to t', tau is summed over
  +\frac{1}{2} \int {\rm d}{\tau} \Big[
  + F^{*db}_{BB 22}(\tau-\tau_{2}) \Big( 
  G^{*f}_{BB \downarrow \uparrow }(\tau'-\tau) {\mathfrak G }_{BB \uparrow \uparrow }(\tau-\tau')
  - G^{*f}_{BB \uparrow \downarrow }(\tau'-\tau) {\mathfrak  G}_{BB \downarrow \downarrow }(\tau-\tau')
  \Big)
  \nonumber \\ & \qquad
  + F^{*db}_{BB 12}(\tau-\tau_{2}) \Big( 
  G^{*f}_{BB \uparrow \uparrow }(\tau'-\tau) {\mathfrak  G}_{BB \downarrow \downarrow }(\tau-\tau')
  + G^{*f}_{BB \downarrow \downarrow }(\tau'-\tau) {\mathfrak G }_{BB \uparrow \uparrow }(\tau-\tau')
  \Big)
  \nonumber \\ & \qquad
  + G^{db}_{BB 22}(\tau-\tau_{2}) \Big(  
  - G^{*f}_{BB \uparrow \uparrow }(\tau'-\tau) {\mathfrak F}_{BB \uparrow \downarrow }(\tau-\tau') 
  + G^{*f}_{BB \downarrow \downarrow }(\tau'-\tau) {\mathfrak F}_{BB \downarrow \uparrow }(\tau-\tau') 
  \Big)
  \nonumber \\ & \qquad
  + G^{db}_{BB 12}(\tau-\tau_{2}) \Big(  
  - G^{*f}_{BB \downarrow \uparrow }(\tau'-\tau) {\mathfrak F}_{BB \downarrow \uparrow }(\tau-\tau') 
  - G^{*f}_{BB \uparrow \downarrow }(\tau'-\tau) {\mathfrak F}_{BB \uparrow \downarrow }(\tau-\tau') 
  \Big)
  \nonumber \\ & \qquad
  + F^{*db}_{AB 22}(\tau-\tau_{2}) \Big( 
  F^{*f}_{BA \downarrow \uparrow }(\tau'-\tau) {\mathfrak G}_{AB \uparrow \uparrow }(\tau-\tau') 
  - F^{*f}_{BA \uparrow \downarrow }(\tau'-\tau) {\mathfrak G}_{AB \downarrow \downarrow }(\tau-\tau') 
  \Big)
  \nonumber \\ & \qquad
  + F^{*db}_{AB 12}(\tau-\tau_{2}) \Big(  
  - G^{*f}_{BA \uparrow \uparrow }(\tau'-\tau) {\mathfrak G}_{AB \downarrow \downarrow }(\tau-\tau') 
  - G^{*f}_{BA \downarrow \downarrow }(\tau'-\tau) {\mathfrak G}_{AB \uparrow \uparrow }(\tau-\tau') 
  \Big)
  \nonumber \\ & \qquad
  + G^{db}_{AB 22}(\tau-\tau_{2}) \Big(  
  - G^{*f}_{BA \uparrow \uparrow }(\tau'-\tau) {\mathfrak F}_{AB \uparrow \downarrow }(\tau-\tau') 
  + G^{*f}_{BA \downarrow \downarrow }(\tau'-\tau) {\mathfrak F}_{AB \downarrow \uparrow }(\tau-\tau') 
  \Big)
  \nonumber \\ & \qquad
  + G^{db}_{AB 12}(\tau-\tau_{2}) \Big( 
  F^{*f}_{BA \downarrow \uparrow }(\tau'-\tau) {\mathfrak F}_{AB \downarrow \uparrow }(\tau-\tau') 
  + F^{*f}_{BA \uparrow \downarrow }(\tau'-\tau) {\mathfrak F}_{AB \uparrow \downarrow }(\tau-\tau') 
  \Big)
  \nonumber \\ & \qquad
  + F^{*bb}_{BB 22}(\tau-\tau_{2}) \Big( 
  G^{*f}_{BB \downarrow \uparrow }(\tau'-\tau) {\mathfrak F}_{BB \downarrow \uparrow }(\tau-\tau')
  + G^{*f}_{BB \uparrow \downarrow }(\tau'-\tau) {\mathfrak F}_{BB \uparrow \downarrow }(\tau-\tau') 
  \Big)
  \nonumber \\ & \qquad
  + F^{*bb}_{BB 12}(\tau-\tau_{2}) \Big(  
  - G^{*f}_{BB \uparrow \uparrow }(\tau'-\tau) {\mathfrak F}_{BB \uparrow \downarrow }(\tau-\tau') 
  + G^{*f}_{BB \downarrow \downarrow }(\tau'-\tau) {\mathfrak F}_{BB \downarrow \uparrow }(\tau-\tau') 
  \Big)
  \nonumber \\ & \qquad
  + G^{bb}_{BB 22}(\tau-\tau_{2}) \Big(  
  - G^{*f}_{BB \uparrow \uparrow }(\tau'-\tau) {\mathfrak  G}_{BB \downarrow \downarrow }(\tau-\tau')
  - G^{*f}_{BB \downarrow \downarrow }(\tau'-\tau) {\mathfrak G }_{BB \uparrow \uparrow }(\tau-\tau')
  \Big)
  \nonumber \\ & \qquad
  + G^{bb}_{BB 12}(\tau-\tau_{2}) \Big( 
  G^{*f}_{BB \downarrow \uparrow }(\tau'-\tau) {\mathfrak G }_{BB \uparrow \uparrow }(\tau-\tau')
  - G^{*f}_{BB \uparrow \downarrow }(\tau'-\tau) {\mathfrak  G}_{BB \downarrow \downarrow }(\tau-\tau')
 \Big)
  \nonumber \\ & \qquad
  + F^{*bb}_{AB 22}(\tau-\tau_{2}) \Big(  
  - F^{*f}_{BA \downarrow \uparrow }(\tau'-\tau) {\mathfrak F}_{AB \downarrow \uparrow }(\tau-\tau') 
  - F^{*f}_{BA \uparrow \downarrow }(\tau'-\tau) {\mathfrak F}_{AB \uparrow \downarrow }(\tau-\tau') 
  \Big)
  \nonumber \\ & \qquad
  + F^{*bb}_{AB 12}(\tau-\tau_{2}) \Big(  
  - G^{*f}_{BA \uparrow \uparrow }(\tau'-\tau) {\mathfrak F}_{AB \uparrow \downarrow }(\tau-\tau') 
  + G^{*f}_{BA \downarrow \downarrow }(\tau'-\tau) {\mathfrak F}_{AB \downarrow \uparrow }(\tau-\tau') 
  \Big)
  \nonumber \\ & \qquad
  + G^{bb}_{AB}(\tau-\tau_{2}) \Big( 
  G^{*f}_{BA \uparrow \uparrow }(\tau'-\tau) {\mathfrak G}_{AB \downarrow \downarrow }(\tau-\tau') 
  + G^{*f}_{BA \downarrow \downarrow }(\tau'-\tau) {\mathfrak G}_{AB \uparrow \uparrow }(\tau-\tau') 
  \Big)
  \nonumber \\ & \qquad
  + G^{bb}_{AB 12}(\tau-\tau_{2}) \Big( 
  F^{*f}_{BA \downarrow \uparrow }(\tau'-\tau) {\mathfrak G}_{AB \uparrow \uparrow }(\tau-\tau') 
  - F^{*f}_{BA \uparrow \downarrow }(\tau'-\tau) {\mathfrak G}_{AB \downarrow \downarrow }(\tau-\tau') 
  \Big)
  \Big]=\delta(\tau-\tau_{2}).
  \end{align}

\subsection{Equation of motion for $d_{A 1}$ particles  }

%   \subsection{Equation of motion for $d_{1 A}$ particle}

 \begin{align}
%   derivative
  & \partial_{ \tau} G^{dd}_{AA 11}(\tau-\tau_{2})
  +\lambda_{A}^{-} G^{dd}_{AA 21}(\tau-\tau_{2})
  +\lambda_{A}^{z} G^{dd}_{AA 11}(\tau-\tau_{2}) 
  -\frac{U}{2} n^{Ff}_{AA \uparrow \downarrow }(\tau\Big) 
  G^{dd}_{AA 21}(\tau-\tau_{2})
  -\frac{U}{2} G^{dd}_{AA 11}(\tau-\tau_{2}) 
  \nonumber \\ &
  \frac{v_{1}}{2} \Big[
  + G^{dd}_{BA 21}(\tau-\tau_{2}) \Big( 
  n^{Ff}_{AB \uparrow \downarrow }(\tau) 
  - n^{Ff}_{AB \downarrow \uparrow }(\tau) 
  \Big)
  + G^{dd}_{BA 11}(\tau-\tau_{2}) \Big(  
  - -n^{f}_{BA \uparrow \uparrow }(\tau) 
  - -n^{f}_{BA \downarrow \downarrow}(\tau) 
  \Big)
  \nonumber \\ &
  + F^{*bd}_{BA 21}(\tau-\tau_{2}) \Big( 
  -n^{f}_{BA \uparrow \uparrow }(\tau) 
  + -n^{f}_{BA \downarrow \downarrow}(\tau) 
  \Big)
  + F^{*bd}_{BA 11}(\tau-\tau_{2}) \Big( 
  n^{Ff}_{AB \uparrow \downarrow }(\tau) 
  - n^{Ff}_{AB \downarrow \uparrow }(\tau)
    \Big)
  \Big]
  \nonumber \\ &
%   derivative with respect to t, tau' summed over
\frac{1}{2} \int {\rm d}{\tau'} \Big[
  + F^{*dd}_{BA 21}(\tau'-\tau_{2}) \Big(  
  - G^{*f}_{BA \uparrow \uparrow }(\tau'-\tau) {\mathfrak F}_{AB \downarrow \uparrow }(\tau-\tau')
  + G^{*f}_{BA \downarrow \downarrow}(\tau'-\tau) {\mathfrak F}_{AB \uparrow \downarrow }(\tau-\tau')
  \Big)
\nonumber \\ & \qquad
+ F^{*dd}_{BA 11}(\tau'-\tau_{2}) \Big(  
- F^{f}_{BA \downarrow \uparrow }(\tau'-\tau) {\mathfrak F}_{AB \downarrow \uparrow }(\tau-\tau')
- F^{f}_{BA \uparrow \downarrow }(\tau'-\tau) {\mathfrak F}_{AB \uparrow \downarrow }(\tau-\tau')
\Big)
\nonumber \\ & \qquad
+ G^{dd}_{BA 21}(\tau-\tau'_{2}) \Big(  
- F^{f}_{BA \downarrow \uparrow }(\tau'-\tau) {\mathfrak G}_{AB \downarrow \downarrow }(\tau-\tau')
+ F^{f}_{BA \uparrow \downarrow }(\tau'-\tau) {\mathfrak G }_{AB \uparrow \uparrow }(\tau-\tau')
\Big)
\nonumber \\ & \qquad
+ G^{dd}_{BA 11}(\tau'-\tau_{2}) \Big( 
G^{*f}_{BA \uparrow \uparrow }(\tau'-\tau) {\mathfrak G}_{AB \downarrow \downarrow }(\tau-\tau')
+ G^{*f}_{BA \downarrow \downarrow}(\tau'-\tau) {\mathfrak G }_{AB \uparrow \uparrow }(\tau-\tau')
\Big)
\nonumber \\ & \qquad
+ F^{*dd}_{AA 21}(\tau'-\tau_{2}) \Big(  
- G^{*f}_{AA \uparrow \uparrow }(\tau'-\tau) {\mathfrak F}_{AA \downarrow \uparrow }(\tau-\tau')
+ G^{*f}_{AA \downarrow \downarrow }(\tau'-\tau) {\mathfrak F}_{AA \uparrow \downarrow }(\tau-\tau')
\Big)
\nonumber \\ & \qquad
+ F^{*dd}_{AA 11}(\tau-\tau_{2}) \Big( 
F^{f}_{AA \downarrow \uparrow }(\tau'-\tau) {\mathfrak F}_{AA \downarrow \uparrow }(\tau-\tau')
+ F^{f}_{AA \uparrow \downarrow }(\tau'-\tau) {\mathfrak F}_{AA \uparrow \downarrow }(\tau-\tau')
\Big)
\nonumber \\ & \qquad
+ G^{dd}_{AA 21}(\tau'-\tau_{2}) \Big(  
- F^{f}_{AA \downarrow \uparrow }(\tau'-\tau) {\mathfrak G}_{AA \downarrow \downarrow }(\tau-\tau')
+ F^{f}_{AA \uparrow \downarrow }(\tau'-\tau) {\mathfrak G}_{AA \uparrow \uparrow }(\tau-\tau')
\Big)
\nonumber \\ & \qquad
+ G^{dd}_{AA 11}(\tau'-\tau_{2}) \Big(  
- G^{*f}_{AA \uparrow \uparrow }(\tau'-\tau) {\mathfrak G}_{AA \downarrow \downarrow }(\tau-\tau')
- G^{*f}_{AA \downarrow \downarrow}(\tau'-\tau) {\mathfrak G}_{AA \uparrow \uparrow }(\tau-\tau')
\Big)
\nonumber \\ & \qquad
+ F^{*bd}_{BA 21}(\tau'-\tau_{2}) \Big(  
- G^{*f}_{BA \uparrow \uparrow }(\tau'-\tau) {\mathfrak G}_{AB \downarrow \downarrow }(\tau-\tau')
- G^{*f}_{BA \downarrow \downarrow}(\tau'-\tau) {\mathfrak G }_{AB \uparrow \uparrow }(\tau-\tau')
\Big)
\nonumber \\ & \qquad
+ F^{*bd}_{BA 11}(\tau'-\tau_{2}) \Big(  
- F^{f}_{BA \downarrow \uparrow }(\tau'-\tau) {\mathfrak G}_{AB \downarrow \downarrow }(\tau-\tau')
+ F^{f}_{BA \uparrow \downarrow }(\tau'-\tau) {\mathfrak G }_{AB \uparrow \uparrow }(\tau-\tau')
\Big)
\nonumber \\ & \qquad
+ G^{bd}_{BA 21}(\tau'-\tau_{2}) \Big( 
F^{f}_{BA \downarrow \uparrow }(\tau'-\tau) {\mathfrak F}_{AB \downarrow \uparrow }(\tau-\tau')
+ F^{f}_{BA \uparrow \downarrow }(\tau'-\tau) {\mathfrak F}_{AB \uparrow \downarrow }(\tau-\tau')
\Big)
\nonumber \\ & \qquad
+ G^{bd}_{BA 11}(\tau'-\tau_{2}) \Big(  
- G^{*f}_{BA \uparrow \uparrow }(\tau'-\tau) {\mathfrak F}_{AB \downarrow \uparrow }(\tau-\tau')
+ G^{*f}_{BA \downarrow \downarrow}(\tau'-\tau) {\mathfrak F}_{AB \uparrow \downarrow }(\tau-\tau')
\Big)
\nonumber \\ & \qquad
+ F^{*bd}_{AA 21}(\tau'-\tau_{2}) \Big( 
G^{*f}_{AA \uparrow \uparrow }(\tau'-\tau) {\mathfrak G}_{AA \downarrow \downarrow }(\tau-\tau')
+ G^{*f}_{AA \downarrow \downarrow }(\tau'-\tau) {\mathfrak G}_{AA \uparrow \uparrow }(\tau-\tau')
\Big)
\nonumber \\ & \qquad
+ F^{*bd}_{AA 11}(\tau'-\tau_{2}) \Big(  
- F^{f}_{AA \downarrow \uparrow }(\tau'-\tau) {\mathfrak G}_{AA \downarrow \downarrow }(\tau-\tau')
+ F^{f}_{AA \uparrow \downarrow }(\tau'-\tau) {\mathfrak G}_{AA \uparrow \uparrow }(\tau-\tau')
\Big)
\nonumber \\ & \qquad
+ G^{bd}_{AA 21}(\tau'-\tau_{2}) \Big(  
- F^{f}_{AA \downarrow \uparrow }(\tau'-\tau) {\mathfrak F}_{AA \downarrow \uparrow }(\tau-\tau')
- F^{f}_{AA \uparrow \downarrow }(\tau'-\tau) {\mathfrak F}_{AA \uparrow \downarrow }(\tau-\tau')
\Big)
\nonumber \\ & \qquad
+ G^{bd}_{AA 11}(\tau'-\tau_{2}) \Big(  
- G^{*f}_{AA \uparrow \uparrow }(\tau'-\tau) {\mathfrak F}_{AA \downarrow \uparrow }(\tau-\tau')
+ G^{*f}_{AA \downarrow \downarrow }(\tau'-\tau) {\mathfrak F}_{AA \uparrow \downarrow }(\tau-\tau')
\Big)
    \Big]
=\delta(\tau-\tau_{2}).
  \end{align}

\subsection{Equation of motion for $d_{2A}$ particles}

%   \subsection{Equation of motion for $d_{2 A}$ particle}

 \begin{align}
%   derivative
  & \partial_{ \tau}  G^{dd}_{AA 22}(\tau-\tau_{2}) 
   +\lambda_{A}^{+} G^{dd}_{AA 12}(\tau-\tau_{2})
   -\lambda_{A}^{z} G^{dd}_{AA 22}(\tau-\tau_{2}) 
  -\frac{U}{2} n^{*F}_{AA \downarrow \uparrow }(\tau) G^{dd}_{AA 12}(\tau-\tau_{2})
  \nonumber \\ &
 - \frac{U}{2}\Big[
  + G^{dd}_{AA 22}(\tau-\tau_{2}) \Big(  
  + n^{F}_{AA \uparrow \downarrow }(\tau) n^{*F}_{AA \downarrow \uparrow }(\tau)
  + n^{f}_{AA \uparrow \uparrow }(\tau) n^{f}_{AA \downarrow \downarrow }(\tau)
  + n^{f}_{AA \downarrow \downarrow }(\tau) n^{f}_{AA \uparrow \uparrow }(\tau)
  + n^{*F}_{AA \downarrow \uparrow }(\tau) n^{Ff}_{AA \uparrow \downarrow }(\tau)
  \Big)
  \Big]
  \nonumber \\ &
 + \frac{v_{1}}{2} \Big[
  + G^{dd}_{BA 22}(\tau-\tau_{2}) \Big( 
  n^{f}_{AB \uparrow \uparrow }(\tau) 
  + n^{f}_{AB \downarrow \downarrow }(\tau) 
  \Big)
  + G^{dd}_{BA 12}(\tau-\tau_{2}) \Big( 
  n^{*F}_{AB \uparrow \downarrow }(\tau)
  - n^{*F}_{AB \downarrow \uparrow }(\tau) 
  \Big)
 \nonumber \\ &
  + F^{*bd}_{BA 22}(\tau-\tau_{2}) \Big(  
  - n^{*F}_{AB \uparrow \downarrow }(\tau)
  + n^{*F}_{AB \downarrow \uparrow }(\tau) 
  \Big)
  + F^{*bd}_{BA 12}(\tau-\tau_{2}) \Big( 
  n^{f}_{AB \uparrow \uparrow }(\tau) 
  + n^{f}_{AB \downarrow \downarrow }(\tau) 
  \Big)
  \Big]
  \nonumber \\ &
%   derivative with t, tau' should be summed over
+\frac{1}{2} \int {\rm d}{\tau'} \Big[
  + F^{*dd}_{BA 22}(\tau'-\tau_{2}) \Big(  
  - G^{*f}_{BA \downarrow \uparrow }(\tau'-\tau) {\mathfrak F}_{AB \uparrow \downarrow }(\tau-\tau')
  - F^{*f}_{BA \uparrow \downarrow }(\tau'-\tau) {\mathfrak F}_{AB \downarrow \uparrow }(\tau-\tau')
    \Big)
  \nonumber \\ & \qquad
  + F^{*dd}_{BA 12}(\tau'-\tau_{2}) \Big( 
  G^{f}_{BA \uparrow \uparrow }(\tau'-\tau) {\mathfrak F}_{AB \uparrow \downarrow }(\tau-\tau') 
  - G^{f}_{BA \downarrow \downarrow }(\tau'-\tau) {\mathfrak F}_{AB \downarrow \uparrow }(\tau-\tau') 
  \Big)
  \nonumber \\ & \qquad
  + G^{dd}_{BA}(\tau'-\tau_{2}) \Big(  
  - G^{f}_{BA \uparrow \uparrow }(\tau'-\tau) {\mathfrak G}_{AB \uparrow \uparrow }(\tau-\tau') 
  - G^{f}_{BA \downarrow \downarrow }(\tau'-\tau) {\mathfrak G}_{AB \downarrow \downarrow }(\tau-\tau')
  \Big)
  \nonumber \\ & \qquad
  + G^{dd}_{BA 12}(\tau'-\tau_{2}) \Big(  
  - G^{*f}_{BA \downarrow \uparrow }(\tau'-\tau) {\mathfrak G}_{AB \uparrow \uparrow }(\tau-\tau') 
  + F^{*f}_{BA \uparrow \downarrow }(\tau'-\tau) {\mathfrak G}_{AB \downarrow \downarrow }(\tau-\tau')
  \Big)
  \nonumber \\ & \qquad
  + F^{*dd}_{AA 22}(\tau'-\tau_{2}) \Big(  
  - F^{*f}_{AA \downarrow \uparrow }(\tau'-\tau) {\mathfrak F}_{AA \uparrow \downarrow }(\tau-\tau')
  - F^{*f}_{AA \uparrow \downarrow }(\tau'-\tau) {\mathfrak F}_{AA \downarrow \uparrow }(\tau-\tau') 
  \Big)
  \nonumber \\ & \qquad
  + F^{*dd}_{AA 12}(\tau'-\tau_{2}) \Big(  
  - G^{f}_{AA \uparrow \uparrow }(\tau'-\tau) {\mathfrak F}_{AA \uparrow \downarrow }(\tau-\tau') 
  + G^{f}_{AA \downarrow \downarrow }(\tau'-\tau) {\mathfrak F}_{AA \downarrow \uparrow }(\tau-\tau') 
  \Big)
  \nonumber \\ & \qquad
  + G^{dd}_{AA 22}(\tau'-\tau_{2}) \Big(  
  - G^{f}_{AA \uparrow \uparrow }(\tau'-\tau) {\mathfrak G}_{AA \uparrow \uparrow }(\tau-\tau')
  - G^{f}_{AA \downarrow \downarrow }(\tau'-\tau) {\mathfrak G}_{AA \downarrow \downarrow }(\tau-\tau') 
  \Big)
  \nonumber \\ & \qquad
  + G^{dd}_{AA 12}(\tau'-\tau_{2}) \Big( 
  F^{*f}_{AA \downarrow \uparrow }(\tau'-\tau) {\mathfrak G}_{AA \uparrow \uparrow }(\tau-\tau')
  - F^{*f}_{AA \uparrow \downarrow }(\tau'-\tau) {\mathfrak G}_{AA \downarrow \downarrow }(\tau-\tau') 
  \Big)
  \nonumber \\ & \qquad
  + F^{*bd}_{BA 22}(\tau'-\tau_{2}) \Big( 
  G^{*f}_{BA \downarrow \uparrow }(\tau'-\tau) {\mathfrak G}_{AB \uparrow \uparrow }(\tau-\tau') 
  - F^{*f}_{BA \uparrow \downarrow }(\tau'-\tau) {\mathfrak G}_{AB \downarrow \downarrow }(\tau-\tau')
  \Big)
  \nonumber \\ & \qquad
  + F^{*bd}_{BA 12}(\tau'-\tau_{2}) \Big(  
  - G^{f}_{BA \uparrow \uparrow }(\tau'-\tau) {\mathfrak G}_{AB \uparrow \uparrow }(\tau-\tau') 
  - G^{f}_{BA \downarrow \downarrow }(\tau'-\tau) {\mathfrak G}_{AB \downarrow \downarrow }(\tau-\tau')
  \Big)
  \nonumber \\ & \qquad
  + G^{bd}_{BA 22}(\tau'-\tau_{2}) \Big(  
  - G^{f}_{BA \uparrow \uparrow }(\tau'-\tau) {\mathfrak F}_{AB \uparrow \downarrow }(\tau-\tau') 
  + G^{f}_{BA \downarrow \downarrow }(\tau'-\tau) {\mathfrak F}_{AB \downarrow \uparrow }(\tau-\tau') 
  \Big)
  \nonumber \\ & \qquad
  + G^{bd}_{BA 12}(\tau'-\tau_{2}) \Big(  
  - G^{*f}_{BA \downarrow \uparrow }(\tau'-\tau) {\mathfrak F}_{AB \uparrow \downarrow }(\tau-\tau') 
  - F^{*f}_{BA \uparrow \downarrow }(\tau'-\tau) {\mathfrak F}_{AB \downarrow \uparrow }(\tau-\tau') 
  \Big)
  \nonumber \\ & \qquad
  + F^{*bd}_{AA 22}(\tau'-\tau_{2}) \Big(  
  - F^{*f}_{AA \downarrow \uparrow }(\tau'-\tau) {\mathfrak G}_{AA \uparrow \uparrow }(\tau-\tau')
  + F^{*f}_{AA \uparrow \downarrow }(\tau'-\tau) {\mathfrak G}_{AA \downarrow \downarrow }(\tau-\tau') 
  \Big)
  \nonumber \\ & \qquad
  + F^{*bd}_{AA 12}(\tau'-\tau_{2}) \Big(  
  - G^{f}_{AA \uparrow \uparrow }(\tau'-\tau) {\mathfrak G}_{AA \uparrow \uparrow }(\tau-\tau')
  - G^{f}_{AA \downarrow \downarrow }(\tau'-\tau) {\mathfrak G}_{AA \downarrow \downarrow }(\tau-\tau') 
  \Big)
  \nonumber \\ & \qquad
  + G^{bd}_{AA 22}(\tau'-\tau_{2}) \Big( 
  G^{f}_{AA \uparrow \uparrow }(\tau'-\tau) {\mathfrak F}_{AA \uparrow \downarrow }(\tau-\tau') 
  - G^{f}_{AA \downarrow \downarrow }(\tau'-\tau) {\mathfrak F}_{AA \downarrow \uparrow }(\tau-\tau') 
  \Big)
  \nonumber \\ & \qquad
  + G^{bd}_{AA 12}(\tau'-\tau_{2}) \Big(  
  - F^{*f}_{AA \downarrow \uparrow }(\tau'-\tau) {\mathfrak F}_{AA \uparrow \downarrow }(\tau-\tau') 
  - F^{*f}_{AA \uparrow \downarrow }(\tau'-\tau) {\mathfrak F}_{AA \downarrow \uparrow }(\tau-\tau') 
  \Big)
  \Big]
=\delta(\tau-\tau_{2}).
 \end{align}

\subsection{Equation of motion for $d_{1B}$ particles}

%   \subsection{Equation of motion for $d_{1 B}$ particle}

  \begin{align}
%   derivative
  & \partial_{ \tau} G^{dd}_{BB 11}(\tau-\tau_{2})
  + \lambda^{-}_{B} G^{dd}_{BB 21}(\tau-\tau_{2}) 
  + \lambda^{z}_{B} G^{dd}_{BB 11}(\tau-\tau_{2}) 
  +\frac{U}{2} n^{Ff}_{BB \uparrow \downarrow }(\tau) G^{dd}_{BB 21}(\tau-\tau_{2}) 
  -\frac{U}{2} G^{dd}_{BB 11}(\tau-\tau_{2}) 
  \nonumber \\ &
 + \frac{v_{1}}{2} \Big[
  + G^{dd}_{AB 21}(\tau-\tau_{2}) \Big(  
  - n^{F}_{BA \uparrow \downarrow }(\tau)
  + n^{Ff}_{BA \downarrow \uparrow }(\tau)
    \Big)
  + G^{dd}_{AB 11 }(\tau-\tau_{2}) \Big( 
  - -n^{f}_{AB \uparrow \uparrow }(\tau) 
  - -n^{f}_{AB \downarrow \downarrow }(\tau) 
  \Big)
  \nonumber \\ & \qquad 
  + F^{*bd}_{AB 21}(\tau-\tau_{2}) \Big( 
  -n^{f}_{AB \uparrow \uparrow }(\tau) 
  + -n^{f}_{AB \downarrow \downarrow }(\tau) 
  \Big)
  + F^{*bd}_{AB 11 }(\tau-\tau_{2}) \Big(  
  - n^{F}_{BA \uparrow \downarrow }(\tau)
  + n^{Ff}_{BA \downarrow \uparrow }(\tau) 
  \Big)
  \Big]
  \nonumber \\ & 
%   derivative with respect to t, \tau' is summbed over
+\frac{1}{2} \int {\rm d}{\tau'} \Big[
+ F^{*dd}_{BB 21}(\tau'-\tau_{2}) \Big( 
G^{*f}_{BB \uparrow \uparrow }(\tau'-\tau) {\mathfrak F}_{BB \downarrow \uparrow }(\tau-\tau') 
- G^{*f}_{BB \downarrow \downarrow }(\tau'-\tau) {\mathfrak F}_{BB \uparrow \downarrow }(\tau-\tau') 
\Big)
  \nonumber \\ & \qquad
+ F^{*dd}_{BB 11}(\tau'-\tau_{2}) \Big( 
F^{f}_{BB \downarrow \uparrow }(\tau'-\tau) {\mathfrak F}_{BB \downarrow \uparrow }(\tau-\tau') 
+ F^{f}_{BB \uparrow \downarrow }(\tau'-\tau) {\mathfrak F}_{BB \uparrow \downarrow }(\tau-\tau') 
\Big)
  \nonumber \\ & \qquad
+ G^{dd}_{BB 21}(\tau'-\tau_{2}) \Big( 
F^{f}_{BB \downarrow \uparrow }(\tau'-\tau) {\mathfrak G}_{BB \downarrow \downarrow }(\tau-\tau') 
- F^{f}_{BB \uparrow \downarrow }(\tau'-\tau) {\mathfrak G}_{BB \uparrow \uparrow }(\tau-\tau')
\Big)
  \nonumber \\ & \qquad
+ G^{dd}_{BB 11}(\tau'-\tau_{2}) \Big(  
- G^{*f}_{BB \uparrow \uparrow }(\tau'-\tau) {\mathfrak G}_{BB \downarrow \downarrow }(\tau-\tau') 
- G^{*f}_{BB \downarrow \downarrow }(\tau'-\tau) {\mathfrak G}_{BB \uparrow \uparrow }(\tau-\tau')
\Big)
  \nonumber \\ & \qquad
+ F^{*dd}_{AB 21}(\tau'-\tau_{2}) \Big( 
G^{*f}_{AB \uparrow \uparrow }(\tau'-\tau) {\mathfrak F}_{BA \downarrow \downarrow }(\tau-\tau') 
- G^{f}_{AB \downarrow \downarrow }(\tau'-\tau) {\mathfrak F}_{BA \uparrow \downarrow }(\tau-\tau') 
\Big)
  \nonumber \\ & \qquad
+ F^{*dd}_{AB 11 }(\tau'-\tau_{2}) \Big(  
- F^{f}_{AB \downarrow \uparrow }(\tau'-\tau) {\mathfrak F}_{BA \downarrow \downarrow }(\tau-\tau') 
- F^{f}_{AB \uparrow \downarrow }(\tau'-\tau) {\mathfrak F}_{BA \uparrow \downarrow }(\tau-\tau') 
\Big)
  \nonumber \\ & \qquad
+ G^{dd}_{AB 21}(\tau'-\tau_{2}) \Big( 
F^{f}_{AB \downarrow \uparrow }(\tau'-\tau) {\mathfrak G}_{BA \downarrow \downarrow }(\tau-\tau') 
- F^{f}_{AB \uparrow \downarrow }(\tau'-\tau) {\mathfrak G}_{BA \uparrow \uparrow }(\tau-\tau')
\Big)
  \nonumber \\ & \qquad
+ G^{dd}_{AB 11 }(\tau'-\tau_{2}) \Big( 
G^{*f}_{AB \uparrow \uparrow }(\tau'-\tau) {\mathfrak G}_{BA \downarrow \downarrow }(\tau-\tau') 
+ G^{f}_{AB \downarrow \downarrow }(\tau'-\tau) {\mathfrak G}_{BA \uparrow \uparrow }(\tau-\tau')
\Big)
  \nonumber \\ & \qquad
+ F^{*bd}_{BB 21}(\tau'-\tau_{2}) \Big( 
G^{*f}_{BB \uparrow \uparrow }(\tau'-\tau) {\mathfrak G}_{BB \downarrow \downarrow }(\tau-\tau') 
+ G^{*f}_{BB \downarrow \downarrow }(\tau'-\tau) {\mathfrak G}_{BB \uparrow \uparrow }(\tau-\tau')
\Big)
  \nonumber \\ & \qquad
+ F^{*bd}_{11 BA}(\tau'-\tau_{2}) \Big( 
F^{f}_{BB \downarrow \uparrow }(\tau'-\tau) {\mathfrak G}_{BB \downarrow \downarrow }(\tau-\tau') 
- F^{f}_{BB \uparrow \downarrow }(\tau'-\tau) {\mathfrak G}_{BB \uparrow \uparrow }(\tau-\tau')
\Big)
  \nonumber \\ & \qquad
+ G^{bd}_{21 BA}(\tau'-\tau_{2}) \Big(  
- F^{f}_{BB \downarrow \uparrow }(\tau'-\tau) {\mathfrak F}_{BB \downarrow \uparrow }(\tau-\tau') 
- F^{f}_{BB \uparrow \downarrow }(\tau'-\tau) {\mathfrak F}_{BB \uparrow \downarrow }(\tau-\tau') 
\Big)
  \nonumber \\ & \qquad
+ G^{bd}_{BB 11}(\tau'-\tau_{2}) \Big( 
G^{*f}_{BB \uparrow \uparrow }(\tau'-\tau) {\mathfrak F}_{BB \downarrow \uparrow }(\tau-\tau') 
- G^{*f}_{BB \downarrow \downarrow }(\tau'-\tau) {\mathfrak F}_{BB \uparrow \downarrow }(\tau-\tau') 
\Big)
  \nonumber \\ & \qquad
+ F^{*bd}_{AB 21}(\tau'-\tau_{2}) \Big(  
- G^{*f}_{AB \uparrow \uparrow }(\tau'-\tau) {\mathfrak G}_{BA \downarrow \downarrow }(\tau-\tau') 
- G^{f}_{AB \downarrow \downarrow }(\tau'-\tau) {\mathfrak G}_{BA \uparrow \uparrow }(\tau-\tau')
\Big)
  \nonumber \\ & \qquad
+ F^{*bd}_{AB 11 }(\tau'-\tau_{2}) \Big( 
F^{f}_{AB \downarrow \uparrow }(\tau'-\tau) {\mathfrak G}_{BA \downarrow \downarrow }(\tau-\tau') 
- F^{f}_{AB \uparrow \downarrow }(\tau'-\tau) {\mathfrak G}_{BA \uparrow \uparrow }(\tau-\tau')
\Big)
  \nonumber \\ & \qquad
+ G^{bd}_{AB 21}(\tau'-\tau_{2}) \Big( 
F^{f}_{AB \downarrow \uparrow }(\tau'-\tau) {\mathfrak F}_{BA \downarrow \downarrow }(\tau-\tau') 
+ F^{f}_{AB \uparrow \downarrow }(\tau'-\tau) {\mathfrak F}_{BA \uparrow \downarrow }(\tau-\tau') 
\Big)
  \nonumber \\ & \qquad
+ G^{bd}_{AB 11 }(\tau'-\tau_{2}) \Big( 
G^{*f}_{AB \uparrow \uparrow }(\tau'-\tau) {\mathfrak F}_{BA \downarrow \downarrow }(\tau-\tau') 
- G^{f}_{AB \downarrow \downarrow }(\tau'-\tau) {\mathfrak F}_{BA \uparrow \downarrow }(\tau-\tau') 
\Big)
  \Big]
=\delta(\tau-\tau_{2}).
  \end{align}

\subsection{Equation of motion for $d_{2B}$ particles}

%   \subsection{Equation of motion for $d_{2 B}$ particle}

  \begin{align}
%   derivative
  & \partial_{ \tau} G^{dd}_{BB 22}(\tau-\tau_{2})
  +\lambda_{B}^{+} G^{dd}_{BB 12}(\tau-\tau_{2}) 
  -\lambda_{B}^{z} G^{dd}_{BB 22}(\tau-\tau_{2}) 
  +\frac{U}{2} n^{*F}_{BB \downarrow \uparrow }(\tau) G^{dd}_{BB 12}(\tau-\tau_{2}) 
  \nonumber \\ &
  -\frac{U}{2} \Big[
  + G^{dd}_{BB 22}(\tau-\tau_{2}) \Big(  
  + n^{Ff}_{BB \uparrow \downarrow }(\tau)n^{*F}_{BB \downarrow \uparrow }(\tau)
  + n^{f}_{BB \uparrow \uparrow }(\tau)n_{BB \downarrow \downarrow }(\tau) 
  + n^{f}_{BB \downarrow \downarrow }(\tau)n_{BB \uparrow \uparrow }(\tau)
  + n^{*F}_{BB \downarrow \uparrow }(\tau)n^{F}_{BB \uparrow \downarrow }(\tau)
  \Big)
  \Big]
  \nonumber \\ &
%   derivative with respect to t, integration over tau'
+\frac{1}{2} \int {\rm d}{\tau'} \Big[
+ F^{*dd}_{BB 22}(\tau'-\tau_{2}) \Big(  
- F^{*f}_{BB \downarrow \uparrow }(\tau'-\tau) {\mathfrak F}_{BB \uparrow \downarrow }(\tau-\tau')
- F^{*f}_{BB \uparrow \downarrow }(\tau'-\tau) {\mathfrak F}_{BB \downarrow \uparrow }(\tau-\tau')
\Big)
  \nonumber \\ & \qquad
+ F^{*dd}_{BB 12}(\tau'-\tau_{2}) \Big( 
G^{f}_{BB \uparrow \uparrow }(\tau'-\tau) {\mathfrak F}_{BB \uparrow \downarrow }(\tau-\tau') 
- G^{f}_{BB \downarrow \downarrow }(\tau'-\tau) {\mathfrak F}_{BB \downarrow \uparrow }(\tau-\tau')
\Big)
  \nonumber \\ & \qquad
+ G^{dd}_{BB 22}(\tau'-\tau_{2}) \Big(  
- G^{f}_{BB \uparrow \uparrow }(\tau'-\tau) {\mathfrak G}_{BB \uparrow \uparrow }(\tau-\tau') 
- G^{f}_{BB \downarrow \downarrow }(\tau'-\tau) {\mathfrak G}_{BB \downarrow \downarrow }(\tau-\tau') 
\Big)
  \nonumber \\ & \qquad
+ G^{dd }_{BB 12}(\tau'-\tau_{2}) \Big(  
- F^{*f}_{BB \downarrow \uparrow }(\tau'-\tau) {\mathfrak G}_{BB \uparrow \uparrow }(\tau-\tau') 
+ F^{*f}_{BB \uparrow \downarrow }(\tau'-\tau) {\mathfrak G}_{BB \downarrow \downarrow }(\tau-\tau') 
\Big)
  \nonumber \\ & \qquad
+ F^{*dd}_{AB 22}(\tau'-\tau_{2}) \Big(  
- F^{*f}_{AB \downarrow \uparrow }(\tau'-\tau) {\mathfrak F}_{BA \uparrow \downarrow }(\tau-\tau')
- F^{*f}_{AB \uparrow \downarrow }(\tau'-\tau) {\mathfrak F}_{BA \downarrow \uparrow }(\tau-\tau') 
\Big)
  \nonumber \\ & \qquad
+ F^{*dd}_{AB 12}(\tau'-\tau_{2}) \Big(  
- G^{f}_{AB \uparrow \uparrow }(\tau'-\tau){\mathfrak F}_{BA \uparrow \downarrow }(\tau-\tau') 
+ G^{f}_{AB \downarrow \downarrow }(\tau'-\tau) {\mathfrak F}_{BA \downarrow \uparrow }(\tau-\tau') 
\Big)
  \nonumber \\ & \qquad
+ G^{dd}_{AB 22}(\tau'-\tau_{2}) \Big(  
- G^{f}_{AB \uparrow \uparrow }(\tau'-\tau){\mathfrak G}_{BA \uparrow \uparrow }(\tau-\tau') 
- G^{f}_{AB \downarrow \downarrow }(\tau'-\tau) {\mathfrak G}_{BA \downarrow \downarrow }(\tau-\tau')
\Big)
  \nonumber \\ & \qquad
+ G^{dd}_{AB 12 }(\tau'-\tau_{2}) \Big( 
F^{*f}_{AB \downarrow \uparrow }(\tau'-\tau) {\mathfrak G}_{BA \uparrow \uparrow }(\tau-\tau') 
- F^{*f}_{AB \uparrow \downarrow }(\tau'-\tau) {\mathfrak G}_{BA \downarrow \downarrow }(\tau-\tau')
\Big)
  \nonumber \\ & \qquad
+ F^{*bd}_{BB 22}(\tau'-\tau_{2}) \Big( 
F^{*f}_{BB \downarrow \uparrow }(\tau'-\tau) {\mathfrak G}_{BB \uparrow \uparrow }(\tau-\tau') 
- F^{*f}_{BB \uparrow \downarrow }(\tau'-\tau) {\mathfrak G}_{BB \downarrow \downarrow }(\tau-\tau') 
\Big)
  \nonumber \\ & \qquad
+ F^{*bd}_{BB 12}(\tau'-\tau_{2}) \Big(  
- G^{f}_{BB \uparrow \uparrow }(\tau'-\tau) {\mathfrak G}_{BB \uparrow \uparrow }(\tau-\tau') 
- G^{f}_{BB \downarrow \downarrow }(\tau'-\tau) {\mathfrak G}_{BB \downarrow \downarrow }(\tau-\tau') 
\Big)
  \nonumber \\ & \qquad
+ G^{bd}_{BB 22}(\tau'-\tau_{2}) \Big(  
- G^{f}_{BB \uparrow \uparrow }(\tau'-\tau) {\mathfrak F}_{BB \uparrow \downarrow }(\tau-\tau') 
+ G^{f}_{BB \downarrow \downarrow }(\tau'-\tau) {\mathfrak F}_{BB \downarrow \uparrow }(\tau-\tau')
\Big)
  \nonumber \\ & \qquad
+ G^{bd}_{BB 12}(\tau'-\tau_{2}) \Big(  
- F^{*f}_{BB \downarrow \uparrow }(\tau'-\tau) {\mathfrak F}_{BB \uparrow \downarrow }(\tau-\tau') 
- F^{*f}_{BB \uparrow \downarrow }(\tau'-\tau) {\mathfrak F}_{BB \downarrow \uparrow }(\tau-\tau')
\Big)
  \nonumber \\ & \qquad
+ F^{*bd}_{AB 22}(\tau'-\tau_{2}) \Big(  
- F^{*f}_{AB \downarrow \uparrow }(\tau'-\tau) {\mathfrak G}_{BA \uparrow \uparrow }(\tau-\tau') 
+ F^{*f}_{AB \uparrow \downarrow }(\tau'-\tau) {\mathfrak G}_{BA \downarrow \downarrow }(\tau-\tau')
\Big)
  \nonumber \\ & \qquad
+ F^{*bd}_{AB 12}(\tau'-\tau_{2}) \Big(  
- G^{f}_{AB \uparrow \uparrow }(\tau'-\tau){\mathfrak G}_{BA \uparrow \uparrow }(\tau-\tau') 
- G^{f}_{AB \downarrow \downarrow }(\tau'-\tau) {\mathfrak G}_{BA \downarrow \downarrow }(\tau-\tau')
\Big)
  \nonumber \\ & \qquad
+ G^{bd}_{AB 22}(\tau'-\tau_{2}) \Big( 
G^{f}_{AB \uparrow \uparrow }(\tau'-\tau){\mathfrak F}_{BA \uparrow \downarrow }(\tau-\tau') 
- G^{f}_{AB \downarrow \downarrow }(\tau'-\tau) {\mathfrak F}_{BA \downarrow \uparrow }(\tau-\tau') 
\Big)
  \nonumber \\ & \qquad
+ G^{bd}_{AB 12}(\tau'-\tau_{2}) \Big(  
- F^{*f}_{AB \downarrow \uparrow }(\tau'-\tau) {\mathfrak F}_{BA \uparrow \downarrow }(\tau-\tau') 
- F^{*f}_{AB \uparrow \downarrow }(\tau'-\tau) {\mathfrak F}_{BA \downarrow \uparrow }(\tau-\tau') 
\Big)
   \Big]
=\delta(\tau-\tau_{2}).
  \end{align}

\subsection{Electron Green's functions in the slave-boson language}\label{Sec:Gel}

% Green's functions

\begin{align}
 G_{ij \uparrow \uparrow }(\tau-\tau') = &
       +\frac{\epsilon_{i}}{2} F^{*f}_{ij \downarrow \uparrow }(\tau-\tau') \Big[
       - F^{*db}_{ji 22}(\tau'-\tau) 
       + G^{*dd}_{ji 21}(\tau'-\tau) 
       - G^{bb}_{ji 12}(\tau'-\tau)
       + F^{bd}_{ji 11}(\tau'-\tau) \Big]
\nonumber \\ & 
       +\frac{\epsilon_{j}}{2} F^{f}_{ij \uparrow \downarrow}(\tau-\tau') \Big[  
       - F^{bd }_{ji 22}(\tau'-\tau) 
       - G^{bb}_{ji 21}(\tau'-\tau) 
       + G^{*dd}_{ji 12}(\tau'-\tau) 
       + F^{*db}_{ji 11}(\tau'-\tau) \Big]
\nonumber \\ & 
       +\frac{\epsilon_{i} \epsilon_{j}}{2} G^{*f}_{ij \downarrow \downarrow}(\tau-\tau') \Big[ 
       F^{bd}_{ji 21}(\tau'-\tau) 
       + F^{*db}_{ji 12}(\tau'-\tau) 
       - G^{bb}_{ji 22}(\tau'-\tau) 
       - G^{*dd}_{ji 11}(\tau'-\tau) \Big]
\nonumber \\ & 
       +\frac{1}{2} G^{f}_{ij \uparrow \uparrow }(\tau-\tau') \Big[ 
       - F^{*db}_{ji 21}(\tau'-\tau) 
       - F^{bd}_{ji 12}(\tau'-\tau) 
       - G^{*dd}_{ji 22}(\tau'-\tau) 
       - G^{bb}_{ji 11}(\tau'-\tau) \Big],
\end{align}
\begin{align}
   G^{*}_{ij \uparrow \uparrow }(\tau-\tau') = &
       +\frac{\epsilon_{j}}{2} F^{*f}_{ij \uparrow \downarrow}(\tau-\tau') \Big[ 
       - F^{*bd}_{ji 22}(\tau'-\tau)
       - G^{*bb}_{ji 21}(\tau'-\tau)
       + G^{dd}_{ji 12}(\tau'-\tau) 
       + F^{db}_{ji 11}(\tau'-\tau) \big]
\nonumber \\ & 
       + \frac{\epsilon_{i}}{2}  F^{f}_{ij \downarrow \uparrow }(\tau-\tau') \Big[ 
       - F^{db}_{ji 22}(\tau'-\tau)
       + G^{dd}_{ji 21}(\tau'-\tau) 
       - G^{*bb}_{ji 12}(\tau'-\tau)
       + F^{*bd}_{ji 11 }(\tau'-\tau) \Big]
\nonumber \\ & 
       + \frac{1}{2}G^{*f}_{ij \uparrow \uparrow }(\tau-\tau') \Big[ 
       - F^{db}_{ji 21}(\tau'-\tau)
       - F^{*bd}_{ji 12}(\tau'-\tau) 
       - G^{dd}_{ji 22}(\tau'-\tau) 
       - G^{*bb}_{ji 11}(\tau'-\tau) \Big]
\nonumber \\ & 
       + \frac{\epsilon_{i} \epsilon_{j}}{2} G^{f}_{ij \downarrow \downarrow}(\tau-\tau') \Big[ 
       F^{*bd}_{ji 21}(\tau'-\tau)
       + F^{db}_{ji 12}(\tau'-\tau)
       - G^{*bb}_{ji 22}(\tau'-\tau)
       - G^{dd}_{ji 11}(\tau'-\tau) \Big],
\end{align}
\begin{align}
   G_{ij \downarrow \downarrow}(\tau-\tau') = &
       +\frac{\epsilon_{i}}{2}  F^{*f}_{ij \uparrow \downarrow}(\tau-\tau') \Big[
       F^{*db}_{ji 22}(\tau'-\tau)
       - G^{*dd}_{ji 21}(\tau'-\tau)
       + G^{bb}_{ji 12}(\tau'-\tau)
       - F^{bd}_{ji 11}(\tau'-\tau)
       \Big]
\nonumber \\ & 
       +\frac{\epsilon_{j}}{2}  F^{f}_{ij \downarrow \uparrow }(\tau-\tau') \Big[ 
       F^{bd }_{ji 22}(\tau'-\tau)
       + G^{bb}_{ji 21}(\tau'-\tau)
       - G^{*dd}_{ji 12}(\tau'-\tau)
       - F^{*db}_{ji 11}(\tau'-\tau)
       \Big]
\nonumber \\ & 
       +\frac{\epsilon_{i} \epsilon_{j}}{2}  G^{*f}_{ij \uparrow \uparrow }(\tau-\tau') \Big[ 
       F^{bd}_{ji 21}(\tau'-\tau)
       + F^{*db}_{ji 12}(\tau'-\tau)
       - G^{bb}_{ji 22}(\tau'-\tau)
       - G^{*dd}_{ji 11}(\tau'-\tau)
       \Big]
\nonumber \\ & 
       +\frac{1}{2} G^{f}_{ij \downarrow \downarrow}(\tau-\tau') \Big[  
       - F^{*db}_{ji 21}(\tau'-\tau)
       - F^{bd}_{ji 12}(\tau'-\tau) 
       - G^{*dd}_{ji 22}(\tau'-\tau) 
       - G^{bb}_{ji 11}(\tau'-\tau) 
       \Big],
\end{align}
\begin{align}
   G^{*}_{ij \downarrow \downarrow}(\tau-\tau') = &
       + \frac{\epsilon_{j}}{2} F^{*f}_{ij \downarrow \uparrow }(\tau-\tau') \Big[
       F^{*bd}_{ji 22}(\tau'-\tau)
       + G^{*bb}_{ji 21}(\tau'-\tau)
       -G^{dd}_{ji 12}(\tau'-\tau)
       - F^{db}_{ji 11}(\tau'-\tau)
       \Big]
\nonumber \\ & 
       +  \frac{\epsilon_{i}}{2}  F^{f}_{ij \uparrow \downarrow}(\tau-\tau') \Big[ 
       F^{db}_{ji 22}(\tau'-\tau)
       - G^{dd}_{ji 21}(\tau'-\tau)
       + G^{*bb}_{ji 12}(\tau'-\tau)
       - F^{*bd}_{ji 11 }(\tau'-\tau)
       \Big]
\nonumber \\ & 
       +\frac{1}{2} G^{*f}_{ij \downarrow \downarrow}(\tau-\tau') \Big[ 
       - F^{db}_{ji 21}(\tau'-\tau)
       - F^{*bd}_{ji 12}(\tau'-\tau) 
       - G^{dd}_{ji 22}(\tau'-\tau) 
       - G^{*bb}_{ji 11}(\tau'-\tau) 
       \Big]
\nonumber \\ & 
       +\frac{\epsilon_{i} \epsilon_{j}}{2} G^{f}_{ij \uparrow \uparrow }(\tau-\tau') \Big[ 
       F^{*bd}_{ji 21}(\tau'-\tau)
       + F^{db}_{ji 12}(\tau'-\tau)
       - G^{*bb}_{ji 22}(\tau'-\tau)
       - G^{dd}_{ji 11}(\tau'-\tau)
       \Big],
\end{align}
\begin{align}
   F_{ij \uparrow \downarrow}(\tau-\tau') = &
       +\frac{\epsilon_{i} \epsilon_{j}}{2} F^{*f}_{ij \downarrow \uparrow }(\tau-\tau') \Big[
       F^{*bb}_{ji 22}(\tau'-\tau)
       - G^{*bd}_{ji 21}(\tau'-\tau)
       - G^{db}_{ji 12}(\tau'-\tau)
       + F^{dd}_{ji 11}(\tau'-\tau)
       \Big]
\nonumber \\ & 
       + \frac{1}{2} F^{f}_{ij \uparrow \downarrow}(\tau-\tau') \Big[ 
       - F^{dd}_{ji 22}(\tau'-\tau)
       - G^{db}_{ji 21}(\tau'-\tau)
       - G^{*bd}_{ji 12}(\tau'-\tau)
       - F^{*bb}_{ji 11}(\tau'-\tau)
       \Big]
\nonumber \\ & 
       + \frac{\epsilon_{i}}{2} G^{*f}_{ij \downarrow \downarrow}(\tau-\tau') \Big[ 
       F^{dd}_{ji 21}(\tau'-\tau)
       - F^{*bb}_{ji 12}(\tau'-\tau)
       - G^{db}_{ji 22}(\tau'-\tau) 
       + G^{*bd}_{ji 11}(\tau'-\tau)
       \Big]
\nonumber \\ & 
       + \frac{\epsilon_{j}}{2}  G^{f}_{ij \uparrow \uparrow }(\tau-\tau') \Big[ 
       F^{*bb}_{ji 21}(\tau'-\tau)
       - F^{dd}_{ji 12}(\tau-\tau')
       + G^{*bd}_{ji 22}(\tau'-\tau)
       - G^{db}_{ji 11}(\tau'-\tau)
       \Big],
\end{align}
\begin{align}
   F^{*}_{ij \uparrow \downarrow}(\tau-\tau') = &
       + \frac{1}{2} F^{*f}_{ij \uparrow \downarrow}(\tau-\tau') \Big[
       - F^{*dd}_{ji 22}(\tau'-\tau)
       - G^{*db}_{ji 21}(\tau'-\tau)
       - G^{bd}_{ji 12}(\tau'-\tau)
       - F^{bb}_{ji 11}(\tau'-\tau)
\nonumber \\ & 
       +\frac{\epsilon_{i}  \epsilon_{j}}{2} F^{f}_{ij \downarrow \uparrow }(\tau-\tau') \Big[ 
       F^{bb}_{ji 22}(\tau'-\tau)
       - G^{bd}_{ji 21}(\tau'-\tau)
       - G^{*db}_{ji 12}(\tau'-\tau)
       + F^{*dd}_{ji 11}(\tau'-\tau)
       \Big]
\nonumber \\ & 
       +\frac{\epsilon_{j}}{2} G^{*f}_{ij \uparrow \uparrow }(\tau-\tau') \Big[ 
       F^{bb}_{ji 21}(\tau'-\tau)
       - F^{*dd}_{ji 12}(\tau'-\tau)
       + G^{bd}_{ji 22}(\tau'-\tau)
       - G^{*db}_{ji 11}(\tau'-\tau) 
       \Big]
\nonumber \\ & 
       + \frac{\epsilon_{i}}{2}  G^{f}_{ij \downarrow \downarrow}(\tau-\tau') \Big[ 
       F^{*dd}_{ji 21}(\tau'-\tau)
       - F^{bb}_{ji 12}(\tau'-\tau)
       - G^{*db}_{ji 22}(\tau'-\tau)
       + G^{bd}_{ji 11}(\tau'-\tau)
       \Big],
\end{align}
\begin{align}
   F_{ij \downarrow \uparrow }(\tau-\tau') = &
       + \frac{\epsilon_{i} \epsilon_{j}}{2} F^{*f}_{ij \uparrow \downarrow}(\tau-\tau') \Big[ 
       F^{*bb}_{ji 22}(\tau'-\tau)
       - G^{*bd}_{ji 21}(\tau'-\tau)
       - G^{db}_{ji 12}(\tau'-\tau)
       + F^{dd}_{ji 11}(\tau'-\tau)
       \Big]
\nonumber \\ & 
       + \frac{1}{2} F^{f}_{ij \downarrow \uparrow }(\tau-\tau') \Big[ 
       - F^{dd}_{ji 22}(\tau'-\tau)
       - G^{db}_{ji 21}(\tau'-\tau)
       - G^{*bd}_{ji 12}(\tau'-\tau)
       - F^{*bb}_{ji 11}(\tau'-\tau)
       \Big]
\nonumber \\ & 
       + \frac{\epsilon_{i}}{2}  G^{*f}_{ij \uparrow \uparrow }(\tau-\tau') \Big[ 
       - F^{dd}_{ji 21}(\tau'-\tau)
       + F^{*bb}_{ji 12}(\tau'-\tau)
       + G^{db}_{ji 22}(\tau'-\tau)
       - G^{*bd}_{ji 11}(\tau'-\tau)
       \Big]
\nonumber \\ & 
       + \frac{\epsilon_{j}}{2}  G^{f}_{ij \downarrow \downarrow}(\tau-\tau') \Big[ 
       - F^{*bb}_{ji 21}(\tau'-\tau)
       + F^{dd}_{ji 12}(\tau-\tau')
       - G^{*bd}_{ji 22}(\tau'-\tau)
       + G^{db}_{ji 11}(\tau'-\tau)
       \Big],
\end{align}
\begin{align}
   F^{*}_{ij \downarrow \uparrow }(\tau-\tau') = &
       + \frac{1}{2} F^{*f}_{ij \downarrow \uparrow }(\tau-\tau') \Big[
       - F^{*dd}_{ji 22}(\tau'-\tau)
       - G^{*db}_{ji 21}(\tau'-\tau)
       - G^{bd}_{ji 12}(\tau'-\tau)
       - F^{bb}_{ji 11}(\tau'-\tau)
       \Big]
\nonumber \\ & 
       + \frac{\epsilon_{i} \epsilon_{j}}{2} F^{f}_{ij \uparrow \downarrow}(\tau-\tau') \Big[ 
       F^{bb}_{ji 22}(\tau'-\tau)
       - G^{bd}_{ji 21}(\tau'-\tau)
       - G^{*db}_{ji 12}(\tau'-\tau)
       + F^{*dd}_{ji 11}(\tau'-\tau)
       \Big]
\nonumber \\ & 
       +\frac{\epsilon_{j}}{2}  G^{*f}_{ij \downarrow \downarrow}(\tau-\tau') \Big[ 
       - F^{bb}_{ji 21}(\tau'-\tau)
       + F^{*dd}_{ji 12}(\tau'-\tau)
       - G^{bd}_{ji 22}(\tau'-\tau)
       + G^{*db}_{ji 11}(\tau'-\tau)
       \Big]
\nonumber \\ & 
       +\frac{\epsilon_{i}}{2} G^{f}_{ij \uparrow \uparrow }(\tau-\tau') \Big[ 
       - F^{*dd}_{ji 21}(\tau'-\tau)
       + F^{bb}_{ji 12}(\tau'-\tau)
       + G^{*db}_{ji 22}(\tau'-\tau)
       - G^{bd}_{ji 11}(\tau'-\tau)
       \Big].
\end{align}

\twocolumngrid

\end{document}